\documentclass[iop,revtex4]{emulateapj}
\usepackage{times}
\usepackage[usenames,dvipsnames]{color}
\usepackage{graphicx}
\usepackage{xspace}
\usepackage{rotating}
\usepackage{lscape}
 \usepackage{amsmath, amsthm, amssymb}
\usepackage{romannum}
\usepackage[inline]{enumitem}
\usepackage{lipsum}

\newcommand{\hii}{${\rm H\textsc{ii}}$}

\renewcommand{\thefootnote}{\fnsymbol{footnote}}

\newcommand{\agem}{$\langle {\rm age} \rangle_{\rm mass}$}

\newcommand{\agelg}{$\langle {\rm age} \rangle_{\lambda,{\rm long}}$}
\newcommand{\agesh}{$\langle {\rm age} \rangle_{\lambda,{\rm short}}$}
\newcommand{\agelam}{$\langle {\rm age} \rangle_{\rm \lambda}$}
\newcommand{\ageg}{$\langle {\rm age} \rangle_{g}$}
\newcommand{\agev}{$\langle {\rm age} \rangle_{V}$}

\newcommand{\logagem}{$\log \langle {\rm age} \rangle_{\rm mass}$}

\newcommand{\logagelg}{$\log \langle {\rm age} \rangle_{\lambda,{\rm long}}$}
\newcommand{\logagesh}{$\log \langle {\rm age} \rangle_{\lambda,{\rm short}}$}

\newcommand{\tma}{t$_{\rm m, 20\%}$}
\newcommand{\tmb}{t$_{\rm m, 80\%}$}

\newcommand{\mhm}{$\langle [\rm M/H]\rangle_{\rm mass}$}

\newcommand{\mhlg}{$\langle {\rm [M/H]} \rangle_{\lambda,{\rm long}}$}
\newcommand{\mhsh}{$\langle {\rm [M/H]} \rangle_{\lambda,{\rm short}}$}
\newcommand{\mhv}{$\langle {\rm [M/H]} \rangle_{V}$}
\newcommand{\mhg}{$\langle {\rm [M/H]} \rangle_{g}$}
\newcommand{\mhlam}{$\langle {\rm [M/H]} \rangle_{\rm \lambda}$}

\newcommand{\clrmtl}{color--$\Upsilon_{\star}$}
\newcommand{\lclrmtl}{color--$\log \Upsilon_{\star}$}

\slugcomment{To appear in The Astrophysical Journal Supplement Series}

\shorttitle{The impact of star formation histories on stellar mass estimation}
\shortauthors{Zhang, Puzia \& Weisz}

\begin{document}

\title{The impact of star formation histories on stellar mass estimation: implications from the Local Group dwarf galaxies}

\author{
Hong-Xin Zhang$^{1,2,3}$, Thomas H. Puzia$^{1}$, Daniel R. Weisz$^{4}$
}
\affil{$^1$ Institute of Astrophysics, Pontificia Universidad Cat\'olica de Chile, Av.~Vicu\~na Mackenna 4860, 7820436 Macul, Santiago, Chile; hzhang@astro.puc.cl, tpuzia@astro.puc.cl}
\affil{$^2$ National Astronomical Observatories, Chinese Academy of Sciences, Beijing 100012, China}
\affil{$^3$ China-Chile Joint Center for Astronomy, Camino El Observatorio 1515, Las Condes, Santiago, Chile}
\affil{$^4$ Department of Astronomy, University of California Berkeley, Berkeley, CA 94720, USA; dan.weisz@berkeley.edu}

\begin{abstract}
Local Group (LG) galaxies have relatively accurate star formation histories (SFHs) and metallicity evolution derived from resolved color-magnitude 
diagram (CMD) modeling, and thus offer a unique opportunity to explore the efficacy of estimating stellar mass $\mathcal{M_{\star}}$ of real 
galaxies based on the integrated stellar luminosities.\ Building on the SFHs and metallicity evolution of 40 LG dwarf galaxies, we carried 
out a comprehensive study of the influence of SFHs, metallicity evolution, and dust extinction on the UV-to-NIR color--mass-to-light-ratio 
(\lclrmtl($\lambda$)) distributions and $\mathcal{M_{\star}}$ estimation of local universe galaxies.\ We find that: 1) The LG galaxies follow 
\lclrmtl($\lambda$) relations that fall in between the ones calibrated by previous studies; 2) Optical \lclrmtl($\lambda$) relations at higher 
[M/H] are generally broader and steeper; 3) The SFH shape parameter ``concentration'' does not significantly affect the \lclrmtl($\lambda$) 
relations; 4) Light-weighted ages \agelam~and metallicities \mhlam~of galaxies are the closest analogs to ages and metallicities of single 
stellar populations, and the two together constrain $\log\Upsilon_{\star}$($\lambda$) with uncertainties ranging from $\lesssim$ 0.1 dex for 
the NIR up to 0.2 dex for the optical passbands; 5) Metallicity evolution induces significant uncertainties to the optical but not NIR 
$\Upsilon_{\star}$($\lambda$) at given \agelam~and \mhlam; 6) The $V$ band is the ideal luminance passband for estimating 
$\Upsilon_{\star}$($\lambda$) from single colors, because the combinations of $\Upsilon_{\star}$($V$) and optical colors such as $B$$-$$V$ 
and $g$$-$$r$ exhibit the weakest systematic dependences on SFHs, metallicities and dust extinction; 7) Without any prior assumption 
on SFHs, $\mathcal{M_{\star}}$ is constrained with biases $\lesssim$ 0.3 dex by the UV- or optical-to-NIR SED fitting.\ Optical passbands 
alone constrain $\mathcal{M_{\star}}$ with biases $\lesssim$ 0.4 dex (or $\lesssim$ 0.6 dex) when dust extinction is fixed (or variable) in SED fitting.\ 
SED fitting with monometallic SFH models tends to underestimate $\mathcal{M_{\star}}$ of real galaxies.\ $\mathcal{M_{\star}}$ tends to 
be overestimated (or underestimated) at the youngest (or oldest) \agem.

%\lipsum[1-2]

\end{abstract}


\section{Introduction}
Stellar mass is among the most fundamental properties of a galaxy, and it traces the galaxy formation and evolution.\ 
Stellar masses of galaxies are typically estimated from the integrated stellar luminosities by adopting appropriate stellar mass-to-light 
ratios, $\Upsilon_{\star}$, which in turn depend on the stellar initial mass function (IMF), stellar evolution/atmosphere models, 
dust extinction, star formation histories (SFHs), and the accompanied metallicity evolution.\

In the context of given stellar evolution/atmosphere models and IMFs, $\Upsilon_{\star}$ of galaxies has 
been estimated based on single broadband colors (pioneered by Bell \& de Jong 2001), modeling of multiwavelength 
broadband spectral energy distributions \citep[SEDs; e.g.][]{sawicki98, brinchmann00}, modeling of 
spectral indices \citep[e.g.][]{kauffmann03} or full moderate- to high-resolution spectra \citep[e.g.][]{heavens00, cid05, panter07}.\ 
All of these methods involve comparison of observations with predictions from evolutionary population synthesis modeling 
of galaxies' SFHs.\ Unfortunately, solving for SFHs based on integrated stellar light is an ill-posed and often ill-conditioned 
inverse problem, in the sense that the solution is generally not unique, partly due to various degeneracies between ages, 
metallicities, dust extinctions, etc.\ \citep[e.g.][]{worthey94}, and that the solution might be very sensitive to noise in the 
observations \citep[e.g.][]{moultaka00, ocvirk06a}.\ $\Upsilon_{\star}$ varies with stellar ages and metallicities.\ 
A small fraction (by mass) of young stars can easily outshine old stars and dominantly contribute to the emergent spectrum 
from a composite stellar population.\ In extreme cases, the light from mass-dominant old stellar populations might be entirely 
``lost'' in the glare of recent bursts of star formation.\ The effort of recovering SFHs and $\Upsilon_{\star}$ is further complicated 
by dust extinction and its {\it uncertain} dependence on wavelength and the age-dependent dust-to-star geometry of galaxies 
\citep[e.g.][]{calzetti94, wild11, reddy15, battisti16}.

With the aforementioned difficulties in robustly constraining SFHs and consequently the potential influence on stellar mass estimation, 
a majority of previous studies, however, suggest that stellar mass appears to be a relatively robust parameter that can be extracted from 
population synthesis modeling of galaxies \citep[e.g.][among many others]{brinchmann00, bell01, papovich01, kauffmann03, borch06, pozzetti07, zibetti09, zhao11, conroy13, maraston13, moustakas13, mobasher15, salim16}.\ This might be attributed to the fact that the various physical parameters, such as age, metallicity, and dust extinction, partially counteract their effects on $\Upsilon_{\star}$.\ Moreover, the observationally ``cheap'' methods of using single broadband colors or multiwavelength broadband SED fitting appear to perform, in nearly all cases, as well (or poor) as fitting to spectroscopic data (i.e.\ spectral indices or full spectrum) when it comes to constraining the $\Upsilon_{\star}$ \citep[e.g.][]{gallazzi09, chen12}.\ Being limited by observing efficiency and sensitivities, spectroscopic information is generally available only for the brightest regions in galaxies and, thus, is not a feasible and optimal choice for a robust stellar mass estimation of galaxies in general.\ With the availability of high-quality multi-wavelength broadband photometric data of large samples of galaxies that are or will be available from various large-area sky surveys, such as the Galaxy Evolution Explorer ($GALEX$) All-Sky Imaging Survey \citep[][]{martin05}, the Sloan Digital Sky Survey \citep[SDSS;][]{york00}, the Two Micron All Sky Survey \citep[2MASS;][]{jarrett00}, the Wide-field Infrared Survey Explorer ($WISE$) mission \citep[][]{wright10}, the Dark Energy Survey \citep[DES;][]{flaugher05}, and the upcoming survey by the Large Synoptic Survey Telescope \citep[LSST;][]{ivezic08}, it has become a standard practice to estimate stellar masses of galaxies based on evolutionary population synthesis (EPS) modeling of broadband photometry.\

While the apparently robust estimation of $\Upsilon_{\star}$ is encouraging, one must be cautioned that a vast 
majority of previous work on stellar mass estimation is based on comparison of observations with predictions from 
template model SFHs that are parameterized by some simple functional (e.g.\ exponentially declining) forms, sometimes 
modulated by randomly added burst events.\ An exponentially declining form of SFHs may reasonably describe 
the cosmic average SFH over a major part of the Hubble time \citep[e.g.][]{madau14}, but it is not necessarily the 
case for individual galaxies or even for specific types of galaxies, which evolve on shorter timescales.\ The uncertainties and biases of stellar mass estimation 
caused by applying the simplified parameterization of SFHs to real galaxies are rarely taken into account.\ Studies of very nearby 
dwarf galaxies, for which relatively accurate SFHs can be recovered through modeling of color-magnitude diagrams (CMD) of 
resolved stars, suggest a large variety of SFHs that are often distinct from the cosmic average SFHs and cannot be described by 
simple SFH models, such as the family of exponentially declining SFHs \citep[e.g.][]{weisz11}.\

By comparing the stellar mass estimates of a large sample of observed disk galaxies with various linear \clrmtl($\lambda$) calibrations 
in the literature, McGaugh \& Schombert (2014) found that many commonly employed calibrations fail to provide self-consistent mass 
estimates for their galaxies when different luminance passbands are used.\ The authors attributed the inconsistencies to a larger 
influence of longer-wavelength passbands by the uncertain treatment of the thermally pulsating asymptotic giant branch (TP-AGB) phase 
of stellar evolution in stellar population models \citep[e.g.][]{baldwin17}.\ We will explore the origin of the inconsistencies in this paper.\ 
The finding by McGaugh \& Schombert (2014) suggests that the color-based stellar-mass estimates can be subject to remarkable systematic 
uncertainties if a single set of \clrmtl($\lambda$) relations were blindly applied to different types of galaxies.

Several previous studies have attempted to address the robustness of stellar mass estimation by applying the typical procedure of SED fitting to 
synthetic photometry of mock galaxies drawn from semianalytic or hydrodynamical galaxy formation simulations \citep{lee09a, wuyts09, 
pforr12, pforr13, mitchell13, michalowski14, hayward15}.\ These studies generally found that a mismatch between the assumed 
and true SFHs can lead to substantial biases and scatters in stellar mass estimation, especially for galaxies with bursty SFHs.\ 
In particular, \cite{pforr12} found that the standard SED fitting with simplified forms of template SFHs can severely underestimate 
(by up to 0.6 dex) the stellar masses of simulated star-forming galaxies at low redshift, largely due to the tendency of SED fitting 
to underestimate the ages.\ \cite{mitchell13} noted that the effects of dust extinction can result in an underestimation of stellar 
masses by up to 0.6 dex.\ These studies are enlightening for understanding the efficacy of SED fitting, but the concern is that the SFHs 
of simulated galaxies may not resemble that of real galaxies at all.\ State-of-the-art cosmological hydrodynamical 
simulations allow the simulation of individual galaxies down to subkiloparsec scales \citep[e.g.][]{vogelsberger14, schaye15}, 
but some crucial small-scale processes related to star formation and feedback are largely 
implemented in a semiempirical way, which limits the predictive power of these galaxy formation simulations 
\citep{naab17}.\

It is desirable to investigate the robustness of stellar mass estimation by resorting to real galaxies with known and temporally resolved SFHs.\
However, as mentioned above, it is generally very difficult to reliably constrain the SFHs of spatially unresolved galaxies, even 
with full-spectrum SED modeling.\ The only exceptions at present are the Local Group (LG) galaxies that are close enough that high-quality 
CMDs of resolved stars can be used to recover the SFHs throughout the galaxy's lifetime \citep[e.g.][]{tolstoy09}.\ In the present work, we take 
as input the CMD-based SFHs of LG dwarf galaxies to study the \lclrmtl~distributions followed by these galaxies and explore the efficacy of 
nonparametric broadband SED fitting in recovering stellar masses.\

The work presented in this paper is carried out under the framework of state-of-the-art single stellar population (SSP) synthesis models.\ 
Nevertheless, one should keep in mind that the current population synthesis models themselves are subject to various degrees of 
uncertainties at different stellar evolution phases.\ Notable examples of the uncertain phases or input physics of stellar evolution 
include the treatment of mass-loss of the horizontal branch (HB) phases \citep[e.g.][]{rich97} which can have a significant contribution 
to the UV to optical emission at old age, the core overshooting problem of intermediate-mass main-sequence stars which dominate 
the UV to optical emission at young to intermediate ages (e.g.\ Pietrinferni et al.\ 2004), the uncertain treatment of many key processes 
(e.g.\ Marigo \& Girardi 2007) of the TP-AGB phases, which can dominate the NIR emission at ages ranging from several hundred Myr to 
2 Gyr \citep[e.g.][]{maraston06}; and the uncertain treatment of convective fluxes of the red giant branch (RGB) phases, which dominate the 
NIR emission for a major part of the lifetime of a stellar system.\ These uncertainties in stellar evolution models can result in quite significant 
uncertainties (by up to a factor of a few in extreme cases) to the physical parameters (e.g.\ ages, stellar masses) inferred from stellar population 
synthesis models \citep[e.g.][]{conroy09, powalka17}.\ Dealing with these uncertainties is well beyond of the scope of the current paper.\

This paper is organized as follows.~We introduce the sample of LG galaxies with SFHs determined from resolved 
CMD modeling, the samples with expanded parameter coverages, and the input ingredients for generating integrated 
broadband SEDs from the SFHs with associated metallicity evolution in Section \ref{input}.\ A nonparametric 
SED fitting method which involves matrix inversion with non-negative least-squares (NNLS) optimization is introduced in 
Section \ref{sec: methodsedfitting}.\ Sections \ref{sec: clrm2l} to \ref{sec: sedfitting} present the main results of this 
paper, including the influence of SFHs, metallicities, metallicity evolution, and dust extinction on the multiband 
\lclrmtl($\lambda$) distributions of our sample galaxies (Section \ref{sec: clrm2l}); the efficacy of the observationally 
accessible light-weighted ages and metallicities in constraining $\Upsilon_{\star}$($\lambda$) and the optimal passband 
combinations for $\Upsilon_{\star}$($\lambda$) estimation (Section \ref{sec: m2l_agezl}); and the efficacy and limitation of 
broadband SED fitting in recovering stellar masses of galaxies (Section \ref{sec: sedfitting}).\ Section \ref{sec: discuss} is 
devoted to a discussion of the discrepancies between various linear \lclrmtl($\lambda$) relations in the literature.\ A summary 
of this paper is given in Section \ref{sec: summary}.

\section{The Input: The Sample and Star Formation Histories}\label{input}
\subsection{SFHs of Local Group Dwarf Galaxies}\label{data}
Within the zero-velocity surface of the LG ($D$ $\approx$ 1.18$\pm$0.15 Mpc; van den Bergh 1999), 
there are 73 known dwarf galaxies (Karachentsev et al.\ 2013).\ Weisz et al.\ (2014; hereafter W14) 
determined SFHs and self-consistent metallicity evolution histories with CMD fitting of 40 of the LG galaxies 
which have deep multiband imaging from the {\it Hubble Space Telescope} ($HST$)/Wide Field Planetary Camera 2 
(WFPC2; Holtzman et al.\ 1995).\ The subsample of 40 galaxies include 27 dwarf spheroidal/ellipticals 
(dSphs/dEs), 5 so-called transition dwarfs (dTrans), and 8 Magellanic dwarf irregulars (dIrrs).\ 
Three-quarters of the galaxies analyzed by W14 were observed with single WFPC2 pointings ($\sim$ L-shaped 
field of view approximated by a 2$\farcm$5 $\times$ 2$\farcm$5 square with one quadrant absent) toward 
the galactic centers, and another one-quarter with multiple pointings.\ The WFPC2 pointings cover a typical 
fractional area of $\sim$ 20\%--30\%, with respect to the area enclosed by one half-light radius of individual galaxies.\

Among the galaxies in the W14 sample, 35\% have deep CMDs that reach the main-sequence turnoffs 
(MSTOs) as old as $\gtrsim$ 12 Gyr, and thus provide relatively precise SFHs throughout the lifetime 
of galaxies.\ For a majority of the remaining two-thirds of galaxies, the CMDs do not reach the ancient MSTOs 
but do reach post-main-sequence phases that are unambiguous probes (e.g.\ blue HBs) of ancient SF.\ 
As claimed by W14, accurate albeit less precise SFHs throughout the lifetime of galaxies may still be 
measured with these relatively shallow WFPC2 observations.\ Of the galaxies without deep WFPC2 
observations, three (Leo A, DDO 210, and LGS 3) have precise SFHs measured (Leo A, Cole et al.\ 2007; 
DDO 210, Cole et al.\ 2014; LGS 3, Hidalgo et al.\ 2011) based on deep CMDs from the $HST$/Advanced 
Camera for Surveys (ACS; Ford et al.\ 1998).\ We use the ACS-based SFHs and metallicity evolution histories 
for the three galaxies.\ 

\subsection{Integrated Broadband SEDs Predicted by the CMD-based SFHs and Metallicity Evolution Histories}
\subsubsection{SSP Models}
The integrated light of composite stellar populations at a given wavelength is essentially a convolution between the temporal evolution 
of SSPs and SFHs.\ As the starting point for constructing integrated SEDs for given SFHs and metallicity evolution histories, there exist  
a variety of SSP models in the literature.\ The popular ones include PEGASE (Fioc \& Rocca-Volmerange 1997), Starburst99 (Leitherer et al.\ 1999), 
BC03 (Bruzual \& Charlot 2003), Maraston (2005), GALEV (Kotulla et al.\ 2009), FSPS (Conroy et al.\ 2009), and BPASS (Eldridge \& Stanway 2009), 
just to name a few.\ Nearly all of these SSP models implement the Padova isochrones, either Padova 1994 (Alongi et al.\ 1993; Bressan et al.\ 1993; 
Fagotto et al.\ 1994a,b; Girardi et al.\ 1996) or Padova 2008 (Girardi et al.\ 2000; Marigo \& Girardi 2007; Marigo et al.\ 2008), as their default input 
isochrones.\ SSP models from different groups are different mainly in the treatment of poorly understood post-main-sequence phases (e.g.\ TP-AGB, 
HB) and stellar spectral libraries.\ The reader is referred to Conroy \& Gunn (2010) and \cite{powalka17} for a recent intercomparison of some popular 
SSP model predictions.\ 

In this work, we adopt the FSPS SSP models (v2.6, 11/17/2015) with the BaSeL spectral library as our fiducial models 
because FSPS uses the Padova 2008 isochrones, which are calculated for a much finer grid of metallicities 
as compared to Padova 1994, and FSPS also includes a more up-to-date empirical calibration of the TP-AGB 
phases as compared to the other SSP models.\ A finer grid of metallicities for SSP models is particularly 
preferred in this work since we take into account the metallicity evolution histories.\ It is worth mentioning that 
both the FSPS models and the classic BC03 models favorably match the optical and NIR colors of star 
clusters in the Magellanic Clouds and post-starburst galaxies (Conroy \& Gunn 2010), which is particularly 
desirable for our current work on low-metallicity dwarf galaxies.\ In addition, the W14 SFHs of the LG dwarfs 
were based on about the same Padova 2008 isochrones as implemented in FSPS.\

\subsubsection{Stellar IMF}\label{sec: imfs}
We adopt the lognormal Chabrier (2003) IMF, which is extremely similar to the broken power-law Kroupa (2001) 
IMF (e.g.\ see Figure 8 in Dabringhausen, Hilker \& Kroupa 2008) used in the CMD modeling of W14.\ 
There has been a growing body of evidence indicating that stellar IMFs might vary with environment or galaxy mass.\ 
In particular, based on star counts of two LG ultrafaint dwarf galaxies (Hercules and Leo IV), Geha et al.\ (2013) claim that 
low-mass dwarf galaxies may have IMFs that are more bottom-light than the Kroupa/Chabrier IMFs, in line with the general trend 
found for massive ETGs (e.g.\ van Dokkum \& Conroy 2012).\ If the IMF varies solely at the faint low-mass end (i.e.\ below 
the oldest MSTO), the integrated luminosities and colors of a given SFH would not be affected significantly, 
but $\Upsilon_{\star}$ at given colors would be subject to a systematic shift.\ If the IMF varies at the usually light-dominant intermediate-mass 
ranges, the integrated luminosities, colors, and $\Upsilon_{\star}$ would all be affected significantly.\ If the IMF varies solely at the 
rapidly evolving high-mass end (say, $\gtrsim$ 8 M$_{\odot}$), the integrated luminosities and colors may appear unaffected for a major 
part of the SFH, but the $\Upsilon_{\star}$ at given colors can be subject to systematic shifts owing to the mass contribution from faint 
stellar remnants.

Nevertheless, stricter investigations on constraining IMFs with direct star counts (e.g.\ El-Badry et al.\ 2017) suggest that the 
Geha et al.\ results are subject to large uncertainties owing to the very narrow stellar mass range studied by that team.\ In fact, 
Wyse et al.\ (2002) concluded that the IMF slope of the Ursa Minor dwarf spheroidal was consistent with that of the Kroupa/Chabrier IMF.\ 
In addition, the findings that more massive ETGs appear to have more bottom-heavy IMFs (e.g.\ van Dokkum \& Conroy 2012) based 
on modeling gravity-sensitive NIR absorption features are usually subject to degeneracies with intrinsic elemental abundance variations 
(e.g.\ McConnell et al.\ 2016), uncertainties on stellar atmosphere models at nonsolar metallicities, and SFHs (e.g.\ Offner 2016).\ 
Some studies even identified massive ETGs with IMFs similar to the Kroupa/Chabrier-like IMFs (e.g.\ Smith et al.\ 2015).\ Even if the bottom-heavy 
IMFs (at stellar masses $<$ 0.5 -- 1 M$_{\odot}$) found for massive ETGs are real, it appears not to apply to the high-mass end of the 
IMFs, which are found to be consistent with the canonical Kroupa/Chabrier IMFs (e.g.\ Peacock et al.\ 2014).

\subsubsection{Dust Extinction}\label{sec: secext}
Following W14, we include an age-dependent differential extinction recipe, which was first introduced in Dolphin et al.\ (2003) in order 
to properly fit the CMDs of LG dwarf galaxies.\ In particular, stars at ages $\leq$ 40 Myr have $V$-band extinction $A_{V, {\rm young}}$ 
fixed to 0.5 mag, while stars older than 40 Myr have an extinction distribution that linearly decreases with age from 0.5 to 0.0 mag at ages 
$\geq$ 100 Myr.\ This dust extinction recipe is physically plausible.\ The relatively high dust extinction at the youngest ages is qualitatively 
in line with a finite lifetime of the dense star-forming clouds, whereas the gradually diminishing dust extinction at older ages is in qualitative 
agreement with the two general observations of disk galaxies, i.e.\ distributions of older stars are vertically more extended than younger stars 
and the distributions of diffuse interstellar dust have a steeper vertical falloff than that of stars (e.g.\ Xilouris et al.\ 1999; Yoachim \& Dalcanton 2006).\ 
In addition, we note that an $A_{V,{\rm young}}$ of 0.5 mag is about consistent with the median reddening of \hii~regions in nearby dwarf irregular 
galaxies (e.g.\ Hunter \& Hoffman 1999), for the average SMC-bar extinction curve (see below).\

For the wavelength dependence of dust extinction, as the default in this work we adopt the average extinction curve ($R_{V}$ = $A_{V}/E$($B$$-$$V$) 
$\simeq$ 2.74) determined for the star-forming bar of the Small Magellanic Cloud (SMC; Gordon et al.\ 2003).\ The SMC-bar extinction curve is 
different from that of the Large Magellanic Cloud and the Milky Way, mainly in the UV wavelength range, where the SMC-bar extinction curve has a 
much weaker 2175\AA~bump and a steeper far-UV (FUV) rise.\ The SMC-bar extinction curve has a steeper wavelength dependence than the famous 
Calzetti et al.\ (1994) dust extinction curve that was calibrated for nearby (unresolved) starburst galaxies.\ Several recent studies suggest that 
the slopes of the effective extinction curves might vary with star formation activities and dust attenuation optical depths, in the sense that 
galaxies with more quiescent star formation and/or smaller dust attenuation optical depth appear to have steeper extinction curves (e.g.\ 
Wuyts et al.\ 2011; Chevallard et al.\ 2013).\ Particularly, through radiative transfer modeling of the light emission from disk galaxies, 
Chevallard et al.\ (2013) show that the effective dust attenuation curves become steeper than the Calzetti et al.\ (1994) curve at attenuation 
optical depths $\tau_{V}$ $\lesssim$ 0.5 mag.\ Moreover, relatively steep dust extinction curves for low-metallicity dwarf galaxies such as the 
SMC might be also expected if these galaxies have a dust grain size distributions that are more abundant in small grains, possibly due to a 
stronger interstellar radiation field.\ We note that the average SMC-bar extinction curve was also used by Johnson et al.\ (2013a) to explore the 
recent star formation rates (SFRs) predicted by CMD-based SFHs of nearby dwarf galaxies.\ Besides using the SMC-bar extinction curve as our default 
choice in this work, we will also explore the effect of shallower extinction curves on the \lclrmtl($\lambda$) distributions.\

\subsubsection{Deriving the Integrated SEDs}\label{sec: sedcalc}
With the input of SFHs, metallicity evolution histories, extinction recipes, and SSP models as described above, we derive the integrated 
present-day monochromatic flux $f_{\lambda,{\rm int}}$ of each galaxy using the following equation:

\begin{equation}\label{eq_sspint}
f_{\lambda,{\rm int}}(t) = \sum_{t_{i}=0}^{t_{i}=t} {\rm SFR}({t-t_{i}}) f_{\lambda,{\rm SSP}}(t_{i},Z[t_{i}]) 10^{-0.4 A_{\lambda}(t_{i})} \Delta t_{i}
\end{equation}
where SFR$({t-t_{i}})$ is the SFR at lookback time $t_{i}$ with respect to a cosmic age of $t$, and $f_{\lambda,{\rm SSP}}(t_{i},Z(t_{i}))$ 
and $A_{\lambda}(t_{i})$ are the monochromatic flux and extinction, respectively, of SSP models with metallicity $Z(t_{i})$ and stellar age $t_{i}$.\ 
The increment $\Delta t_{i}$ is set by the original age resolution of FSPS SSP evolution models (187 logarithmic age steps between 
0.0003 and 14.125 Gyr).\ The sum is from the current cosmic age $t$=13.8 Gyr back to 0.0 Gyr.\ The temporal resolution of SFHs 
derived by W14 (0.1 dex from log($t_{i}$) = 6.6 to 8.7, and 0.05 dex from log($t_{i}$) = 8.7 to 10.15) is lower than that of the SSP 
evolution models; therefore, we use the nearest-neighbor interpolation for assigning the SFR to a given lookback time in Equation 
\ref{eq_sspint}.\ Note that we do not include nebular recombination emission of ionized gas associated with stellar populations 
$\lesssim$ 10 Myr old.\ For extreme starburst galaxies, the recombination continuum emission can have non-negligible contribution 
toward the red-to-NIR wavelength regime, while recombination emission lines may also have a non-negligible contribution in the relevant 
broad passbands.\ As an example, an equivalent width of 100 \AA~of the strongest optical recombination line H$\alpha$ would 
contribute $\simeq$ 7\% of the flux from the SDSS $r$ band (FWHM = 1240 \AA).\

Monochromatic flux density in various filter passbands is derived by convolving the filter transmission curves 
with Equation \ref{eq_sspint}.\ In this work, we choose to use 18 broadband filters to fully explore the efficacy 
of broadband colors and SED fitting in recovering the stellar mass of nearby galaxies.\ The 18 filters include 
GALEX FUV/near-UV (NUV), Johnson $UBVRI$, SDSS $u,g,r,i,z$, LSST $y$, 2MASS $JHK$, 
and $Spitzer$/IRAC 3.6$\mu$m ($L$) and 4.5$\mu$m ($M$).\ These broadband filters are representative of 
filters commonly used in modern all-sky surveys, and they cover the entire wavelength range from FUV to 
near-IR that are generally dominated by direct stellar emission.\ We note that the $Spitzer$/IRAC $L$ and $M$ 
filters are close to the WISE W1 and W2 filters, respectively (Wright et al.\ 2010), as well as the F356W and F444W 
filters to be offered on the upcoming {\it James Webb Space Telescope} ($JWST$).\ In addition, the $J$, $H$ and $K$ 
filters are close to the F115W, F162M, and F210M filters, respectively, to be offered on the $JWST$.\

\begin{figure*}[t]
\centering
\includegraphics[width=0.95\linewidth]{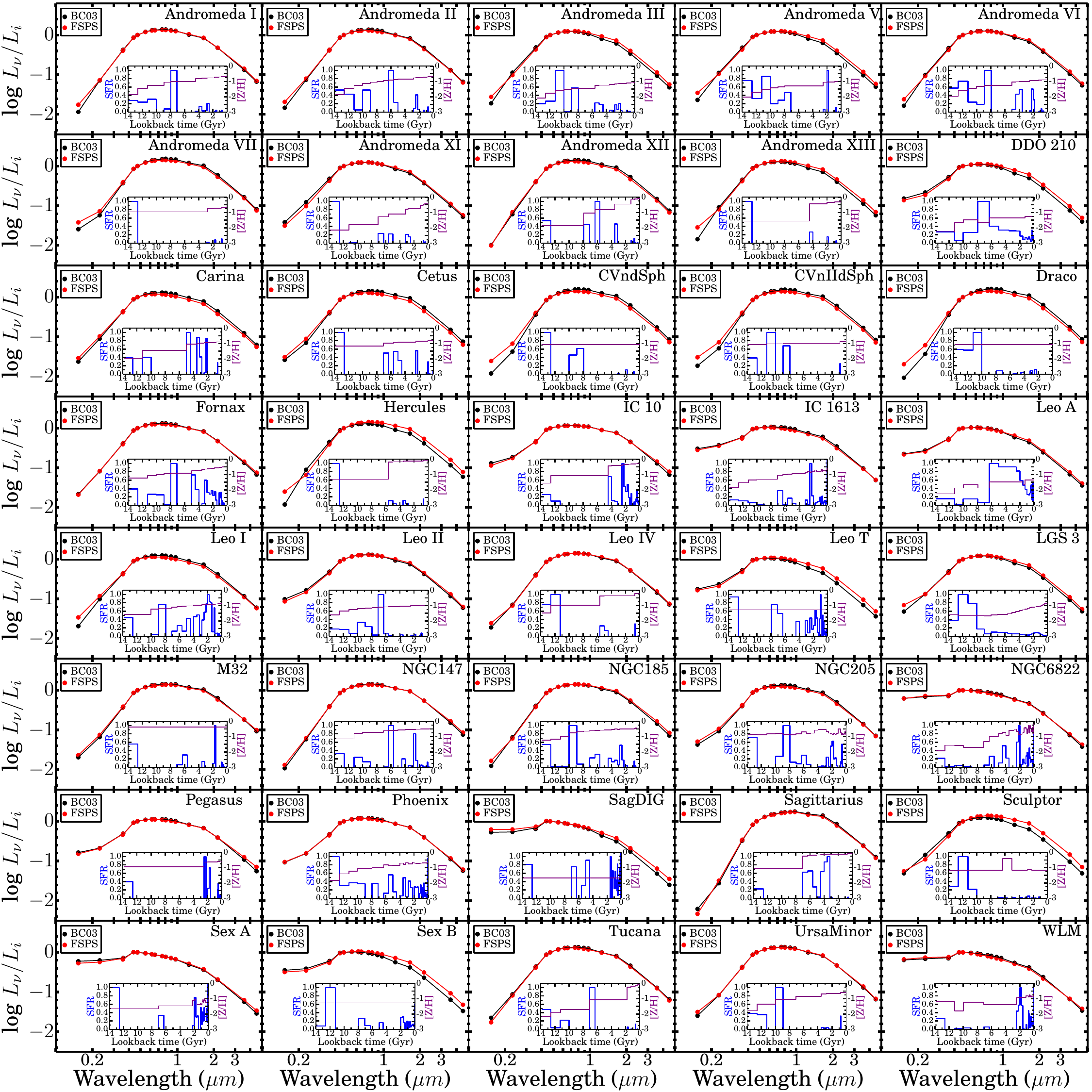}
\caption{
Broadband FUV-to-$M$ SEDs (i.e.\ FUV, NUV, $U$, $u$, $B$, $g$, $V$, $r$, $R$, $i$, $I$, $z$, $y$, $J$, $H$, $K$, $L$, 
and $M$) of the 40 LG dwarf galaxies predicted by the resolved CMD-based SFHs and metallicity evolution histories.\ 
The monochromatic luminosities in all passbands have been normalized by the $i$ band luminosities.\
The SEDs generated based on the BC03 models and the FSPS models are shown as {\it black} and {\it red} dots respectively.\ 
The inset plot in each panel shows the normalized SFHs and metallicity evolution histories as determined by Cole et al.\ (2007, 2014) 
for Leo A and DDO 210, by Hidalgo et al.\ (2011) for LGS 3, and by Weisz et al.\ (2014) for the remaining galaxies.
\label{fig_sedplt_1}}
\end{figure*}

The broadband SEDs predicted by the CMD-based SFHs and metallicity evolution histories of the 40 LG galaxies are shown 
in Figure \ref{fig_sedplt_1}.\ SEDs generated based on either the BC03 models or the FSPS models are presented separately 
in Figure \ref{fig_sedplt_1}, and they generally agree very closely with each other.\ In what follows, unless noted otherwise, the 
Johnson $UBVRI$ magnitudes are on the Vega system, and magnitudes of the other passbands are on the AB system.\ Besides 
the integrated broadband SEDs for each galaxy, we also determine several SFH-related parameters, including the mass-weighted 
ages \agem, monochromatic light-weighted ages \agelam, lookback times \tma~and \tmb~when 20\% and 80\% of the total mass 
was formed, respectively, mass-weighted metallicities \mhm, and monochromatic light-weighted metallicities \mhlam.\

\subsection{Expanding the LG sample by Assigning New Metallicities and Varying the SFHs in the Recent 1 Gyr}

\begin{figure}
\centering
\includegraphics[width=0.95\linewidth]{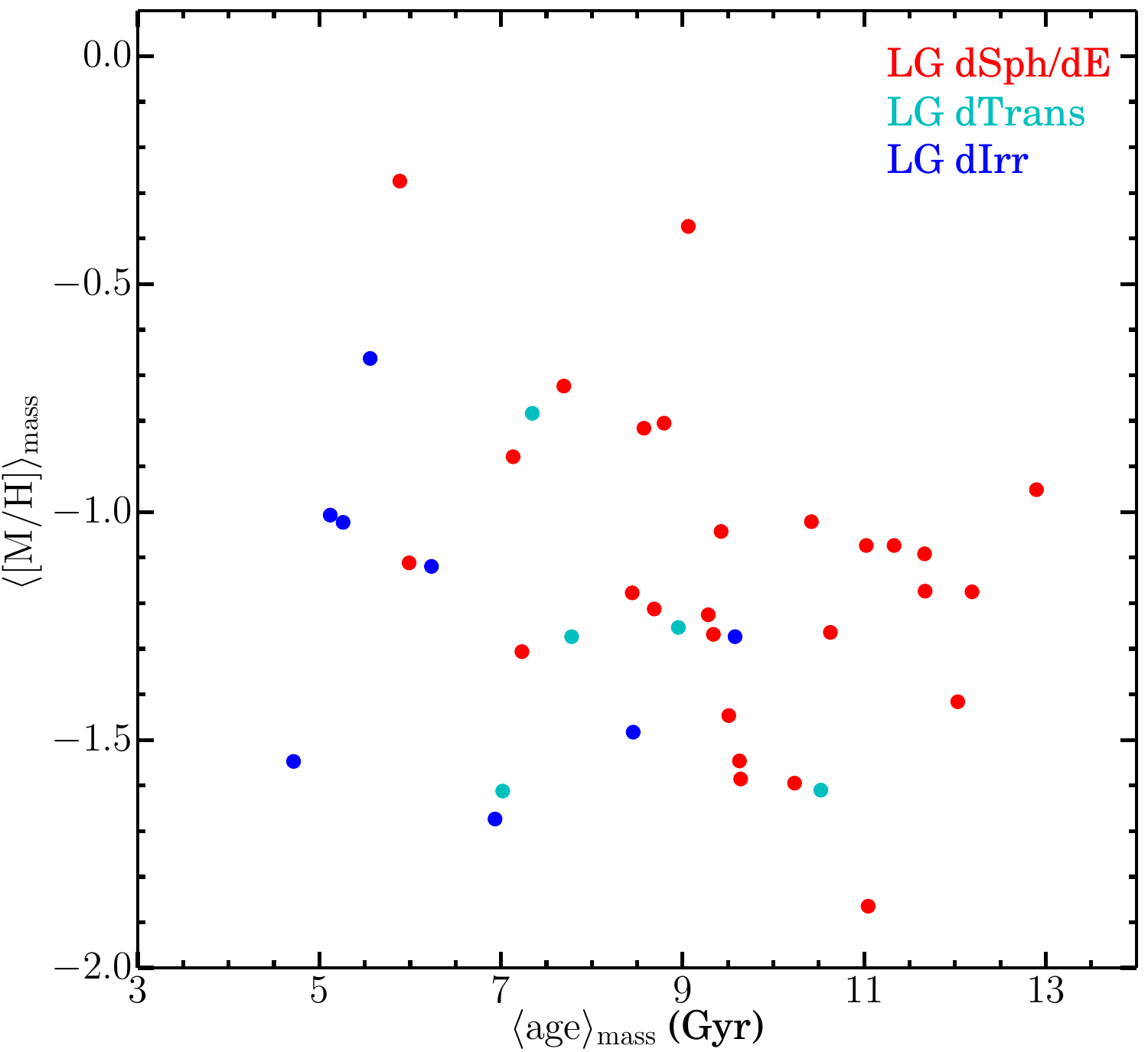}
\caption{
Mass-weighted ages \agem~and mass-weighted metallicities \mhm~of the 40 LG dwarf galaxies studied in this work.
\label{fig_lg_1}}
\end{figure}

The original sample of 40 LG dwarfs introduced above covers a variety of SFHs, as (for instance) 
quantified by the spread of mass-weighted stellar population ages (see Fig.~\ref{fig_lg_1}), and thus 
offers us a unique opportunity to investigate the dependence of stellar mass estimates on real SFHs.\ 
In order to explore the impact of metallicities on stellar mass estimates, starting from the SFH of each 
LG dwarf galaxy, we draw new mass-weighted metallicities from a uniform grid of metallicities with an 
interval of 0.3 dex and a range of $-1.95$ to 0.22 in [M/H] and then create new composite stellar populations 
with the same shape of SFH and metallicity evolution history but with the newly drawn mass-weighted metallicities.\ 
Since metallicities generally increase with time along given metallicity evolution histories, [M/H] of the recently 
formed (or ancient) stellar populations resulting from our expansion of the weighted-[M/H] coverage would be $>$ 0.22 (or 
$<$ $-1.95$) in some cases.\ We exclude these cases from our final expanded sample.\
The exercise of expanding the sample with [M/H] is justified by a lack of significant correlation between SFHs 
and [M/H] of the LG dwarfs (Fig.~\ref{fig_lg_1}).\ By doing so, we expand the original sample size from 40 to 251.\

Our sample of LG dwarf galaxies lacks extreme starbursts.\ The ``burst phase'' of local galaxies can be usually 
attributed to environmental influences, such as galaxy interactions and cosmic gas accretion.\ Starburst 
events are short-lived and thus are not necessarily relevant to the past SFHs of their host galaxies.\ In order 
to explore the influence of recent burst or suppression of star formation on stellar mass estimates, we add new 
composite stellar populations by varying the relative strengths of star formation in the recent 1 Gyr of the 
original SFHs.\ Specifically, for each ``galaxy'' in the above [M/H]-expanded sample, the recent 1 Gyr is divided into 
three age bins that are separated at lookback times of 0.1 and 0.4 Gyr, then the average SFRs 
in each age bin are independently drawn from a grid of seven values: 0.0, 0.1, 0.3, 1.0, 2.0, 5.0, and 10.0 times the 
lifetime-averaged SFR of the SFH in question.\ A maximum average burst strength (a.k.a. birth-rate 
parameter, as defined by Scalo 1986) of 10 over $\sim$100 Myr matches that of the extreme starbursts observed in 
nearby galaxies (e.g.\ McQuinn et al.\ 2010).\ The birth-rate parameters smaller than 1 mimic truncated SFHs expected 
for galaxies that are abruptly stripped off their gas supply in dense environments such as galaxy clusters.\ By doing so, 
our full sample size is increased by another 86,064, and thus the final sample to be analyzed for the major part of this paper 
includes 86,344 ``galaxies'' with different SFHs.\ Note that this exercise of expanding the parameter coverage of recent SFHs does 
not apply to cases where the relative star formation strengths of the original SFHs averaged over the relevant lookback times 
are already in accord with the corresponding values to be drawn from the grid in the expanded parameter space.\
The corresponding broadband SEDs and SFH-related parameters of the expanded sample are derived as with 
the original sample (Section \ref{sec: sedcalc}).\ In the rest of this paper, we will refer to this expanded sample of 
86,344 unique combinations of SFHs and metallicity evolution histories, with the SMC-bar extinction curve and a $A_{V, {\rm young}}$ 
of 0.5 mag, as our default full sample.\

\section{Methodology of SED fitting}\label{sec: methodsedfitting}
\subsection{A Brief Review}
Solving for the constituent stellar populations of different mass-to-light ratios ($\Upsilon_{\star}$) based on observed 
SEDs is an ill-posed problem, as mentioned in the Introduction.\ There are two distinctly different 
techniques for fitting SEDs of galaxies.\ One is the ``Bayesian template fitting'' approach, which generally 
involves Bayesian inferences of physical parameters through comparisons of the observed SEDs and model 
SEDs from a {\it predefined} library of template SFHs.\ The library of template SFHs, being either precomputed or computed on the fly, 
are usually built with simplified parameterization and/or strong prior assumptions of the lifetime SFHs.\
The other one is the ``inversion'' approach, where the observed SEDs are inverted into as many independent SSP 
components as necessary, with the weights assigned to individual components being solved on the fly.\ 
A recent review on different fitting techniques was given by Walcher et al.\ (2011).\

The Bayesian template fitting method is so far the most commonly used one for fitting broadband SEDs (e.g.\ Silva et al.\ 1998; 
Bolzonella et al.\ 2000; Brammer et al.\ 2008; da Cunha, Charlot \& Elbaz 2008; Noll et al.\ 2009; Acquaviva et al.\ 2011;  
Arnouts \& Ilbert 2011; Pacifici et al.\ 2012; Johnson et al.\ 2013b; Han \& Han 2014; Chevallard \& Charlot 2016; 
Iyer \& Gawiser 2017, among many others), but it can be severely limited/biased by the prior assumptions on the form 
(e.g.\ exponentially declining) of template SFHs.\ To avoid a simplified assumption on the parametric form of model SFHs, 
Zhang et al.\ (2012) adopted a non-parametric method by building a library of model SFHs which are characterized by six independent 
logarithmically-spaced star formation periods.\ Leja et al.\ (2017) also presented a non-parametric method through a Monte Carlo Marko 
Chain (MCMC) sampling of the posterior distributions of SFHs characterized by six independent star formation periods.\
The inversion approach, which is realized through either Monte Carlo simulations or direct (non)linear matrix inversion, is nonparametric 
and is in principle the unbiased way to fit for stellar populations of galaxies, as long as the full parameter space can be explored.\ The inversion 
approach has been mostly applied to fitting of spectroscopic data (e.g.\ Ocvirk et al.\ 2006b; Tojeiro et al.\ 2007; Cappellari 2017), with the   
exception of Blanton \& Roweis (2007).\

\subsection{Nonparametric SED fitting: The Matrix Inversion with NNLS Optimization}\label{sec: nnls_method}
One important goal of this work is to explore the efficacy of broadband SED fitting for recovering 
the integrated stellar mass of galaxies, without any prior assumption on the SFHs.\ Here we introduce 
a pure mathematical matrix inversion method for broadband SED fitting.\
 
Any SFHs can be decomposed into a series of SSPs with different ages, metallicities, and masses (or weights).\
As long as the nonlinear dependence on extinction can be treated separately, decomposing the observed 
SEDs into that of a series of SSPs is mathematically an underdetermined non-negative linear matrix inversion 
process (e.g.\ Press et al.\ 1992).\ To implement this approach, we turn to the classical {\it Active-set algorithm}, 
first introduced by Lawson \& Hanson (1974).\ The {\it Active-set algorithm} solves the non-negative matrix inversion 
problem by iteratively minimizing the quadratic residual.\ In our particular case, starting from an ``active set'' that  
contains all SSP components that constitute the {\it design matrix}, at each iteration, the SSP component in the ``active 
set'' that maximizes the negative gradient of the quadratic residual and at the same time leads to a non-negative solution to the 
{\it normal equation} is removed from the ``active set'' and added to a ``passive set'' or solution space that contains 
the SSP components that result in the final non-negative solution vector of weights that minimizes the quadratic residual.\
Recent studies suggest that the non-negative regularized least-squares regression can be remarkably effective 
even in cases where the data sample size is smaller than the dimensions of the parameter space (e.g.\ Meinshausen 2013), 
as long as the solution vector is known to be sparse, i.e.\ only a subset of the solution vector is not equal to zero.\

To apply the above NNLS method to our broadband SED fitting, 
we adopt a base of 180 SSPs, encompassing 15 ages uniformly distributed in log space between 
0.001 and 14 Gyr, and a grid of 12 metallicity $Z$ values ranging from 2$\times$10$^{-4}$ to 0.03 
(or [M/H] from $-$1.95 to 0.22, for $Z_{\odot}$ = 0.018).\ Note that the 12 metallicity values are selected 
to be every other value as provided by the Padova+BaSeL isochrones used in the FSPS models.\ 
We then use the Lawson \& Hanson NNLS algorithm through a Python wrapper of the original 
FORTRAN code (scipy.optimize.nnls) to solve for non-negative weights of SSPs, given any ``observed'' 
input SED.\ The same NNLS code has been used in the {\sc pPXF} software (Cappellari 2017) for full-spectrum fitting.\ 

We consider NNLS SED fitting to our default full sample, with $A_{V, {\rm young}}$ being either fixed to 0.5 mag or a free 
parameter in the fitting, following the same extinction recipes as described in Section \ref{sec: secext}.\ To fit SEDs with $A_{V, {\rm young}}$ 
as a free parameter, we run two successive iterations of NNLS SED fitting with decreasing step sizes of $A_{V, {\rm young}}$ 
in order to find the global minimum.\ In particular, the first iteration of fitting goes through a grid of $A_{V, {\rm young}}$ values 
ranging from 0.0 to 2.0 mag with a step size of 0.1 mag, and then a second iteration goes through a finer grid of $A_{V, {\rm young}}$ 
with a step size of 0.03 mag and centered around the $A_{V, {\rm young}}$ value that leads to the minimum $\chi^{2}$ in the first iteration.\

A vast majority of previous works on SED modeling assumed constant metallicities along individual model SFHs, 
without taking into account metallicity evolution.\ Gallazzi \& Bell (2009) found that stellar masses may be systematically 
over- or underestimated (depending on the average ages) when using monometallic (exponential) model SFHs to fit SEDs 
resulting from variable-metallicity SFHs.\ Here we consider both monometallic and multimetallic SFH solutions in our NNLS fitting.\ 
In particular, for multimetallic SFH solutions, all of the 180 SSP templates are used at the same time to search for the non-negative 
solutions, while for monometallic SFH solutions, 12 candidate monometallic solutions for the 12 metallicity values are first found by 
putting separately each set of SSP templates of the same metallicity in the ``active set'' to search for non-negative solutions, and then 
the final best-fit solution vector is selected as the one leading to the minimum $\chi^{2}$.\

Every SSP component is normalized to a total stellar mass of 1 $M_{\odot}$, so the total mass corresponding to a 
given NNLS solution vector is simply a sum of the weights of individual SSP components.\ For each input SED, the 
NNLS fitting is performed for 500 times with random Gaussian noise added at each time.\ Then the 50th percentile 
of the resulting distribution of the 500 best-fit stellar masses is adopted as the most probable stellar mass, and the 
16th and 84th percentiles as the 1-$\sigma$ confidence intervals.\ The results of our SED fitting to the default full 
sample will be presented in Section \ref{sec: sedfitting}.\

\section{Color--Mass-to-Light Ratio Relations}\label{sec: clrm2l}

\subsection{The Context}

\begin{figure*}[t]
\centering
\includegraphics[width=0.95\linewidth]{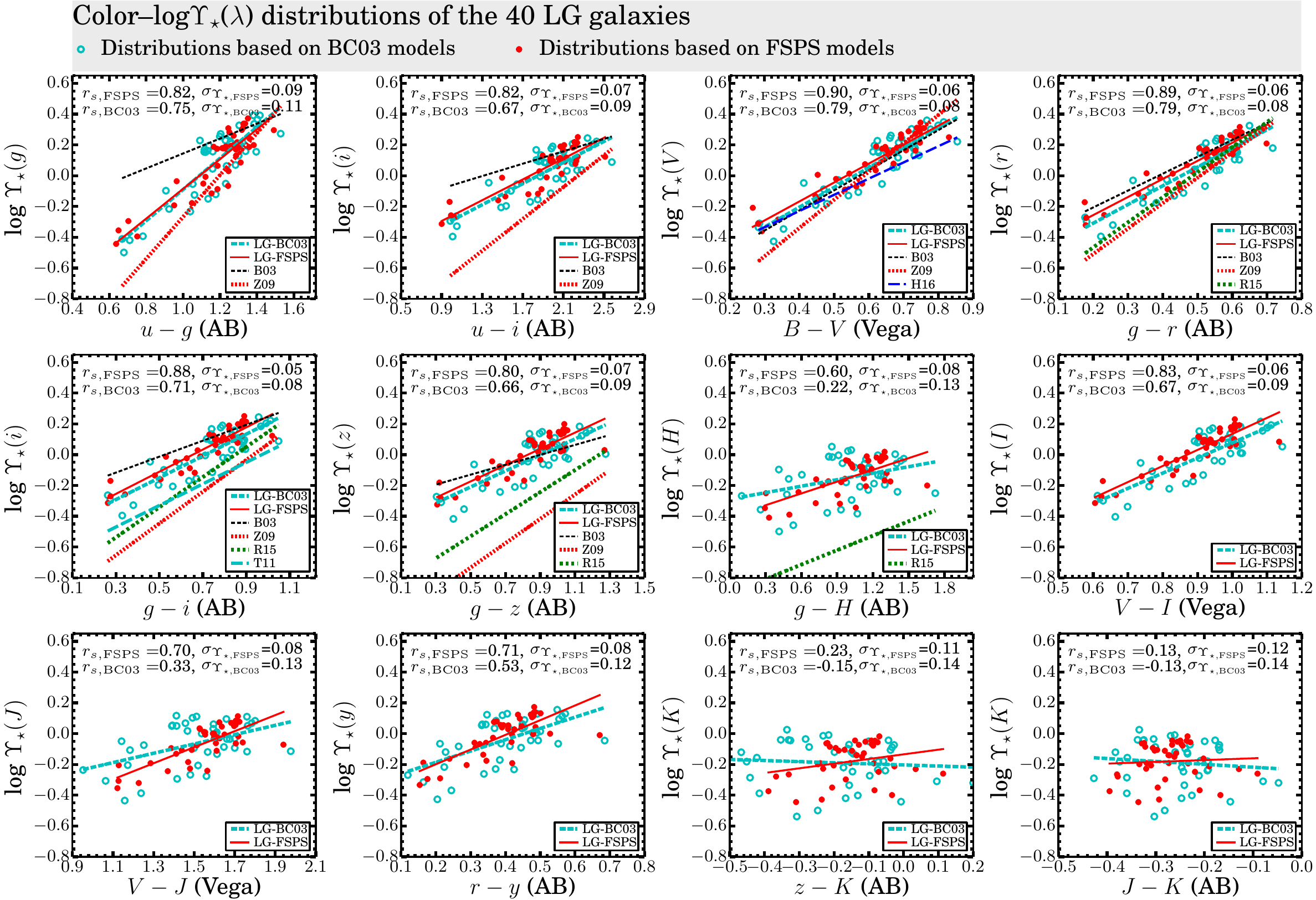}
\caption{
Distributions of the 40 LG dwarf galaxies studied in this work on various color--mass-to-light-ratio (\lclrmtl) diagrams.\
The {\it red filled circles} and {\it cyan open circles} represent the distributions of the 40 galaxies based on SFHs 
reconstructed with the FSPS and BC03 models, respectively.\ The {\it black solid lines} and {\it cyan short-dashed lines} 
represent the linear regression of the \lclrmtl distributions based on the FSPS and BC03 models, respectively.\ Several published 
linear \lclrmtl($\lambda$) relations (whenever available for given passbands) are also plotted for comparison.\ In particular, these published 
relations are respectively from Bell et al.~(2003; B03), Zibetti et al.\ (2009; Z09), Taylor et al.\ (2011; T11), Roediger \& 
Courteau (2015; R15), and Herrmann et al.\ (2016; H16).\ The line styles and colors of these overplotted relations are as 
indicated in the legends of individual panels.\ Except for the B03 relations, which were calibrated based on the PEGASE models, 
all the other relations were based on either the original (i.e.\ T11, R15) or the updated (i.e.\ Z09, H16) BC03 models.\ The PEGASE 
models were based on about the same isochrones as BC03.\ Therefore, the differences between different relations in the optical 
bands primarily reflect the different {\it priors} of SFHs assumed in the different studies.
\label{fig_m2lclr_1}}
\end{figure*}

The \clrmtl($\lambda$)~relations provide a cheap way to estimate stellar masses of galaxies (Bell \& de Jong 2001).\ 
Linear color--$\log \Upsilon_{\star}$($\lambda$) relations of galaxies have been calibrated in different passbands by numerous studies in 
the past.\ To name a few, these studies include Bell \& de Jong (2001), Bell et al.\ (2003, hereafter B03), Portinari et al.\ (2004), 
Zibetti et al.\ (2009, hereafter Z09), Taylor et al.\ (2011, hereafter T11), Into \& Portinari (2013), Roediger \& Courteau (2015, 
hereafter R15), Herrmann et al.\ (2016, hereafter H16), etc.\ These studies calibrated linear \lclrmtl($\lambda$) relations with sometimes significantly 
different slopes and zero points (or intercepts).\ What is even more worrisome, McGaugh \& Schombert (2014) found that all of the commonly 
employed relations fail to provide self-consistent results, in the sense that the same set of relations (i.e.\ the ones from the same authors) 
gives systematically different stellar mass estimates when different passbands are used for given galaxies.

The effectiveness of single colors as indicators of $\Upsilon_{\star}$($\lambda$) may rely on the partially compensating effect of ages, 
metallicities and dust reddening on colors, and therefore $\Upsilon_{\star}$($\lambda$), in the sense that an increase of any of these three parameters 
generally results in a reddening of colors and an increase of $\Upsilon_{\star}$($\lambda$).\ This interpretation is straightforward to understand 
for SSPs, but not for composite stellar populations whereby the integrated light in different passbands may be dominated by 
different stellar populations and the light-dominant populations may not be the mass-dominant ones.\ The location of a galaxy 
on the \clrmtl~diagrams is in principle determined by its SFH, metallicity evolution history, and dust reddening.\ 

The impact of SFHs on colors and $\Upsilon_{\star}$($\lambda$) may be reflected by the mass-weighted ages \agem~and the monochromatic 
light-weighted ages \agelam~in passbands of different wavelengths $\lambda$.\ The impact of metallicity evolution on colors and $\Upsilon_{\star}$($\lambda$) 
may be reflected by the mass-weighted metallicities \mhm~and the monochromatic light-weighted metallicities \mhlam~in different 
passbands.\ In what follows in this section, we will use the mass- and light-weighted quantities mentioned above, together with 
the lookback times when 20\% (\tma) and 80\% (\tmb) of total stellar masses were formed, to investigate how SFHs and metallicity evolution affect 
the \clrmtl($\lambda$) distributions.\ Note that we will use \agelg~and \agesh~to represent \agelam~at the longer- and shorter-wavelength 
passbands involved in a given color, and \mhlg~and \mhsh~to represent monochromatic light-weighted [M/H] at the longer- and shorter-wavelength 
passbands involved in a given color.

\subsection{Distribution of LG Dwarf Galaxies}
A selection of representative \lclrmtl($\lambda$) relations of the LG dwarf galaxies is shown in Figure \ref{fig_m2lclr_1}.\ A majority of 
previous studies of galaxy stellar masses were carried out with the BC03 models.\ Therefore, we generate the \lclrmtl($\lambda$) relations 
of the LG dwarfs based on both our default FSPS models ({\it red filled circles}) and the BC03 models ({\it cyan open circles}), 
in order to facilitate a fair comparison with other studies.\ Following the common practice, we conducted linear least-squares regression 
to the color--$\log \Upsilon_{\star}$($\lambda$) distributions.\ The fitted linear relations are overplotted on the distributions shown in Figure 
\ref{fig_m2lclr_1}.\ We also calculate the Spearman's rank correlation coefficients $r_{s,{\rm FSPS}}$ and $r_{s,{\rm BC03}}$ 
between colors and $\log \Upsilon_{\star}$($\lambda$), and the standard deviations $\sigma_{\Upsilon_{\star},{\rm FSPS}}$ and 
$\sigma_{\Upsilon_{\star},{\rm BC03}}$ of $\log \Upsilon_{\star}$($\lambda$) around the best-fit linear relations.\ Both the Spearman's rank 
correlation coefficients and the standard deviations are listed at the top of each panel of Figure \ref{fig_m2lclr_1}.

From Figure \ref{fig_m2lclr_1}, we can see that SFHs constructed with the FSPS and BC03 models give \lclrmtl($\lambda$) relations 
(e.g.\ the slopes and intercepts of the linear fits) that are in decent agreement across the whole wavelength range.\ Nevertheless, 
we notice that the FSPS models give significantly tighter \lclrmtl($\lambda$) relations than the BC03 models, which is quantified by both the 
systematically higher correlation coefficients and the overall smaller scatters for the FSPS models.\ These differences might be 
primarily attributed to the different versions of Padova isochrones used by the two models.\

Several linear \lclrmtl($\lambda$) relations calibrated by previous studies are also overplotted in Figure \ref{fig_m2lclr_1} for a visual 
comparison.\ The references to these literature relations are given in the caption of Figure \ref{fig_m2lclr_1}.\ These plotted 
linear relations are the ones in the literature that were calibrated based on either the original or updated BC03 models.\ 
Here we just point out that the LG dwarfs follow \lclrmtl($\lambda$) relations that fall in between the extreme ones calibrated by previous 
studies.\ A more detailed discussion about these relations and their differences will be given in Section \ref{sec: clrmtl_orig}.\ 
In the Appendix section, we present the linear least-squares fitting to various optical \lclrmtl($\lambda$) relations of the 40 LG 
dwarf galaxies and the expanded samples in Tables \ref{tab_clrmtl_fsps} and \ref{tab_clrmtl_bc03}.

\subsection{Impact of SFHs on \lclrmtl($\lambda$) Relations}\label{sec: impact_sfh}
We use the (expanded) default full sample to explore the impact of SFHs on \lclrmtl($\lambda$) distributions.\ In what follows in this subsection, 
we show the \lclrmtl($\lambda$) distributions of the default full sample, which are grouped by different ranges of aforementioned parameters, 
including \agem~(Fig.~\ref{fig_m2lclr_2}), $\log$(\tma/\tmb) (Fig.~\ref{fig_m2lclr_3}), \agelg~(Fig.~\ref{fig_m2lclr_4}), and (\logagesh$-$\logagem) 
(Fig.~\ref{fig_m2lclr_5}).\ In addition, Figure~\ref{fig_m2lclr_4_v} presents the \lclrmtl($V$) distributions, which are grouped by \agev~in order to 
demonstrate the variation of given $\log \Upsilon_{\star}$($\lambda$) as functions of different colors.\ To generate the distributions shown in 
Figures \ref{fig_m2lclr_2} to \ref{fig_m2lclr_5}, the \lclrmtl($\lambda$) two-dimensional histograms of number densities $N$ of data points with bin 
sizes of 0.02$\times$0.02 are first created, and then contours enclosing the relevant area with $N>0$ are plotted.\

\subsubsection{Correlations with \agem}

As a zero-order representation of SFHs, \agem~is barely correlated with either colors or $\log \Upsilon_{\star}$($\lambda$)  (see Fig.~\ref{fig_m2lclr_2}).\ 
However, \agem~appears to be positively correlated with the lower and especially upper bounds of $\log \Upsilon_{\star}$($\lambda$)  for given color 
values, and this \agem-dependent effect is more significant for redder color values, leading to more positive {\it average} slopes of the \lclrmtl($\lambda$) 
relations for galaxies with larger \agem.\ Moreover, the slope differences for different \agem~are larger for optical colors covering longer-wavelength 
baselines.\ For instance, the slope differences between subsamples with \agem~= 11--13 Gyr and \agem~=  3--5 Gyr (Fig.~\ref{fig_m2lclr_2}) 
are $\sim$ 0.01 and $\sim$ 0.11, respectively, for the ($B$$-$$V$)--$\log \Upsilon_{\star}$($V$) and ($g$$-$$z$)--$\log \Upsilon_{\star}$($z$) relations.\
We also note that subsamples with younger \agem~have tighter and more linear \lclrmtl($\lambda$) relations than those with older \agem.\
This implies that $\Upsilon_{\star}$($\lambda$) of higher-redshift galaxies is generally better constrained as long as the redshift information 
is used to restrict the oldest possible \agem.\

\subsubsection{Correlations with SFH ``concentration''}
In analogy to the definition of concentration of galaxy light profiles (e.g.\ Bershady et al.\ 2000), log(\tma/\tmb) can be used to quantify the 
``concentration'' of SFHs in time (Figure \ref{fig_m2lclr_3}).\ Subsamples with larger $\log$(\tma/\tmb) (i.e.\ more extended SFHs) cover smaller 
ranges of colors and $\Upsilon_{\star}$($\lambda$).\ However, subsamples with different $\log$(\tma/\tmb) follow \lclrmtl($\lambda$) correlations 
with very similar slopes, and except for those with the smallest $\log$(\tma/\tmb) (i.e.\ $\sim$ 0.0 -- 0.3, the most concentrated SFHs), similar intercepts.\ 

\subsubsection{Correlations with \agelam}
The monochromatic light-weighted ages, e.g.\ \agelg~(Figures \ref{fig_m2lclr_4} and \ref{fig_m2lclr_4_v}), are broadly correlated with 
$\log \Upsilon_{\star}$($\lambda$).\ Similar to the findings of Worthey (1994) for SSPs, we point out that the correlation coefficients between 
\agelam~and $\log \Upsilon_{\star}$($\lambda$) peak in $J$ band ($r_{s}\!=\!0.96$), with the correlation being poorer toward both longer- 
and shorter-wavelength luminance passbands.\ In addition, for a given combination of color and $\log \Upsilon_{\star}$($\lambda$), the 
relations defined by subsamples of different \agelg~run largely parallel to each other, and the perpendicular (i.e.\ normal to the running 
directions of color--$\log \Upsilon_{\star}$($\lambda$) correlations) offset between them becomes smaller for shorter-wavelength luminance 
passband and optical colors that involve the $g$ or $B$ band as the shorter-wavelength passband and at the same time cover shorter-wavelength 
baselines (e.g.\ $B$$-$$V$, $g$$-$$r$).~In particular, we note that the ($B$$-$$V$)--$\log \Upsilon_{\star}(V)$ relation is the one with the 
least ``perpendicular offset'' between subsamples of different \agelg, whereas the \lclrmtl($\lambda$) relations involving NIR passbands are 
the ones with the largest ``perpendicular offsets.'' A smaller ``perpendicular offset'' suggests a smaller {\it systematic} dependence on \agelam.\

\subsubsection{Mismatches between \agelam~and \agem}
Mismatches between light- and mass-weighted ages, e.g.\ \logagesh$-$\logagem~(Figure \ref{fig_m2lclr_5}) reflect the degree of the 
``outshining'' effect of younger light-dominant populations over older mass-dominant populations.\ Subsamples with \logagesh$-$\logagem\ 
$\lesssim$ $-0.6$ tend to have slightly steeper \lclrmtl($\lambda$) correlations, and this steepening trend is sharper for colors covering 
longer-wavelength baselines.\ The effect of mismatches between \logagesh~and \logagelg~(not shown in this paper) is very similar to that 
between \logagesh~and \logagem.\

\begin{figure*}
\centering
\includegraphics[width=0.9\linewidth]{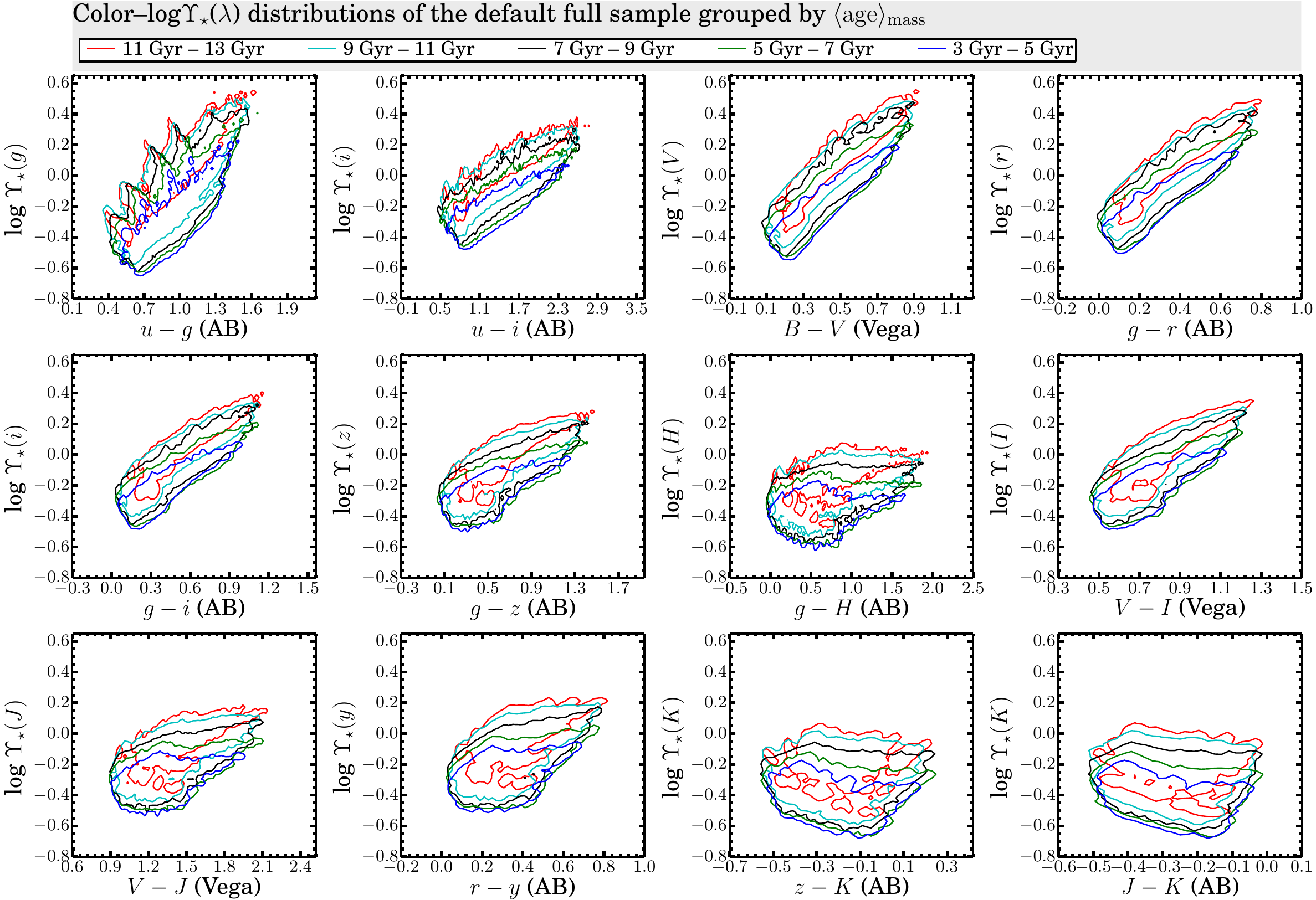}
\caption{
Impact of mass-weighted ages, \agem,~on \lclrmtl($\lambda$)~distributions of the default full sample.\ 
The solid contours of different colors in each panel enclose the distributions of five subsamples with different ranges of \agem, 
as indicated in the figure title.
\label{fig_m2lclr_2}}
\end{figure*}

\begin{figure*}
\centering
\includegraphics[width=0.9\linewidth]{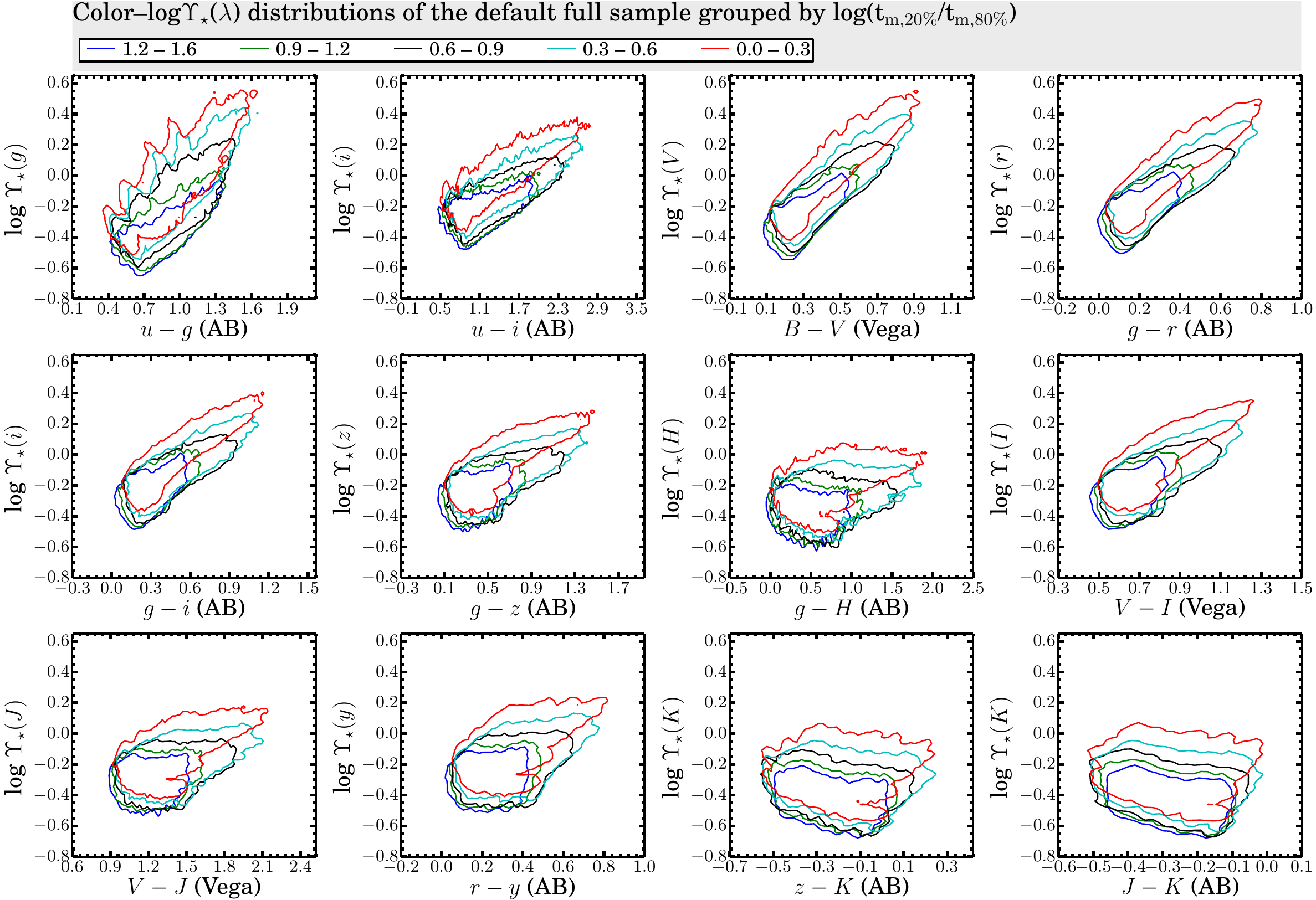}
\caption{
Impact of ``concentration'' of SFHs (in time) on \lclrmtl($\lambda$) distributions of the default full sample.\
Here the sample is grouped into subsamples of different $\log$(\tma/\tmb), where \tma~and \tmb~are the lookback times when 20\% and 
80\% of the total stellar masses were formed, respectively.\ The solid contours of different colors in each panel enclose the distributions of five subsamples 
with different ranges of $\log$(\tma/\tmb), as indicated in the figure title.
\label{fig_m2lclr_3}}
\end{figure*}

\begin{figure*}
\centering
\includegraphics[width=0.9\linewidth]{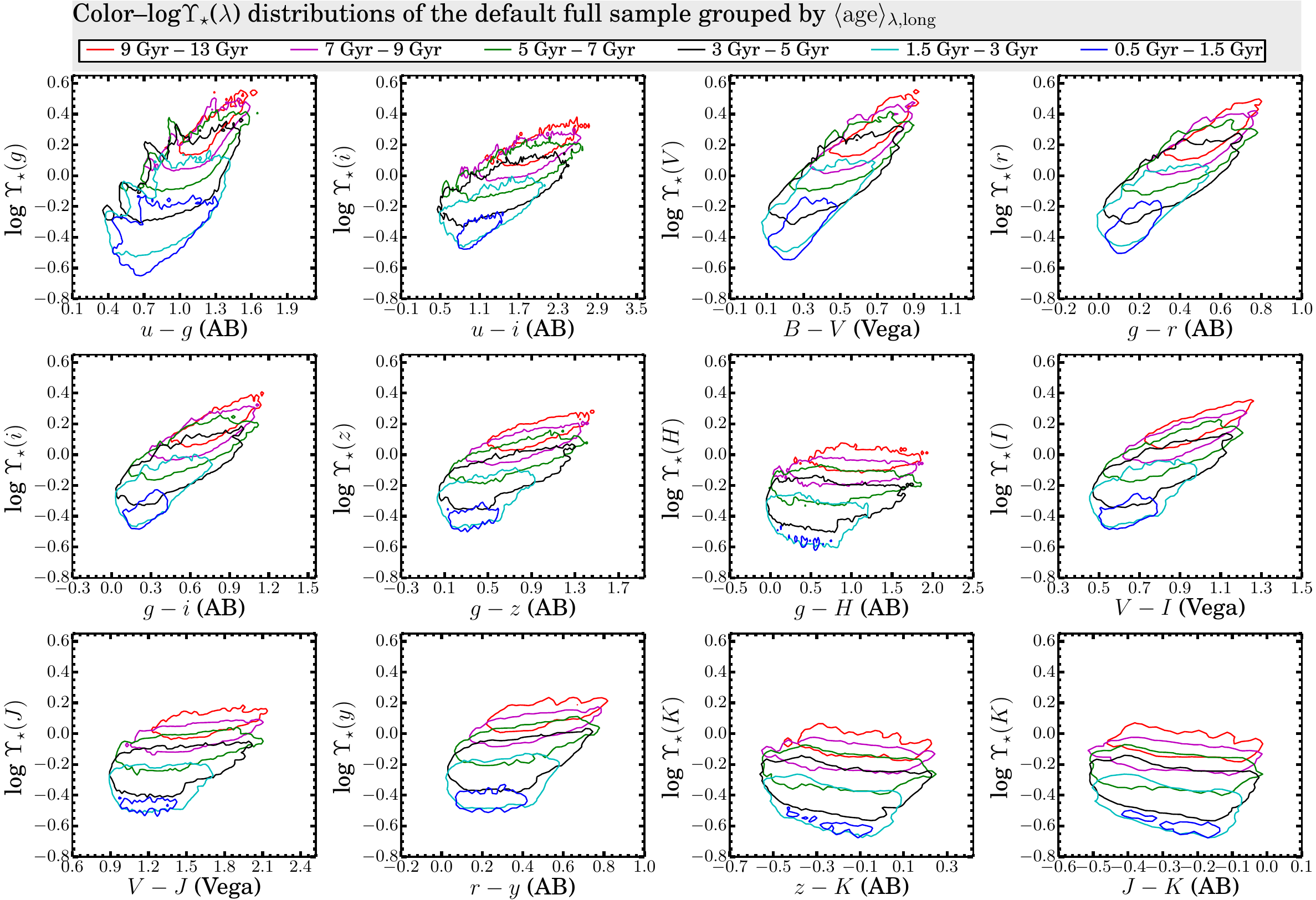}
\caption{
Impact of monochromatic light-weighted ages on \lclrmtl($\lambda$) distributions of the default full sample.\
Here the sample is grouped into subsamples of different \agelg which represents the monochromatic light-weighted ages of the longer-wavelength 
passband involved in each color.\ The solid contours of different colors in each panel enclose the distributions of five subsamples with different \agelg, 
as indicated in the figure title.
\label{fig_m2lclr_4}}
\end{figure*} 

\begin{figure*}
\centering
\includegraphics[width=0.9\linewidth]{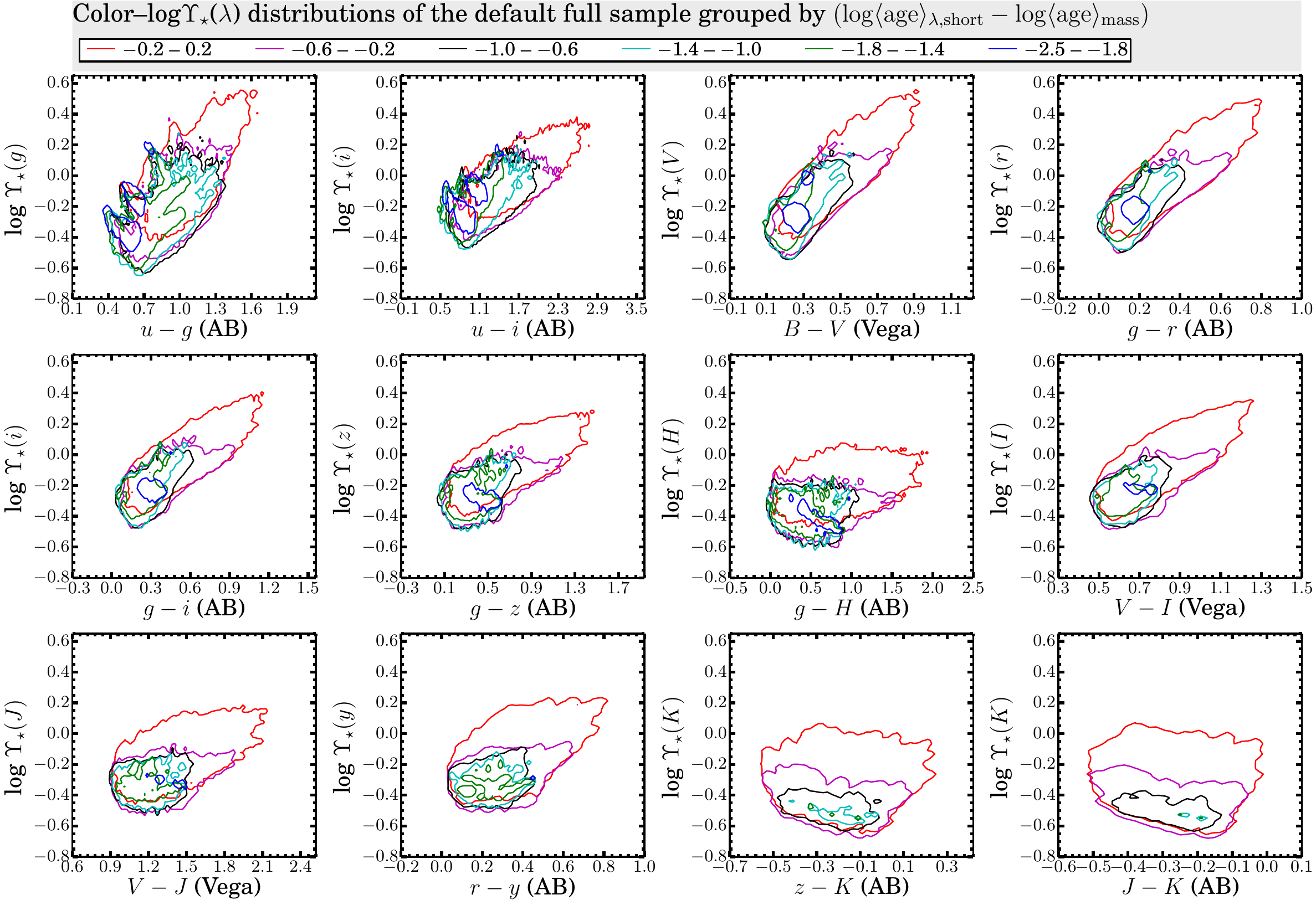}
\caption{
Impact of the ``outshining'' effect of the light-dominant young populations over mass-dominant old populations on \lclrmtl($\lambda$) distributions of the 
default sample.\ Here the full sample is grouped into subsamples of different ranges of log\agesh$-$log\agem, where \agem~represents 
the mass-weighted age and \agesh~represents the monochromatic light-weighted ages of the shorter-wavelength passband involved in a 
given color.\ The solid contours in each panel enclose the distributions of subsamples with different log\agesh$-$log\agem, as indicated 
in the figure title.
\label{fig_m2lclr_5}}
\end{figure*}

\begin{figure*}
\centering
\includegraphics[width=0.9\linewidth]{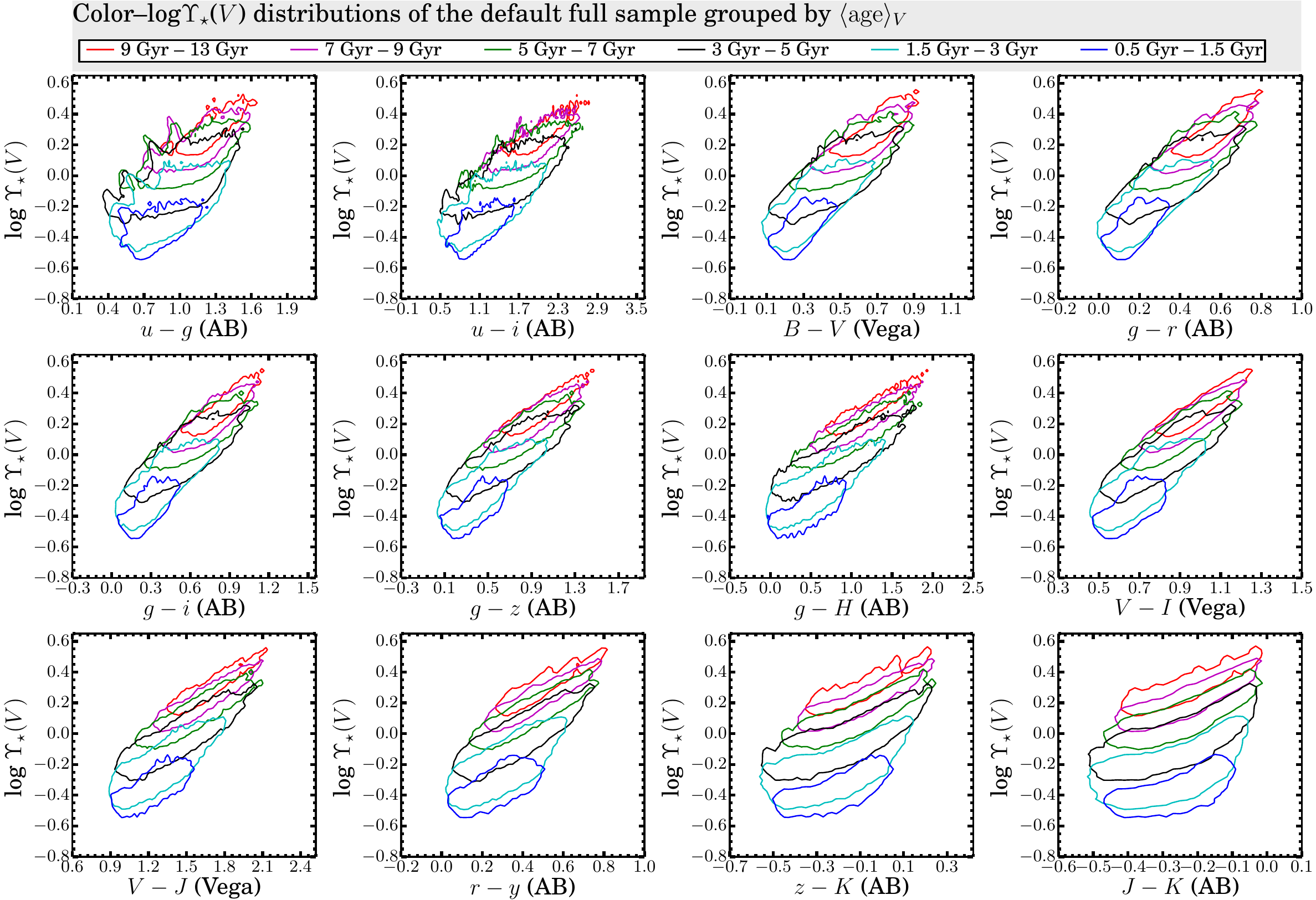}
\caption{
Same as Figure \ref{fig_m2lclr_4}, but here for color--$\log \Upsilon_{\star}$($V$) distributions.
\label{fig_m2lclr_4_v}}
\end{figure*} 

\begin{figure*}
\centering
\includegraphics[width=0.9\linewidth]{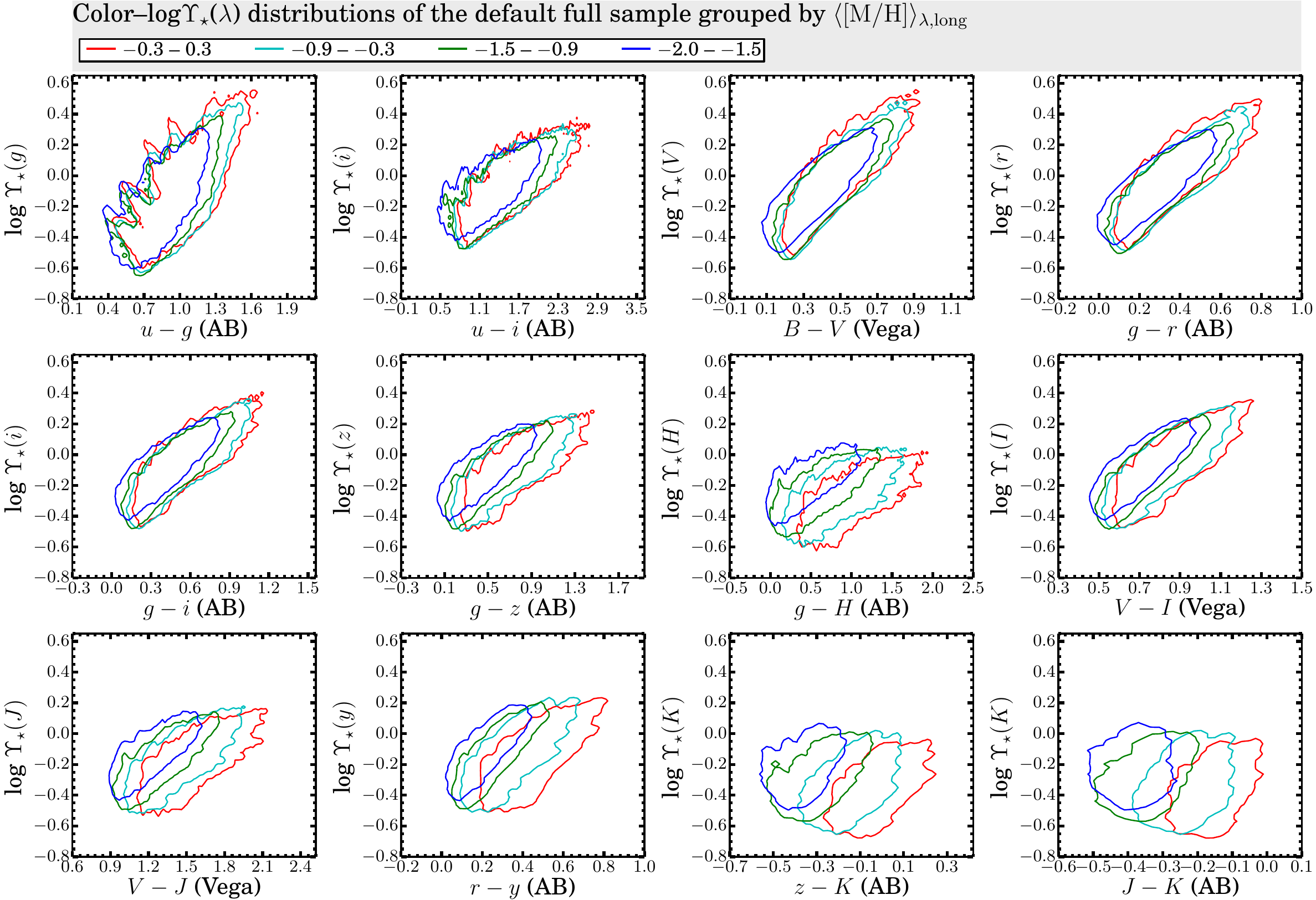}
\caption{
Impact of light-weighted metallicities on \lclrmtl($\lambda$) relations of the full sample.\
Here the sample is grouped into subsamples of different ranges of \mhlg.\ \mhlg~represents the monochromatic light-weighted 
[M/H] of the longer-wavelength passband involved in each color.\ The solid contours in each panel enclose the distributions of 
four subsamples with different ranges of \mhlg, as indicated in the figure title.
\label{fig_m2lclr_7}}
\end{figure*}

\subsection{Impact of metallicity evolution on \lclrmtl($\lambda$) Relations}
\subsubsection{Impact of Average Metallicities}\label{sec: impact_met}
As in Figures \ref{fig_m2lclr_2} -- \ref{fig_m2lclr_5}, \lclrmtl($\lambda$) distributions of the default full sample grouped by light-weighted 
metallicities \mhlg~are shown in Figure \ref{fig_m2lclr_7}.\ The distributions with respect to \mhm~are very similar to those with respect to 
\mhlg, so they are not shown here.\
 
Subsamples with higher \mhlg~follow \lclrmtl($\lambda$) relations that are slightly steeper and have a higher degree of nonlinearity 
than those with lower \mhlg, owing to the fact that more metal-rich stellar atmospheres leave a stronger and nonlinear imprint on 
specific SED regions.\ Moreover, subsamples with higher \mhlg~generally have a larger spread of $\Upsilon_{\star}$($\lambda$) 
at given color values.\ The ``perpendicular offsets'' between the \lclrmtl($\lambda$) relations of subsamples with different \mhlg~are larger 
for colors involving longer-wavelength passbands, in line with the {\it well-known} higher sensitivity of colors covering longer wavelength 
baselines to stellar atmosphere metallicities.\ 

$\Upsilon_{\star}$($\lambda$) has relatively tight correlations with optical--optical colors but not optical--NIR or (especially) NIR-NIR 
colors.\ This is because both $\Upsilon_{\star}$($\lambda$) and optical--optical colors have a stronger correlation with \agelam~than 
\mhlam, whereas optical--NIR colors depend about equally on both \agelam~and \mhlam, and NIR--NIR colors primarily on \mhlam.\

\subsubsection{Impact of the Metallicity Evolution}\label{sec: impact_meh}
\begin{figure*}
\centering
\includegraphics[width=0.9\linewidth]{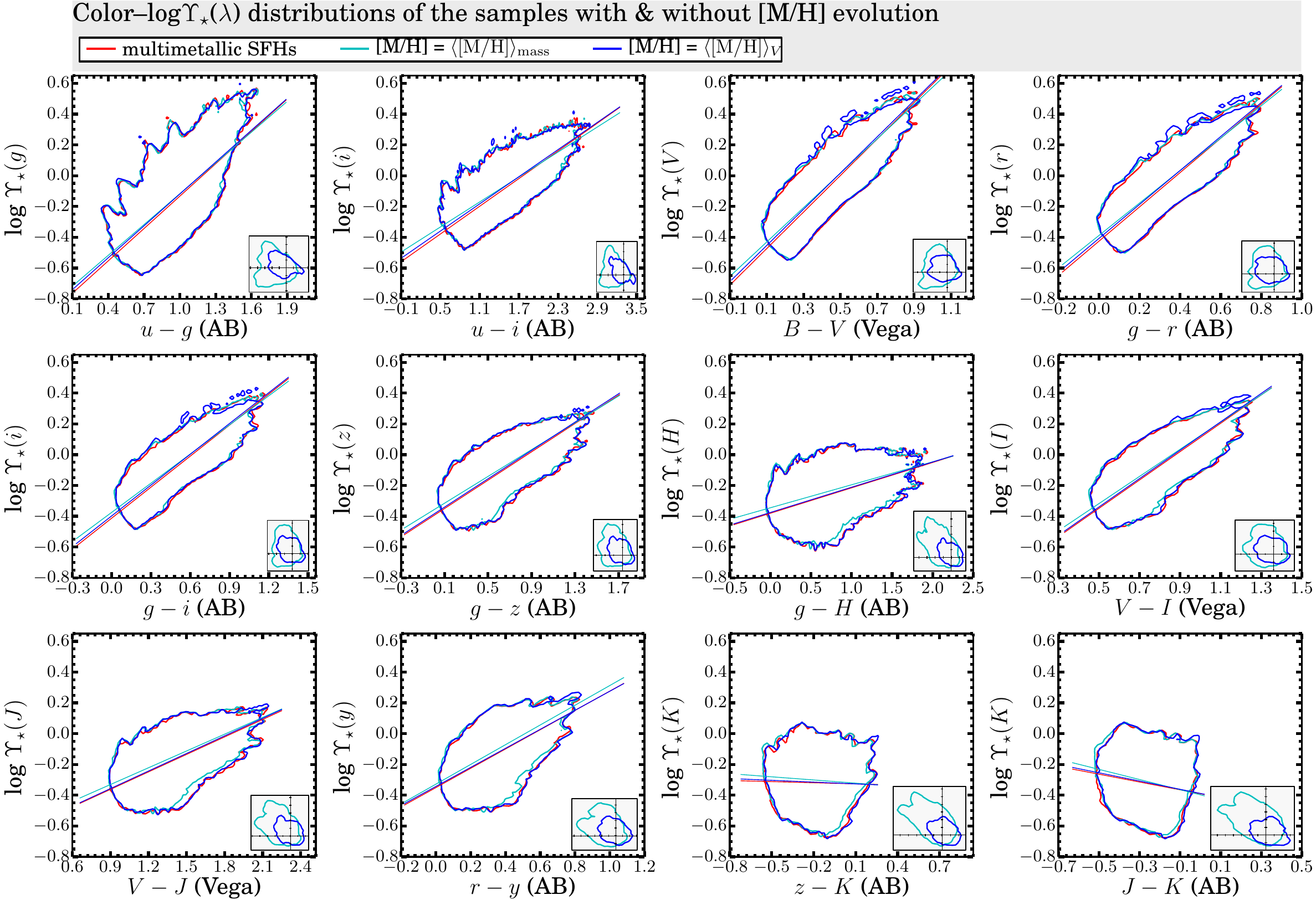}
\caption{
Comparison of \lclrmtl($\lambda$) distributions resulting from SFHs with and without accompanied metallicity evolution, 
i.e.\ multimetallic and monometallic SFHs.\ The {\it red} contours mark coverage of our {\it default} \lclrmtl($\lambda$) distributions with 
multimetallic SFHs, whereas the {\it cyan} and {\it blue} contours mark the \lclrmtl($\lambda$) coverage of distributions resulting from monometallic 
SFHs with the [M/H] being fixed to \mhm~and \mhv, respectively.\ The linear least-squares regression to the three distributions 
is represented by the solid lines, with the same color scheme as for the contours.\ The inset plot at the lower right corner of each 
panel shows contours ({\it cyan} for [M/H]=\mhm; {\it blue} for [M/H]=\mhv) enclosing the distributions of $\Delta$(color) (horizontal 
axis) vs.\ $\Delta$log($\Upsilon_{\star}$($\lambda$)) (vertical axis), where $\Delta$(color) and $\Delta$log($\Upsilon_{\star}$($\lambda$)) are, respectively,  
the color differences and $\log$($\Upsilon_{\star}$($\lambda$)) differences between the same SFHs without and with metallicity evolution.\ The position of 
$\Delta$(color) = 0 and $\Delta$log($\Upsilon_{\star}$($\lambda$)) = 0 coincides with the horizontal and vertical lines (cross-hair) in the inset plot of each 
panel.\ The inset and main plots in each panel share the same axis scale.
\label{fig_m2lclr_meh}}
\end{figure*}

Nearly all previous studies on stellar mass estimates do not take into account metallicity evolution when building their model SFH 
libraries.\ Here we explore the differences between \lclrmtl($\lambda$) distributions resulting from the multi\-metallic SFHs (the {\it default} 
in this work) and the {\it unrealistic} monometallic SFHs.\ To this end, for every ``galaxy'' in the default full sample, we determine two 
additional sets of multiband SEDs resulting from the same SFH but with a single [M/H] fixed to either \mhm~or the $V$-band light-weighted 
\mhv.\ The \lclrmtl($\lambda$) distributions for the samples with multimetallic and monometallic SFHs are shown separately in 
Figure~\ref{fig_m2lclr_meh}.\ In addition, we also derive the differences of colors ($\Delta$[color]) and $\log \Upsilon_{\star}$($\lambda$) 
($\Delta$[$\log \Upsilon_{\star}$($\lambda$)]) between the monometallic and multimetallic ``versions'' of each galaxy, and the corresponding 
contours enclosing the $\Delta$(color) vs.\ $\Delta$($\log \Upsilon_{\star}$($\lambda$)) distributions of the samples are shown as inset plots of 
Figure~\ref{fig_m2lclr_meh}.\ We note that the conclusions drawn below do not change if monometallic SFHs with [M/H] fixed to light-weighted 
[M/H] in passbands other than $V$ are used.\ 

The multimetallic and monometallic SFHs result in about the same coverage on the \lclrmtl($\lambda$) planes.\ The \lclrmtl($\lambda$) 
relations of monometallic SFHs are generally slightly shallower (especially for the ones with [M/H] = \mhm) but still in good agreement 
with that of multimetallic SFHs (see the {\it solid} linear least-squares regression lines in each panel of Figure \ref{fig_m2lclr_meh}).\ 
Nevertheless, the multimetallic SFHs match the monometallic SFHs with [M/H] = \mhv~better than those with [M/H] = \mhm.\ In particular, 
the relative slope differences of monometallic SFHs of [M/H] = \mhv~with respect to the multimetallic SFHs are ($\lesssim$ 0.3\% -- 3\%) 
generally one order of magnitude smaller than those of monometallic SFHs of [M/H] = \mhm~($\sim$ 5\% -- 30\%).\ Note that the relative slope 
differences for the NIR \lclrmtl($\lambda$) relations are larger than for the optical \lclrmtl($\lambda$) relations.\ Moreover, the half max--min 
ranges of $\Delta$$\log \Upsilon_{\star}$($\lambda$) and $\Delta$(colors) resulting from the monometallic and multimetallic versions of any 
given SFHs are also generally smaller when [M/H] is fixed to \mhv.

As shown in the inset plots of Figure \ref{fig_m2lclr_meh}, $\Delta$$\log \Upsilon_{\star}$($\lambda$) distributions extend to about 2$\times$ 
larger ranges (by $\gtrsim$ 0.05 dex and 0.10 dex for [M/H] = \mhv~and [M/H] = \mhm, respectively) above than below 
$\Delta$$\log \Upsilon_{\star}$($\lambda$) = 0, suggesting that the monometallic versions of given SFHs tend to predict higher 
$\log \Upsilon_{\star}$($\lambda$) than the multimetallic versions.\ Because the differences between stellar masses resulting from the 
monometallic and multimetallic SFHs are negligible ($<$ 0.01 dex), the tendency of monometallic SFHs having higher 
$\log \Upsilon_{\star}$($\lambda$) can be explained if the relatively old stars have higher [M/H] and thus lower luminosities in the monometallic 
SFHs than in the multimetallic SFHs.\ In addition, we note that the half max--min ranges of $\Delta$$\log \Upsilon_{\star}$($\lambda$) for the 
monometallic SFHs with [M/H] = \mhv~are generally $\lesssim$ 0.1 dex, and the half max--min ranges of $\Delta$(color) are $\lesssim$ 0.1 for 
optical--optical/NIR--NIR colors, $\lesssim$ 0.2 for UV--optical colors, and $\lesssim$ 0.3 for optical--NIR colors.\ 

Lastly, we note that subsamples with larger \agem~have relatively larger slope differences between \lclrmtl($\lambda$) relations of monometallic 
and multimetallic SFHs, and they also have larger half max--min ranges of $\Delta$$\log \Upsilon_{\star}$($\lambda$) and $\Delta$(colors) than 
subsamples with smaller \agem.\ We mention that the nearly 2$\times$ larger ranges of $\Delta$$\log \Upsilon_{\star}$($\lambda$) above than 
below 0 mentioned above exist independently of \agem.\ The relative slope differences for optical \lclrmtl($\lambda$) relations vary from $\sim$ 5\% 
for \agem~=11 -- 14 Gyr to $\lesssim$ 1\% for \agem~=3 -- 5 Gyr.\

%The multimetallic version of a given SFH generally has older \agelam~than the monometallic version, this is because older stellar populations 
%in the multimetallic version have relatively lower metallicities than the younger populations and thus have proportionally larger contribution 
%to the integrated light than those in the monometallic version.

%Bershady, M.A., Jangren, J.A., Conselice, C.J. 2000, \aj, 119, 2645

\subsection{Impact of Dust Extinction and Dust Extinction Curves on \lclrmtl($\lambda$) Relations}\label{sec: impact_dust}
Dust extinction and reddening have been generally regarded as at most a secondary effect on stellar mass estimation, due to the 
partially counteracting effect of dust reddening and stellar aging on the \lclrmtl($\lambda$) planes (e.g.\ Bell \& de Jong 2001).\ 
However, as we will show below, the influence of dust extinction and reddening on \lclrmtl($\lambda$) distributions of real galaxies 
depends on several factors, including the dust extinction curves, the amount of dust extinction, the wavelength ranges, and SFHs.\

Here we explore how the \lclrmtl($\lambda$) distributions of our samples vary with dust extinction and dust extinction curves.\ To this end, 
for every ``galaxy'' in our default full sample, we derive additional sets of multiband SEDs by applying $A_{V,{\rm young}}$ values other than our 
default 0.5 mag and using the relatively shallow Charlot \& Fall (2000; CF00) dust extinction curve where $A_{\lambda}$ $\propto$ $\lambda^{-0.7}$.\ 
For reference, the SMC-bar dust extinction curve can be fitted with a power-law index of $\simeq$ $-$1.2 in the optical wavelength range.\ 
The resultant \lclrmtl($\lambda$) distributions of the $A_{V,{\rm young}}$-expanded and the default full samples are shown in Figure \ref{fig_m2lclr_dust}.\

\subsubsection{Impact of the Amount of Dust Extinction}

In the age-dependent dust extinction recipe adopted in this work, the free parameter is the $V$-band extinction suffered by   
populations younger than 40 Myr ($A_{V,{\rm young}}$).\ Extinction linearly decreases with stellar ages from the maximum 
at ages $\leq$ 40 Myr to zero at ages $\geq$ 100 Myr.\ Therefore, for the same $A_{V,{\rm young}}$, galaxies with higher fractions 
of young stellar populations ($<$ 100 Myr) are subject to higher extinction, and due to the proportionally larger light contribution of 
young populations at shorter wavelength than at longer wavelength, they have steeper {\it effective} extinction curves.\ Moreover, for 
the same galaxy, as $A_{V,{\rm young}}$ increases, the {\it effective} extinction curves become progressively shallower, due to the 
proportionally greater degree of decrement in contribution from extinction-affected young populations at shorter wavelengths than 
longer wavelengths.\ At sufficiently large $A_{V,{\rm young}}$, the colors gradually ``saturate,'' but $\Upsilon_{\star}$($\lambda$) 
are systematically higher than those of extinction-free composite stellar populations with only stars older than 100 Myr.\

The \lclrmtl($\lambda$) distributions resulting from our default SMC-type dust extinction curve are illustrated as {\it red} contours in 
Figure \ref{fig_m2lclr_dust}.\ Dust extinction and reddening result in \lclrmtl($\lambda$) distributions with slightly steeper envelopes 
at the blue end and systematic scatter toward lower $\Upsilon_{\star}$ at red colors.\ As discussed above, an increase of $A_{V,{\rm young}}$ 
results in shallower {\it effective} extinction curves and thus steeper dust extinction/reddening vectors on the \lclrmtl($\lambda$) planes.\ 
Therefore, as $A_{V,{\rm young}}$ increases, the regions on the \lclrmtl($\lambda$) planes with extinction-induced scatter move from 
the bottom right to middle right toward redder colors, and as a result, the resultant average optical and NIR \lclrmtl($\lambda$) relations 
first steepen at small $A_{V,{\rm young}}$ ($\lesssim$ 1.0 mag for our sample; Figure \ref{fig_m2lclr_dust}), and then flatten.\ If a variable 
$A_{V,{\rm young}}$ ranging from 0 to 5 mag is considered, the net effect of dust extinction and reddening is to flatten the average slopes 
of \lclrmtl($\lambda$) distributions and increase the median scatter of $\log \Upsilon_{\star}$($\lambda$) for given color values.\ 

The above-mentioned effect of dust extinction/reddening is generally stronger on \lclrmtl($\lambda$) distributions that cover longer-wavelength 
baselines, as is demonstrated in Figure \ref{fig_m2lclr_dust}.\ As examples, for the ($g$$-$$z$)--$\log \Upsilon_{\star}$($z$) distributions affected by 
the SMC-bar dust extinction curve, the average slopes decrease from 0.34 at $A_{V,{\rm young}}$ = 0 mag to 0.23 for a variable $A_{V,{\rm young}}$  
of 0 -- 5 mag, and the median scatter of $\log \Upsilon_{\star}$($z$) increases from 0.18 to 0.27 dex, whereas for the ($B$$-$$V$)--$\log \Upsilon_{\star}$($V$) 
distributions, the average slopes remain the same (0.98) and the median scatter of $\log \Upsilon_{\star}$($V$) increases marginally from 0.26 to 
0.27 dex.\ 

\subsubsection{Impact of the Slopes of Dust Extinction Curves}

\begin{figure*}
\centering
\includegraphics[width=0.9\linewidth]{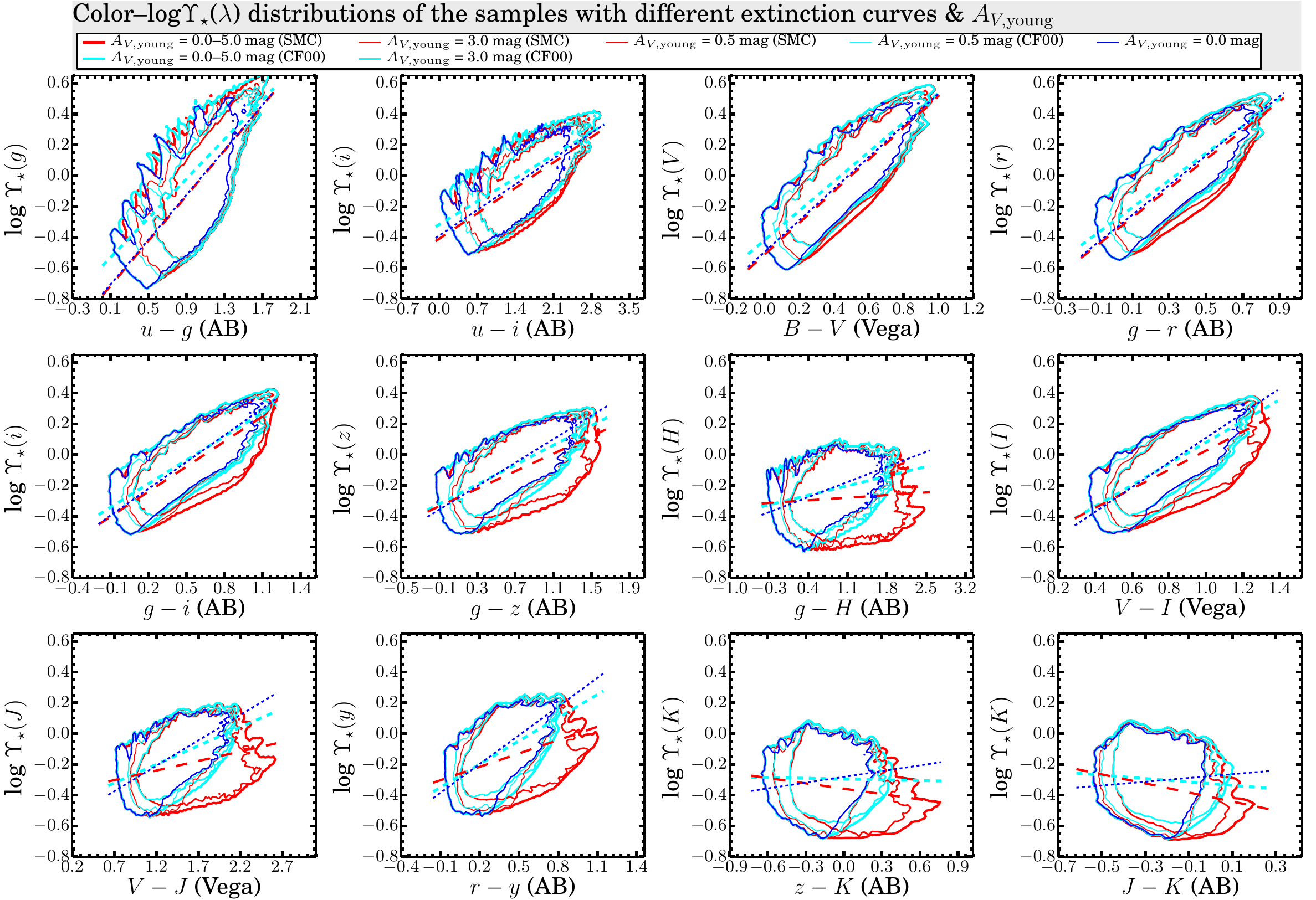}
\caption{
Effect of dust extinction and dust extinction curves on the \lclrmtl($\lambda$) distributions of the full samples.\ Here our default samples are expanded 
in parameter coverage of $A_{V,{\rm young}}$ by adopting a range of $A_{V,{\rm young}}$ from 0 to 5 mag (with a 0.25 mag step size), and 
they are also expanded by adopting the relatively shallow CF00 dust extinction curve.\ The contours in each panel enclose the distributions of the 
samples with different ranges of $A_{V,{\rm young}}$ and/or dust extinction curves of different slopes, as indicated in the figure title.\ The 
SMC-type dust extinction curve has a power-law index of $\simeq$ $-1.2$ in the optical wavelength range, while the CF00 extinction curve 
has a power-law index of $-0.7$.\ In each panel, the {\it blue dotted line} represents the linear least-squares regression to the full samples 
generated with $A_{V,{\rm young}}$ = 0.0 mag, and the {\it short-dashed cyan line} and {\it long-dashed red line} respectively represents the linear 
regression to the full samples resulting from CF00 and SMC-type dust extinction curves, respectively, with a variable $A_{V,{\rm young}}$ of 0 -- 5 mag.
\label{fig_m2lclr_dust}}
\end{figure*}

The above discussion primarily concerns the \lclrmtl($\lambda$) distributions resulting from the relatively steep SMC-bar dust extinction 
curve.\ The \lclrmtl($\lambda$) distributions resulting from the shallow CF00 $\lambda^{-0.7}$ extinction curve are shown as {\it cyan contours} 
in Figure \ref{fig_m2lclr_dust}.\ It can be seen that the CF00 extinction curve induces much smaller variations of the slopes and scatters of the 
\lclrmtl($\lambda$) distributions than the SMC-bar extinction curve.\ As examples, for the ($g$$-$$z$)--$\log \Upsilon_{\star}$($z$) distributions, the 
CF00 dust extinction curve results in negligible slope variation ($<$ 0.01) and a small increase of median scatters from 0.18 to 0.22 dex, whereas 
for the ($g$$-$$r$)--$\log \Upsilon_{\star}$($r$) and ($B$$-$$V$)--$\log \Upsilon_{\star}$($V$) distributions, the CF00 extinction curve results in negligible 
variations of slopes and median scatters.\ Given the different influences of dust extinction curves with different slopes on \lclrmtl($\lambda$) 
distributions, $\log \Upsilon_{\star}$($\lambda$) have the tendency of being overestimated or underestimated if, respectively, erroneously shallower or steeper 
dust extinction curves are used for real galaxies (see also Lo Faro et al.\ 2017).

The age-dependent dust extinction recipe results in steeper {\it effective} extinction curves than does an age-independent dust extinction 
recipe (see also Wild et al.\ 2011).\ While the age-dependent extinction recipe adopted here is physically plausible and provides self-consistent 
solutions for CMD-based SFHs of nearby dwarf galaxies, it might not be unreasonable to expect a nonzero extinction for populations older than 
100 Myr, especially for galaxies with relatively high ISM metallicities.\ Previous modeling of panchromatic SEDs from ultraviolet to far-IR of nearby 
spiral galaxies, although being subject to substantial uncertainties, indicates a non-negligible dust heating by old stellar populations (e.g.\ 
Popescu et al.\ 2011; Bendo et al.\ 2015).\ Compared to our default extinction recipe, a non-negligible amount of extinction suffered by stellar 
populations older than 100 Myr would result in steeper dust extinction/reddening vectors on the \lclrmtl($\lambda$) planes.\

Lastly, we point out that our general conclusion about the impact of SFHs, metallicities, and metallicity evolution on \lclrmtl($\lambda$) 
distributions in previous sections does not change when a variable $A_{V,{\rm young}}$ or a different dust extinction curve is adopted.\ 
In particular, at low dust extinctions (i.e.\ $A_{V,{\rm young}}$ $\lesssim$  0.5 mag), there are only marginal differences between 
\lclrmtl($\lambda$) distributions resulting from the steep SMC-bar and the shallow CF00 extinction curves.\

\section{$\Upsilon_{\star}$($\lambda$) Estimation with Weighted Ages, Metallicities and Colors}\label{sec: m2l_agezl}

As is found in this work, \agelam~and \mhlam~of composite stellar populations are probably the closest analogs to ages and metallicities 
of SSPs, as far as the \lclrmtl($\lambda$) relations are concerned.\ In particular, $\Upsilon_{\star}$($\lambda$) has the strongest correlation 
with \agelam.\ For our default full sample, the Spearman's rank correlation coefficients $r_{s}$ between \agelam\ and $\Upsilon_{\star}$($\lambda$) 
peak at 0.96 for the $y$ and $J$ bands and decrease toward both shorter- and longer-wavelength passbands (0.88 for $g$, 0.94 for $M$).\ Ignoring 
variations of chemical abundance patterns, ages and metallicities are the two parameters that uniquely determine $\Upsilon_{\star}$($\lambda$) of 
SSPs.\ Then the question is, to what extent can the \agelam~and \mhlam~constrain $\Upsilon_{\star}$($\lambda$) of galaxies with a wide diversity 
of SFHs? Understanding the limitation of \agelam, \mhlam, and colors in constraining $\Upsilon_{\star}$($\lambda$) is particularly important, because 
they are the relevant parameters for planing future observing campaigns and fundamental physical observables that can be readily obtained from 
spectroscopic or photometric observations.\

\subsection{SFHs with Associated Metallicity Evolution}
The uncertainties of the $\log \Upsilon_{\star}$($\lambda$) estimates, as quantified as the maximum of the half max--min ranges of 
$\log \Upsilon_{\star}$($\lambda$) within given small intervals of the relevant parameter values, in different passbands are shown in 
Figure \ref{fig_m2l_agez} (see the caption for details).\ With \agelam~alone, $\log \Upsilon_{\star}$($\lambda$) in $J$ band is predicted 
with an uncertainty of $\simeq$ 0.12 dex, and the uncertainties increase up to 0.25 in the $g$ band and 0.16 in the $M$ band.\ With 
additional constraints from \mhlam, the estimation of $\log \Upsilon_{\star}$($\lambda$) improves for all passbands, with the degree of 
improvement being larger for passbands either longer or shorter than $z$ or $y$.\ A stronger sensitivity to \agelam~and a weaker sensitivity 
to \mhlam~for $\Upsilon_{\star}$($\lambda$) of passbands around the boundary between optical and NIR wavelength ranges are analogous 
to similar findings for SSPs (Worthey 1994).\ \agelam~and \mhlam~together predict $\log \Upsilon_{\star}$($\lambda$) with uncertainties of 
$\simeq$ 0.10 dex for $J$ and longer-wavelength passbands, while shortward of $J$, the uncertainties become larger at shorter-wavelength 
passbands.\ We note that the uncertainties reported in Figure \ref{fig_m2l_agez} remain about the same if, say, 2$\times$ smaller binning sizes 
for \agelam~and \mhlam~are used.\

Therefore, at a given \agelam~and \mhlam, the scatter of $\log \Upsilon_{\star}$($\lambda$) induced by the diversity of SFHs and metallicity 
evolution histories are on the order of $\lesssim$ 0.10 dex for $J$ and longer-wavelength passbands, and the induced scatter becomes 
increasingly larger toward shorter-wavelength passbands.\ In particular, the scatter in $g$ band is nearly 2$\times$ larger than in NIR.\ 
By adding an additional constraint from \agem, the estimation of $\log \Upsilon_{\star}$($\lambda$) improves by $\sim$ 0.03 dex for nearly 
all passbands.\ The dependence of $\log \Upsilon_{\star}$($\lambda$) on \agem~for given \agelam\ and \mhlam~is further demonstrated in 
Figure \ref{fig_m2l_agez_con}, whereby one can see a negative correlation between $\log \Upsilon_{\star}$($\lambda$) and \agem, especially 
at relatively young \agelam.\ This negative correlation indicates that \agem~can induce systematic uncertainties to the $\log \Upsilon_{\star}$($\lambda$) 
estimation of up to $\sim$ 0.2 dex.\ No single color is particularly sensitive to \agem~in general.\ However, at a given \agelam~and \mhlam, 
colors have negative correlations with \agem~(e.g.\ Figure \ref{fig_m2l_agezclr_con}).\ By combining \agelam, \mhlam~and colors, 
$\log \Upsilon_{\star}$($\lambda$) can be constrained with uncertainties ({\it bottom} panel of Figure \ref{fig_m2l_agez}) similar to that by 
\agelam, \mhlam~and \agem.\ We point out that various optical colors perform similarly well in this regard.\

The above discussion concerns the ideal cases where $\log \Upsilon_{\star}$($\lambda$) can be estimated through \agelam~and \mhlam~in 
the relevant passbands.\ However, in practice, \agelam~and \mhlam~are usually estimated only for a limited wavelength range.\ For example, 
people usually use the standard Lick indices (e.g.\ Worthey et al.\ 1994) defined in the optical wavelength range to estimate (light-weighted) 
ages and metallicities.\ The few Lick indices with the strongest sensitivities to ages (e.g.\ H$\beta$) and metallicities (e.g.\ Fe5270, Fe5335, Mg$_2$) 
are mainly located in the wavelength range covered by the $g$ band.\ Figure \ref{fig_m2l_agez} also shows the uncertainties (in {\it red} colors) of 
$\log \Upsilon_{\star}$($\lambda$) as predicted solely by $g$-band light-weighted ages \ageg~and metallicities \mhg.\ Compared to using 
\agelam~and \mhlam, the uncertainties of $\log \Upsilon_{\star}$($\lambda$) estimation increase by as much as $\sim$ 0.05 dex for 
longer-wavelength passbands.\ At a given \ageg~and \mhg, the scatter of $\log \Upsilon_{\star}$($\lambda$) induced by the diversity of SFHs 
and metallicity evolution histories is on the order of $\lesssim$ 0.15 dex for passbands with wavelengths longward of $g$.\ It is worth pointing out that, 
by using \ageg~and \mhg, the advantage of NIR over shorter-wavelength passbands (e.g.\ $V$) in stellar mass estimation becomes very 
weak ($\sim$ 0.01 -- 0.03 dex).\

\subsection{Monometallic SFHs}
The panels on the right-hand side of Figure \ref{fig_m2l_agez} show the expected uncertainties of $\log \Upsilon_{\star}$($\lambda$) estimates  
with \agelam, \mhlam, and colors, in the context of monometallic SFHs.\ Here [M/H] of each individual monometallic SFH in our sample is 
fixed to the $V$-band light-weighted [M/H], i.e.~\mhv, as determined from our default multimetallic ``version'' of SFHs.\ Note that the results, including 
the values annotated in Figure \ref{fig_m2l_agez}, do not depend on how we set [M/H] of the monometallic SFHs.\

Compared to the cases of multimetallic SFHs discussed above, \agelam~predicts $\log \Upsilon_{\star}$($\lambda$) to very similar accuracies.\ 
By including additional constraints from \mhlam, \agem, or colors, the uncertainties of $\log \Upsilon_{\star}$($\lambda$) estimates are also 
very similar to those of the multimetallic cases for $J$ and longer wavelength passbands, but shortward of $J$, the uncertainties of 
$\log \Upsilon_{\star}$($\lambda$), which are about the same as those in $J$ and longer-wavelength passbands, are $\sim$ 2$\times$ smaller 
than that in the multimetallic cases.\ By combining \agelam, [M/H], and colors, the uncertainties of $\log \Upsilon_{\star}$($\lambda$) estimates  
are well below 0.10 dex in all but the $M$ bands, and the uncertainties are actually larger in the NIR passbands than the optical passbands.

Therefore, by using monometallic SFH models to interpret \ageg~and \mhg~determined from optical spectra of real galaxies, one tends to 
underestimate the uncertainties of $\log \Upsilon_{\star}$($\lambda$) by $\lesssim$ 0.1 dex for the optical passbands and $\lesssim$ 0.05 dex 
for the NIR passbands.

\subsection{Scatter Induced by Dust Extinction and Reddening}
The above two subsections concern the uncertainties of $\log \Upsilon_{\star}$($\lambda$) estimates induced by SFHs and metallicity evolution.\
Here we explore the additional uncertainties induced by a variable $A_{V, {\rm young}}$, under the assumption of an SMC-bar dust 
extinction curve that has a larger influence on \lclrmtl($\lambda$) distributions than shallower ones (Section \ref{sec: impact_dust}).\

\begin{figure*}
\centering
\includegraphics[width=0.90\linewidth]{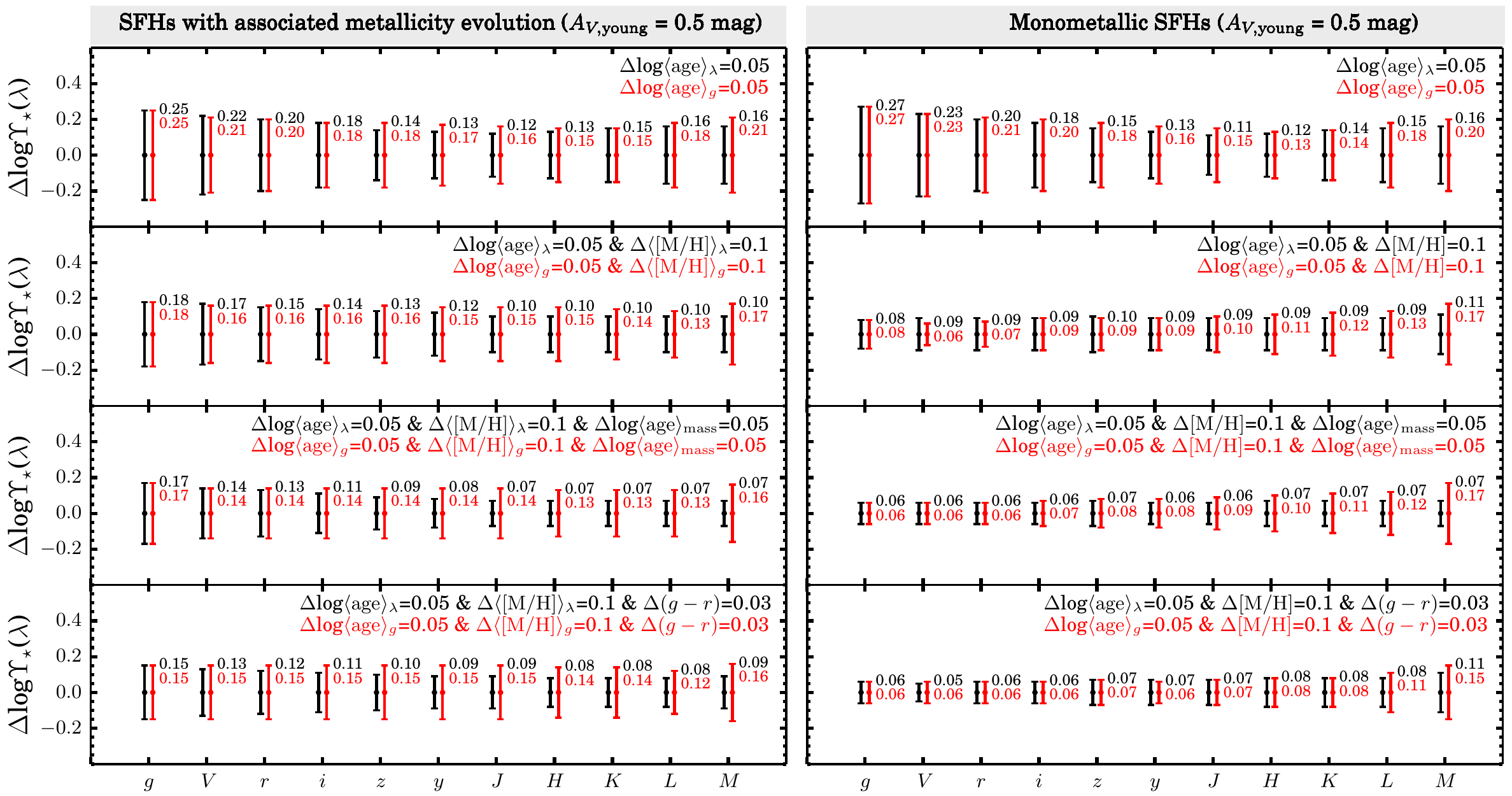}
\caption{
Maximum of half max--min ranges of $\log \Upsilon_{\star}$($\lambda$) constrained by light-weighted ages \agelam, light-weighted 
metallicities \mhlam, \agem, and ($g$$-$$r$) colors.\ The panels on the {\it left}-hand side are for the SFHs of the full sample with associated 
metallicity evolution, while the panels on the {\it right}-hand side are for monometallic SFHs of the full sample with [M/H] fixed to their 
respective $V$-band light-weighted [M/H].\ From the top to bottom panels, the results are constrained by log\agelam~(0.05 dex intervals); 
by log\agelam~and \mhlam~(0.05$\times$0.1 dex intervals); by log\agelam, \mhlam, and log\agem~(0.05$\times$0.1$\times$0.05 dex 
intervals); and by log\agelam, \mhlam, and ($g$$-$$r$) colors (0.05$\times$0.1$\times$0.03 intervals).\ In each panel, the {\it black} symbols represent 
results as constrained by light-weighted parameters in passbands $\lambda$ indicated by the x-axis tick labels, while the {\it red} symbols illustrate 
the corresponding results as constrained by $g$-band light-weighted parameters.\ The numerical values of $\Delta$$\log \Upsilon_{\star}$($\lambda$) 
are shown as labels correspondingly.\ See Section \ref{sec: m2l_agezl} for discussions.
\label{fig_m2l_agez}}
\end{figure*}

\begin{figure*}
\centering
\includegraphics[width=0.9\linewidth]{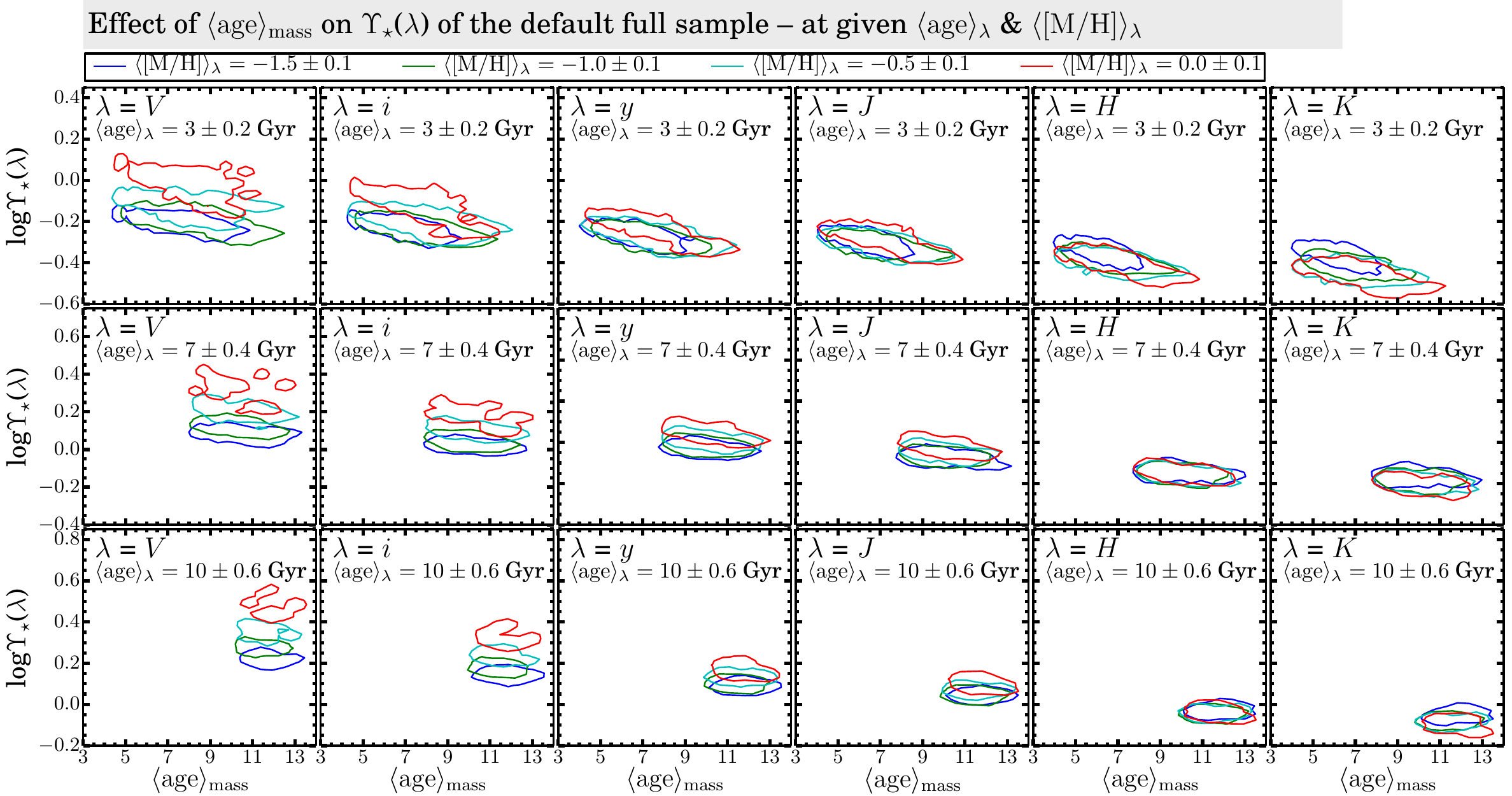}
\caption{
Effect of \agem~on $\log \Upsilon_{\star}$($\lambda$), at given \agelam~and \mhlam.\
Here the effect is illustrated for several representative passbands as indicated in the figure.\
See Section \ref{sec: m2l_agezl} for discussions.
\label{fig_m2l_agez_con}}
\end{figure*}

\begin{figure*}
\centering
\includegraphics[width=0.9\linewidth]{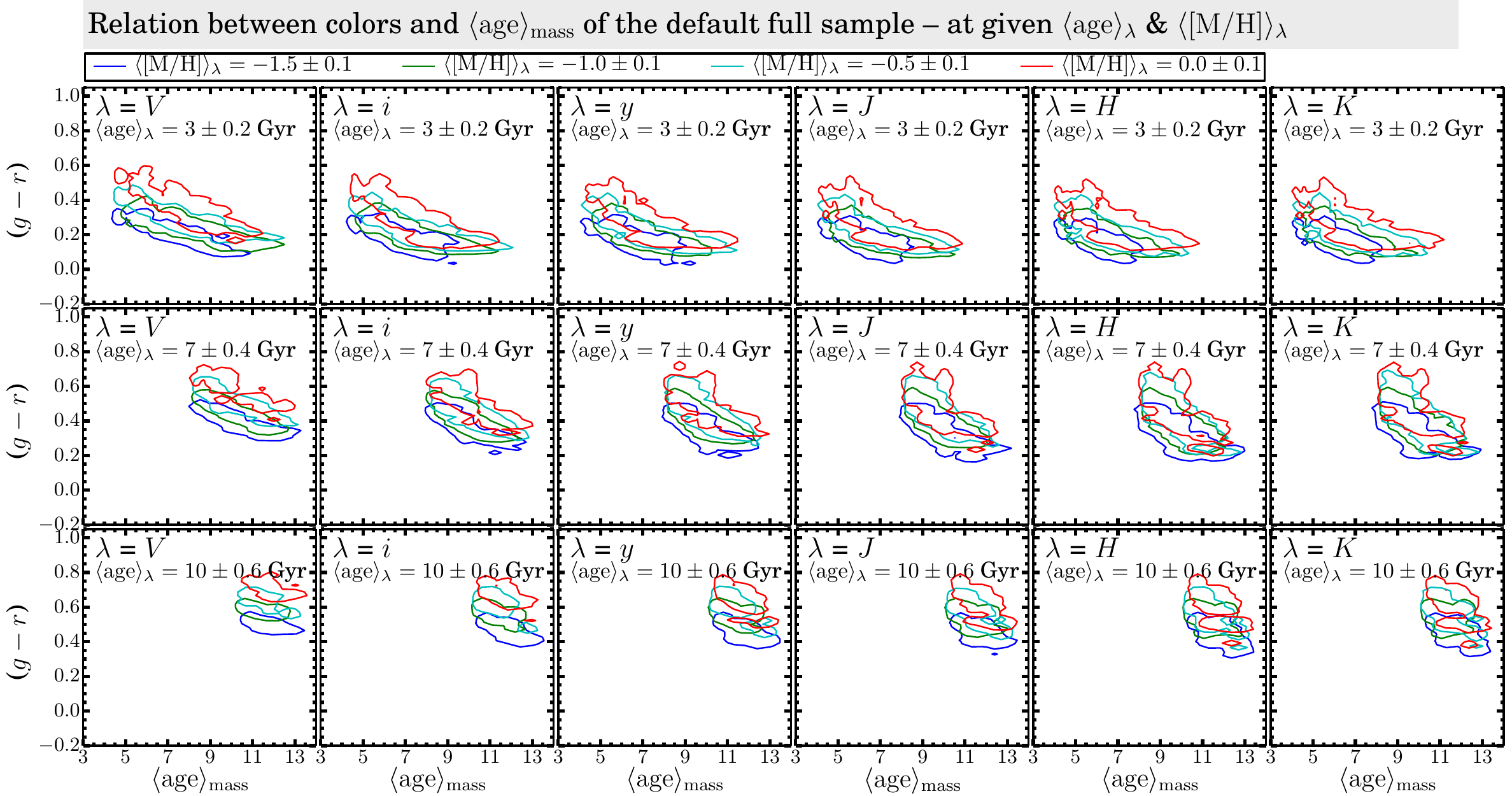}
\caption{
Relation between ($g$$-$$r$) and \agem~at given \agelam~and \mhlam.\
Here the effect is illustrated for \agelam~and \mhlam~in several representative passbands as indicated in the figure.\
See Section \ref{sec: m2l_agezl} for discussions.
\label{fig_m2l_agezclr_con}}
\end{figure*}

\begin{figure*}
\centering
\includegraphics[width=0.90\linewidth]{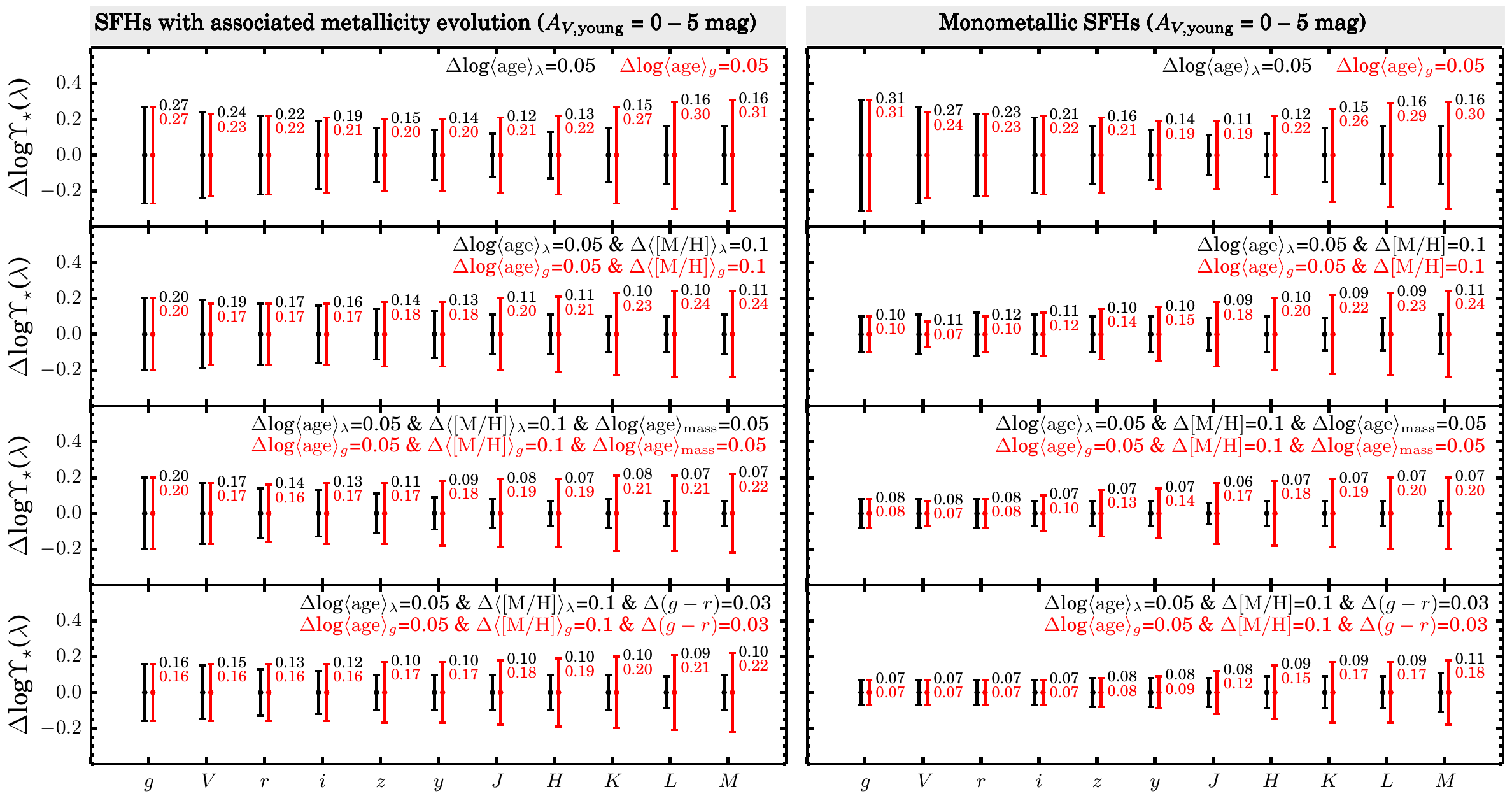}
\caption{
Same as Figure \ref{fig_m2l_agez}, but for our full samples expanded with variable $A_{V, {\rm young}}$ ranging from 0.0 to 0.5 mag.\
See Section \ref{sec: m2l_agezl} for discussions.
\label{fig_m2l_agez_av}}
\end{figure*}

Like in Figure \ref{fig_m2l_agez}, the results for the samples expanded with variable $A_{V, {\rm young}}$ from 0 to 5 mag are shown 
in Figure \ref{fig_m2l_agez_av}.\ If \agelam~and \mhlam~are invoked to estimate $\log \Upsilon_{\star}$($\lambda$) for the case of 
variable $A_{V, {\rm young}}$, $\log \Upsilon_{\star}$($\lambda$) can be constrained with uncertainties very similar to those of the low 
dust extinction case.\ Nevertheless, if \ageg~and \mhg~are used, $\log \Upsilon_{\star}$($\lambda$) estimation is subject to uncertainties 
of $\gtrsim$ 0.20 dex for nearly all passbands, and the uncertainties for the NIR passbands are slightly larger than those for the optical passbands.\

\subsection{The Optimal Passbands and Colors for $\Upsilon_{\star}$($\lambda$) Estimates}

\subsubsection{The Optimal Single Passbands}\label{sec: optsingle}
\begin{figure}
\centering
\includegraphics[width=0.90\linewidth]{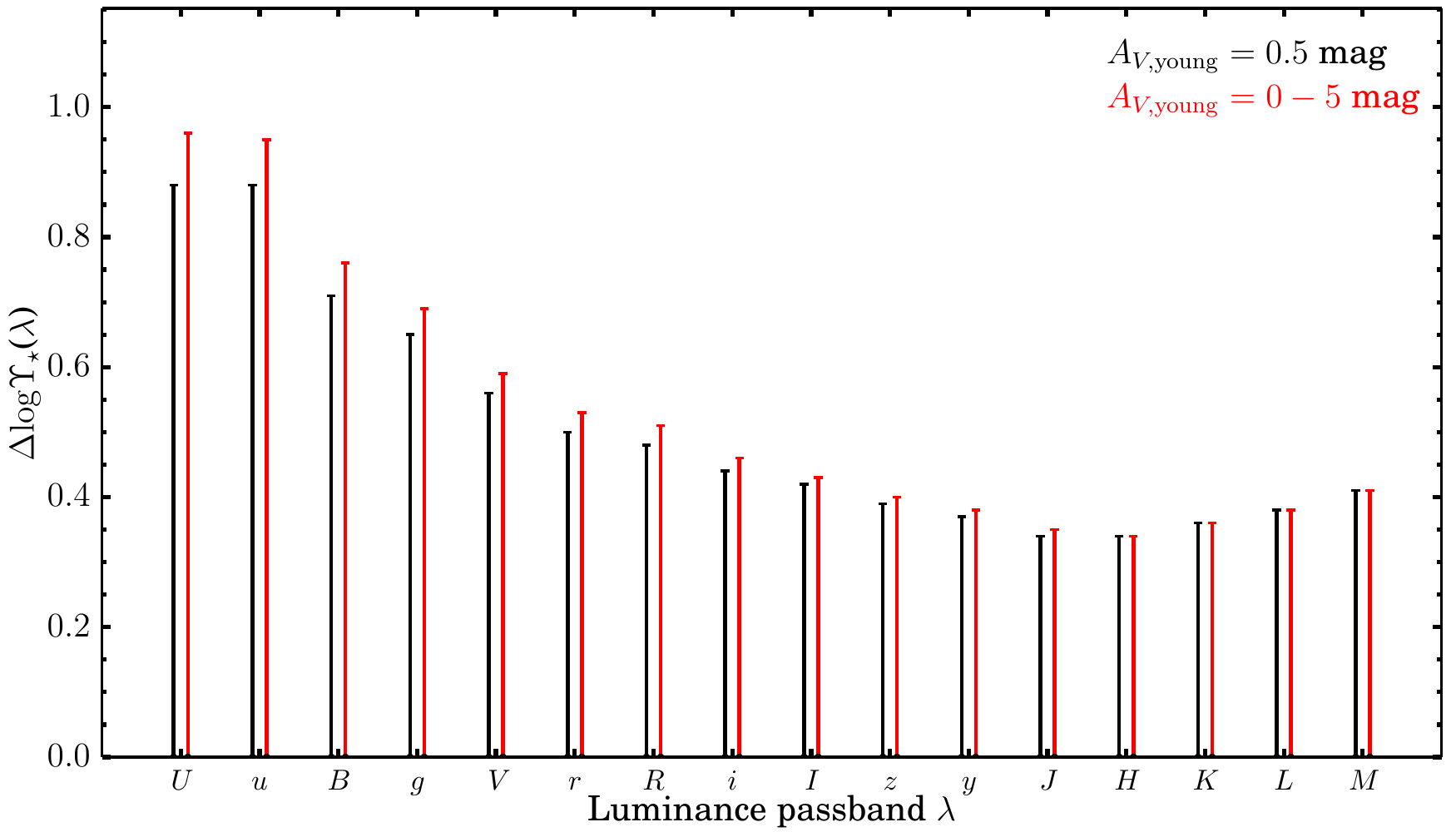}
\caption{
Half max--min ranges of $\log \Upsilon_{\star}$($\lambda$) of different passbands.\
Results for our default full sample of $A_{V, {\rm young}}$ = 0.5 mag ({\it black}) and the full samples with variable $A_{V, {\rm young}}$ 
ranging from 0.0 to 5.0 mag ({\it red}) are shown separately.\
\label{fig_m2l_single}}
\end{figure}

As shown in Figure \ref{fig_m2l_single}, $\log \Upsilon_{\star}$($\lambda$) of the $J$ and $H$ bands have the smallest half max--min ranges 
$\Delta$$\log \Upsilon_{\star}$($\lambda$) ($\simeq$ 0.35 dex) among all passbands, regardless of dust extinction.\ Passbands longward and 
shortward of $JH$ have progressively larger $\Delta$$\log \Upsilon_{\star}$($\lambda$).\ As noted in previous sections, 
$\log \Upsilon_{\star}$($\lambda$) of passbands with wavelength around the $J$ band show the strongest sensitivities to \agelam~and the weakest 
sensitivities to \mhlam, while $\log \Upsilon_{\star}$($\lambda$) of passbands at either longer or shorter wavelengths have increasingly larger sensitivities 
to \mhlam~and smaller sensitivities to \agelam.\ Although we do not deal with systematic uncertainties inherent in stellar evolution models in 
this work, it is worth emphasizing that modeling of NIR emission from stellar populations of several hundred Myr to $\sim$ 3 Gyr is subject to 
substantial uncertainties, due to our poor knowledge of the TP-AGB phase of stellar evolution.

\subsubsection{The Optimal Single \lclrmtl($\lambda$) Relations}\label{sec: singlecolor}
\begin{figure*}
\centering
\includegraphics[width=0.8\linewidth]{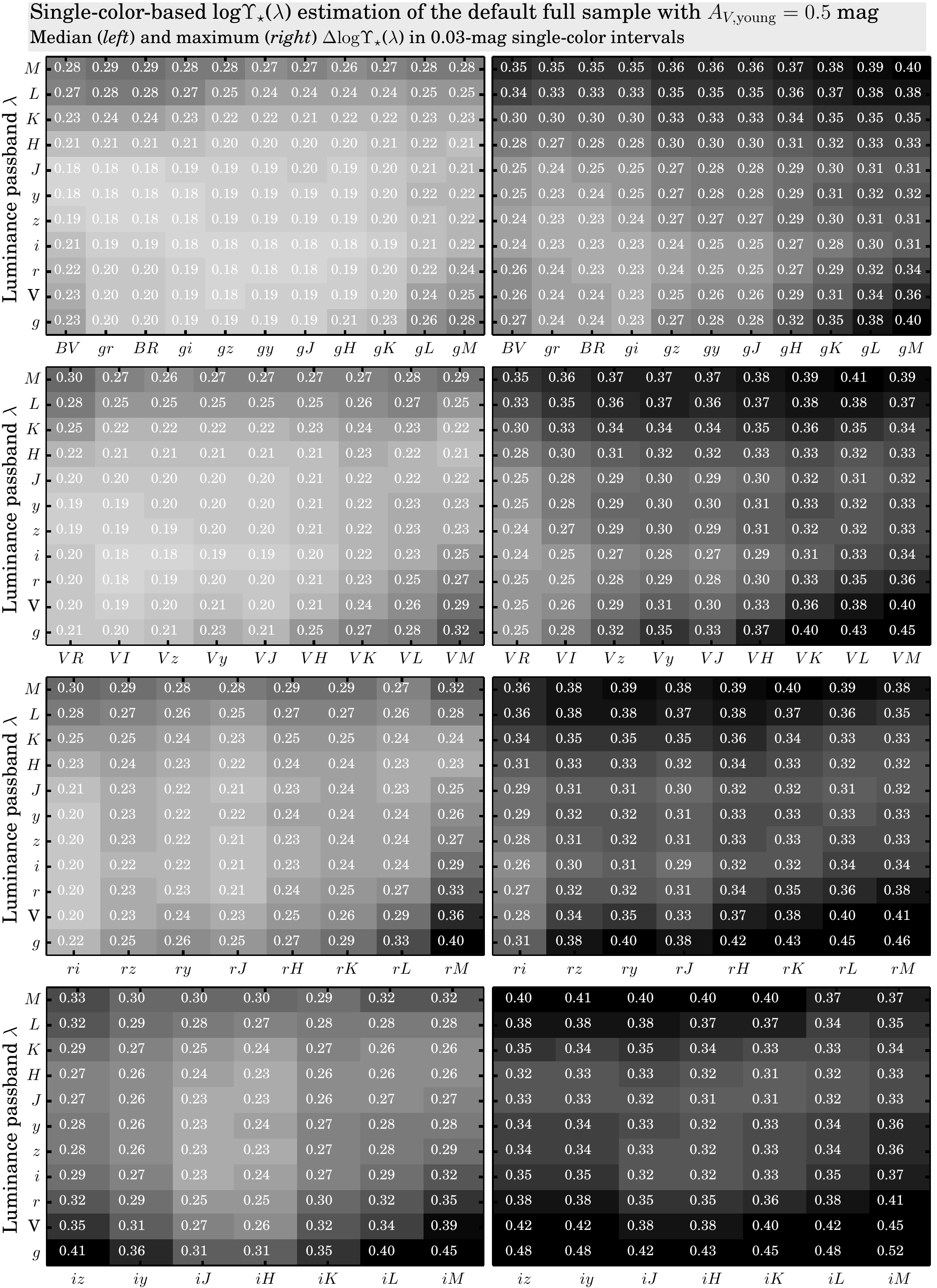}
\caption{
Uncertainties of $\log \Upsilon_{\star}$($\lambda$) estimated by one single color for the our default full sample, 
with $A_{V, {\rm young}}$ = 0.5 mag.\ Colors are labeled on the x axes, and the luminance passband $\lambda$ 
are labeled on the y axes.\ The uncertainties $\Delta$$\log \Upsilon_{\star}$($\lambda$) are quantified as either the 
median ({\it left panels}) or maximum ({\it right panels}) of the half max--min ranges of $\log \Upsilon_{\star}$($\lambda$) 
in 0.03 mag color intervals.\ $\Delta$$\log \Upsilon_{\star}$($\lambda$) is represented in gray scale (larger 
$\Delta$$\log \Upsilon_{\star}$($\lambda$) are represented with darker gray scales) and is also annotated correspondingly 
in each cell.
\label{fig_m2lclr_av05}}
\end{figure*}

\begin{figure*}
\centering
\includegraphics[width=0.8\linewidth]{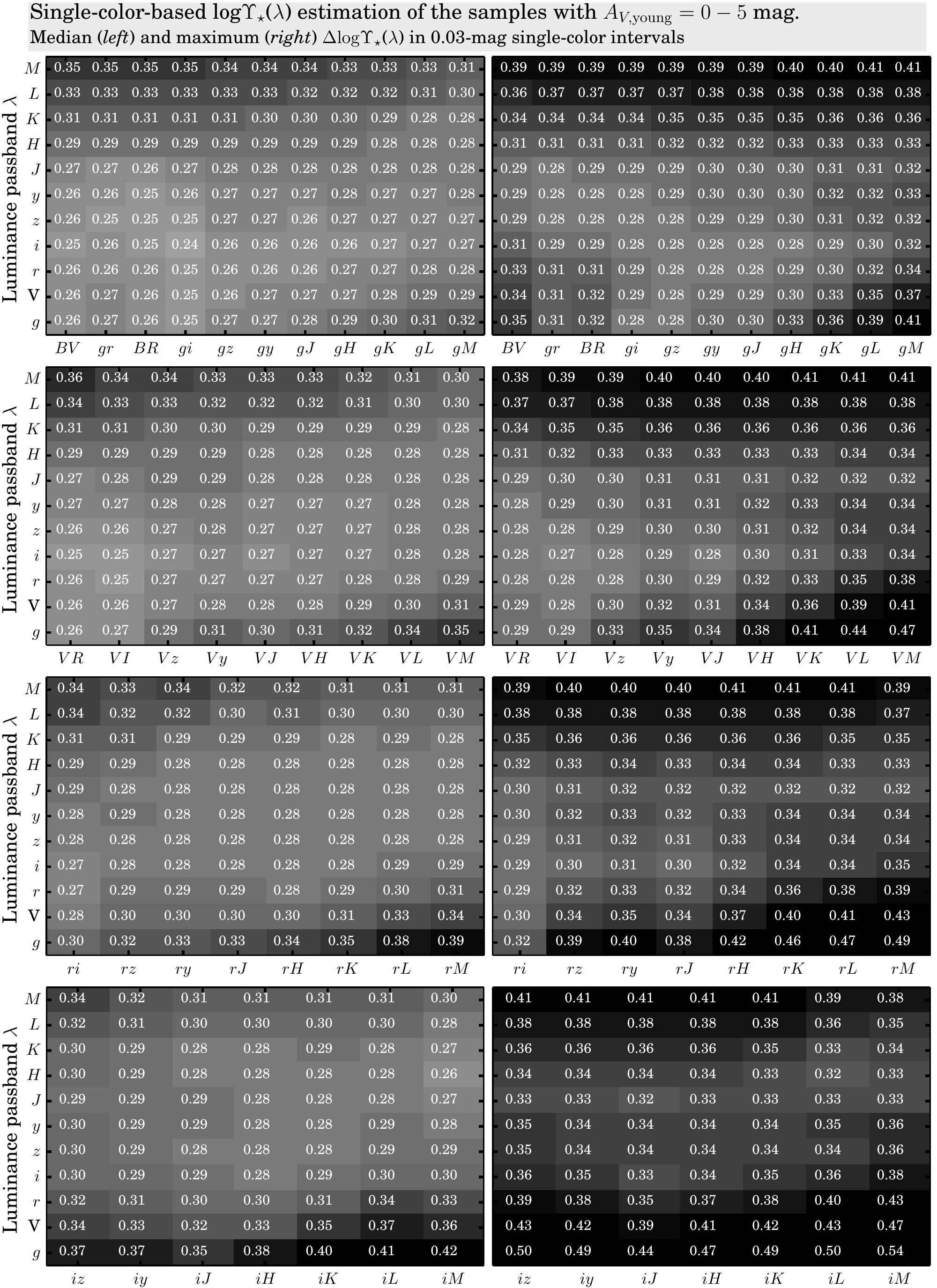}
\caption{
Same as Figure \ref{fig_m2lclr_av05}, but for the full samples with $A_{V, {\rm young}}$ ranging from 0.0 to 5.0 mag.
\label{fig_m2lclr_avall}}
\end{figure*}

Color--$\log \Upsilon_{\star}$($\lambda$) correlations help to reduce the uncertainties of $\log \Upsilon_{\star}$($\lambda$) estimation for all passbands.\
Figure \ref{fig_m2lclr_av05} shows the median and maximum $\Delta$$\log \Upsilon_{\star}$($\lambda$) of various passbands 
$\lambda$ in 0.03 mag single-color intervals of our default full sample with $A_{V, {\rm young}}$ = 0.5 mag.\ Likewise, 
Figure \ref{fig_m2lclr_avall} shows the results for the full sample with variable $A_{V, {\rm young}}$.\ Compared to the cases 
without color information (Section \ref{sec: optsingle}), single colors help to reduce systematic uncertainties in the $\log \Upsilon_{\star}$($\lambda$) estimation 
by factors of $\sim$ 1.6 -- 2.5 for the sample with $A_{V, {\rm young}}$ fixed to 0.5 mag and of $\sim$ 1.3 -- 2.2 for the samples with 
variable $A_{V, {\rm young}}$.\ In particular, $\log \Upsilon_{\star}$($\lambda$) of passbands shortward of $H$ can generally be better 
constrained by single colors than that of longer-wavelength passbands.\ For given $\log \Upsilon_{\star}$($\lambda$), colors covering 
longer wavelength baselines toward the blue end, preferably $B$ or $g$, generally offer tighter constraints on $\log \Upsilon_{\star}$($\lambda$) 
than the others.\

Taking $\Delta$$\log \Upsilon_{\star}$($\lambda$) reported in Figures \ref{fig_m2lclr_av05} and \ref{fig_m2lclr_avall} at face value, 
$\log \Upsilon_{\star}$($i$) can be constrained with the smallest half max--min ranges by using the ($g$$-$$i$) color.\ However, 
$\log \Upsilon_{\star}$($\lambda$) of other passbands, such as $J$, $y$, $z$, $r$ and $V$, can be constrained with very similar   
$\Delta$$\log \Upsilon_{\star}$($\lambda$) (within $\sim$ 0.01 dex).\ We note that ($g$$-$$i$) is the preferred color for $\Upsilon_{\star}$ 
estimation by Z09.\ Gallazzi \& Bell (2009) reached a similar conclusion that ($g$$-$$i$) offers the tightest constraint on $\Upsilon_{\star}$($z$).\ 
Nevertheless, $\Upsilon_{\star}$($\lambda$) of longer-wavelength passbands are subject to larger {\it systematic} uncertainties induced 
by SFHs (e.g.\ \agelam, \agem), and \lclrmtl($\lambda$)~relations (e.g.\ slopes and intercepts) defined by colors covering longer-wavelength 
baselines are subject to larger {\it systematic} influences from metallicities, dust extinction and reddening (Sections \ref{sec: impact_sfh} and 
\ref{sec: impact_dust}).\ The linear slopes and intercepts of ($g$$-$$r$)--$\log \Upsilon_{\star}$($r$) and {\it especially} ($B$$-$$V$)--$\log \Upsilon_{\star}$($V$)
relations have the weakest dependencies on \agelam, \mhlam, \agem, dust extinction, and reddening and thus are the optimal choices for 
a {\it least-biased} estimation of $\Upsilon_{\star}$($\lambda$), if no additional constraints on SFHs and metallicities are available.\ It is noteworthy that, for our 
samples with variable $A_{V, {\rm young}}$, the difference between median $\Delta$$\log \Upsilon_{\star}$($V$) as estimated by ($B$$-$$V$) and 
median $\Delta$$\log \Upsilon_{\star}$($i$) as estimated by ($g$$-$$i$) is only $\simeq$ 0.02 dex.\

\subsubsection{The Optimal Combinations of Two Colors}
\begin{figure}
\centering
\includegraphics[width=0.97\linewidth]{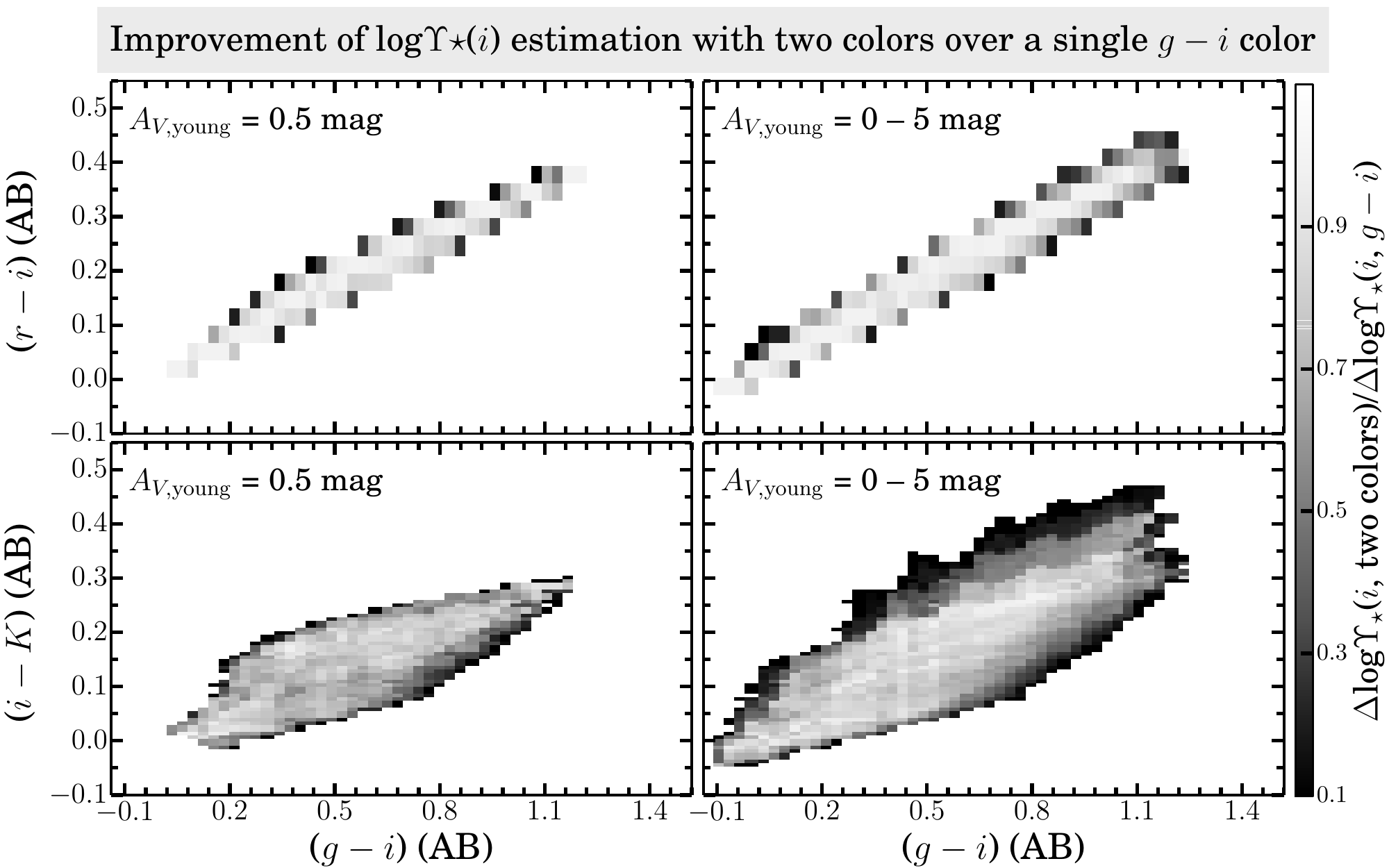}
\caption{
Improvement of $\log \Upsilon_{\star}$($i$) estimation with combinations of ($g$$-$$i$) and ($r$$-$$i$) ({\it top panels}) and of ($g$$-$$i$) 
and ($i$$-$$K$) ({\it bottom panels}) over using a single ($g$$-$$i$) color.\ The distributions for the samples of $A_{V, {\rm young}}$ = 0.5 
mag and of $A_{V, {\rm young}}$ = 0 -- 5 mag are shown in the {\it left} and {\it right panels} respectively.\ The half max--min 
ranges of $\Delta$$\log \Upsilon_{\star}$($i$) are determined in each 0.03$\times$0.03 color interval.\ The gray scales represent the 
ratios of $\Delta$$\log \Upsilon_{\star}$($i$) from combinations of two colors to $\Delta$$\log \Upsilon_{\star}$($i$) from a single ($g$$-$$i$) 
color.
\label{fig_2clr_m2l}}
\end{figure}

\begin{figure}
\centering
\includegraphics[width=0.97\linewidth]{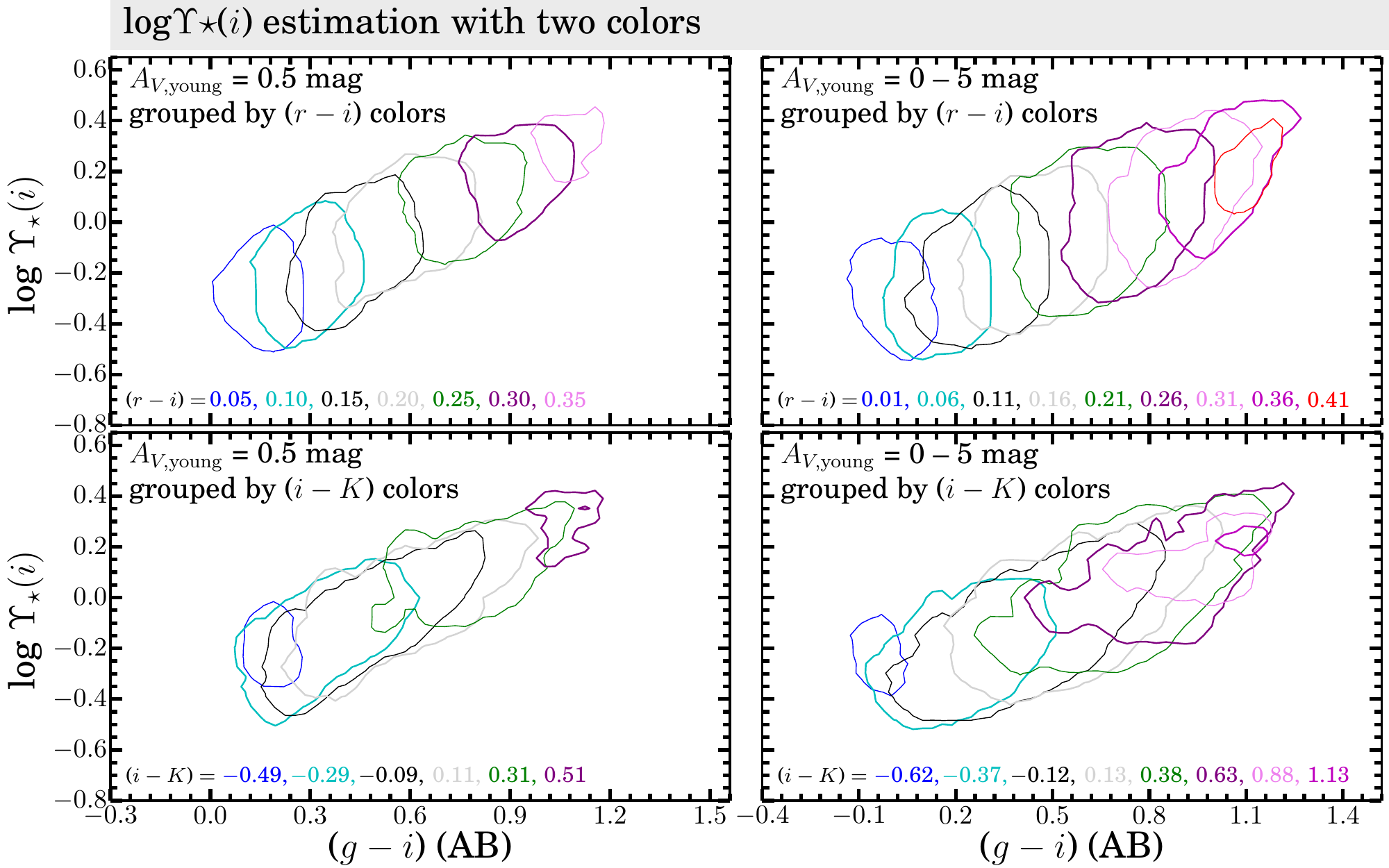}
\caption{
Distributions of subsamples with different $r$$-$$i$ or $i$$-$$K$ colors on the ($g$$-$$i$)--$\log \Upsilon_{\star}$($i$) planes.\ 
The full samples with $A_{V, {\rm young}}$ = 0.5 mag ({\it left panels}) and $A_{V, {\rm young}}$ = 0 -- 5 mag ({\it right panels}) 
are grouped into different $r$$-$$i$ ({\it top panels}) and $i$$-$$K$ ({\it bottom panels}) color values.\ Each contour encloses subsamples 
within a color interval of 0.05 mag around the values annotated at the bottom of each panel.
\label{fig_m2l2clr_con}}
\end{figure}

$\Upsilon_{\star}$($\lambda$) can be better constrained with more than one color.\ As examples, Figure \ref{fig_2clr_m2l} 
shows the color-dependent improvement of $\log \Upsilon_{\star}$($i$) estimation by combining a second optical--optical ($r$$-$$i$) or 
an optical--NIR ($i$$-$$K$) color with the ($g$$-$$i$)--$\log \Upsilon_{\star}$($i$) relations.\ The half max--min ranges of $\log \Upsilon_{\star}$($i$) 
$\Delta$$\log \Upsilon_{\star}$($i$) are generally reduced when ($r$$-$$i$) or {\it especially} ($i$$-$$K$) is included, and the degree of improvement 
is generally larger toward either the bluer or redder edges of colors.\ The efficacy of a second color is further illustrated in Figure 
\ref{fig_m2l2clr_con}, where the distributions of galaxies with different ($r$$-$$i$) or ($i$$-$$K$) color values are distinguished on the 
($g$$-$$i$)--$\log \Upsilon_{\star}$($i$) planes.\ At a low dust extinction of $A_{V, {\rm young}}$ = 0.5 mag (left panels of Figure 
\ref{fig_m2l2clr_con}), the iso-($i$$-$$K$) contours largely follow that of the $i$-band light-weighted [M/H] (Figure \ref{fig_m2lclr_7}) 
on the ($g$$-$$i$)--$\log \Upsilon_{\star}$($i$) planes, suggesting that the effect of an additional optical--NIR color is to partially remove 
the dependence of $\log \Upsilon_{\star}$($i$) on [M/H].\ Due to the age-metallicity degeneracy, galaxies with older \agelam~tend to 
have smaller \mhlam~for given color values.\ Therefore, by partially removing the [M/H] dependence, optical--NIR colors help to improve 
the constraints on \agelam~and thus $\log \Upsilon_{\star}$($\lambda$).\ On the other hand, at high dust extinctions (e.g.\ right panels 
of Figure \ref{fig_m2l2clr_con}), galaxies with higher dust extinction have on average redder optical-NIR colors for given optical color 
values, so optical--NIR colors help to partially remove the positive dependence of $\log \Upsilon_{\star}$($\lambda$) on dust extinction.\

\begin{figure*}
\centering
\includegraphics[width=0.8\linewidth]{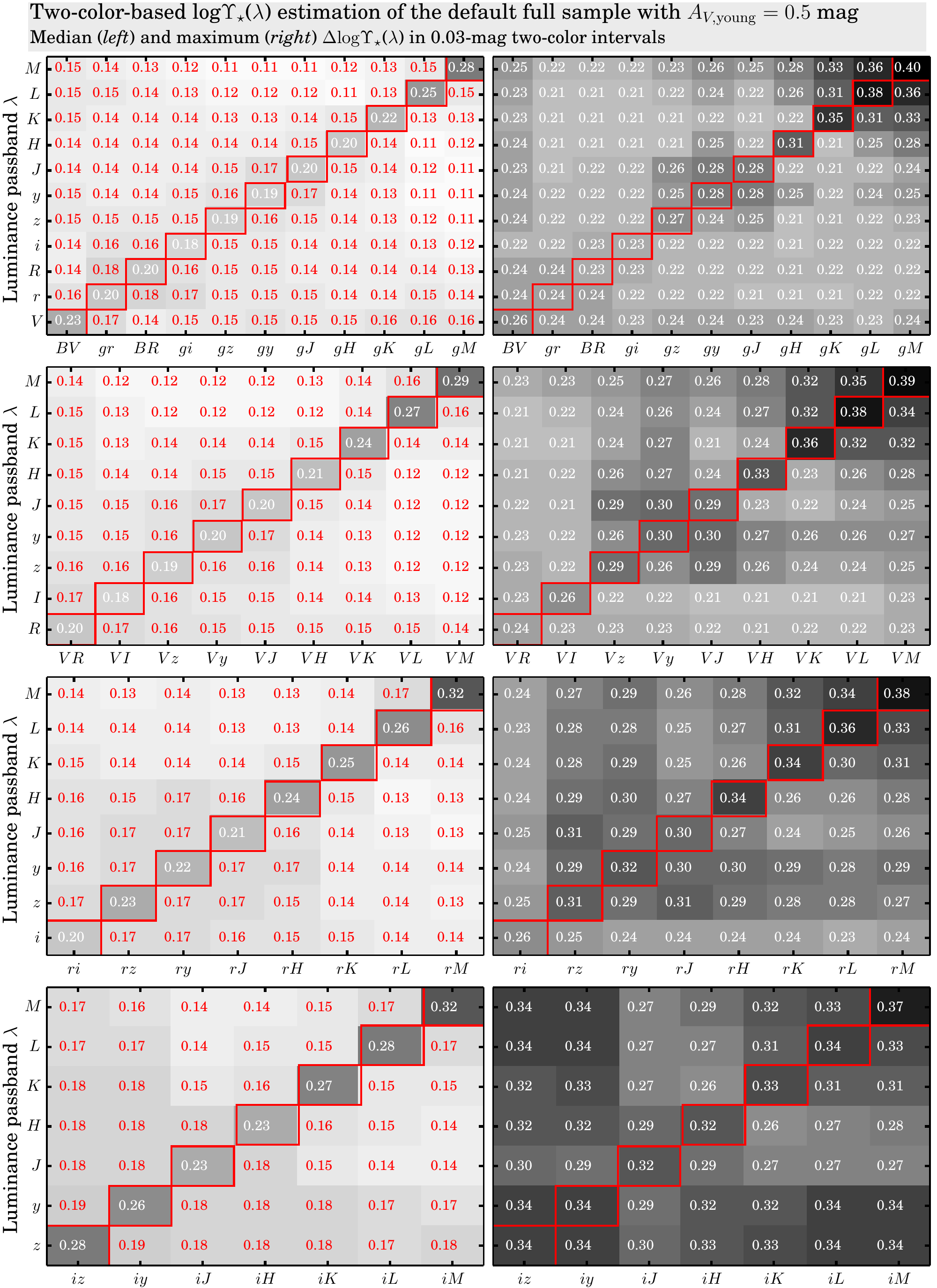}
\caption{
Uncertainties of $\log \Upsilon_{\star}$($\lambda$) estimated by two colors (except for the cells along the diagonal of each panel) 
of the our default full sample, with $A_{V, {\rm young}}$ = 0.5 mag.\ Pairs of passbands as labeled on the x axes, together with 
passbands $\lambda$ as labeled on the y axes, are used to estimate $\log \Upsilon_{\star}$($\lambda$).\ The uncertainties 
$\Delta$$\log \Upsilon_{\star}$($\lambda$) are quantified as either the median ({\it left panels}) or maximum ({\it right panels}) 
of the half max-min ranges of $\log \Upsilon_{\star}$($\lambda$).\ $\Delta$$\log \Upsilon_{\star}$($\lambda$) is represented in  
gray scale (larger $\Delta$$\log \Upsilon_{\star}$($\lambda$) are represented with darker gray scales) and is also annotated 
correspondingly in each cell.\ Note that the text colors, either red or white, are chosen just to increase the contrast with respect 
to the background for a better readability.
\label{fig_m2lclrs_av05}}
\end{figure*}

\begin{figure*}
\centering
\includegraphics[width=0.8\linewidth]{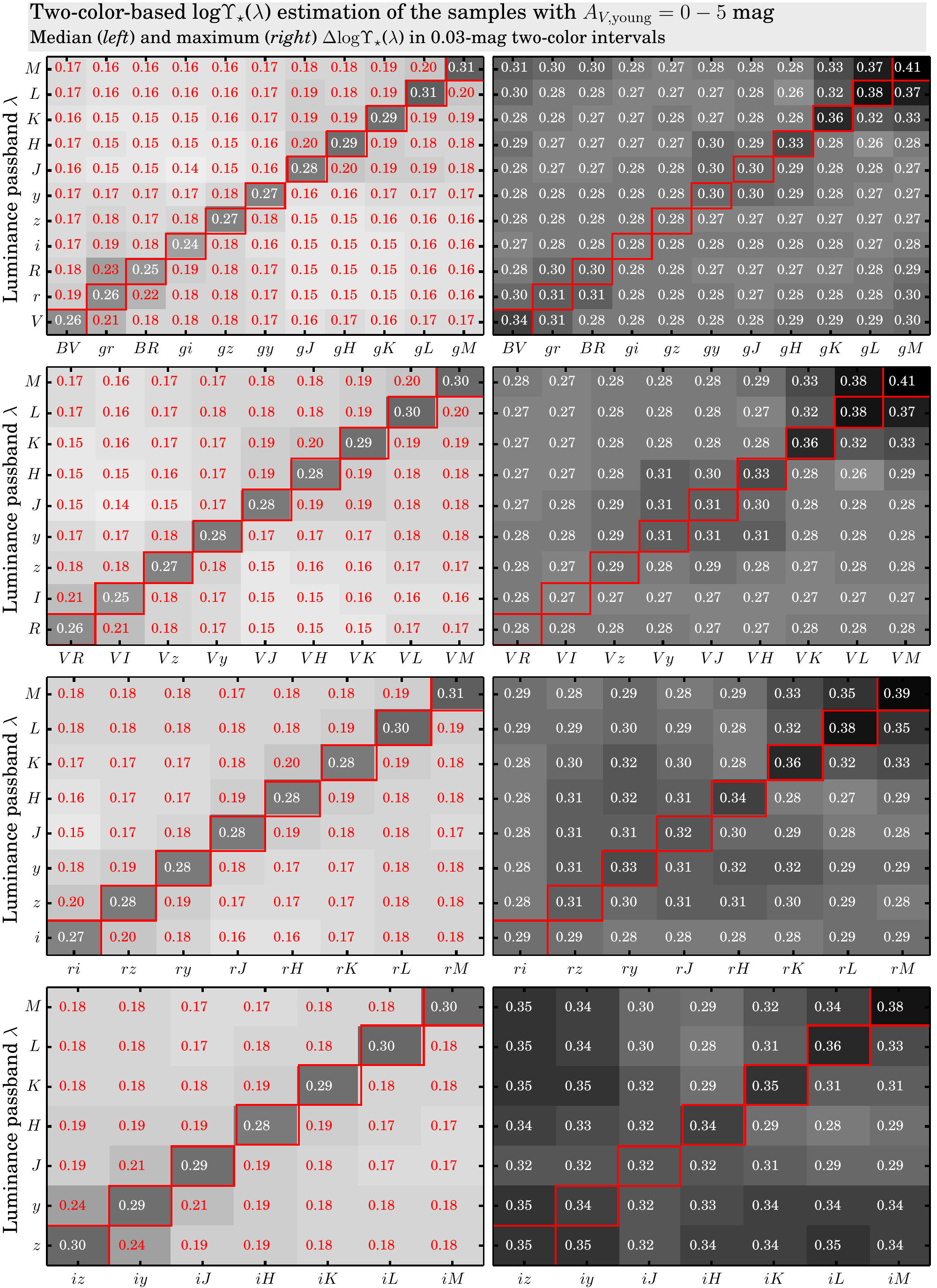}
\caption{
Same as Figure \ref{fig_m2lclrs_av05}, but for the full samples with $A_{V, {\rm young}}$ ranging from 0 to 5 mag.
\label{fig_m2lclrs_avall}}
\end{figure*}

To be complete, we determine the median and maximum $\Delta$$\log \Upsilon_{\star}$($\lambda$) predicted by various 
pairs of colors, and the results are shown in Figures \ref{fig_m2lclrs_av05} and \ref{fig_m2lclrs_avall}.\ The combinations of optical--optical and 
optical--NIR colors covering larger wavelength baselines generally provide tighter constraints on $\log \Upsilon_{\star}$($\lambda$) than the other 
combinations.\ $\log \Upsilon_{\star}$($\lambda$) of the samples with variable $A_{V, {\rm young}}$ are constrained with larger uncertainties 
than that of the sample with a low $A_{V, {\rm young}}$ of 0.5 mag.\ Nevertheless, the degree of improvement of two colors over single colors 
for the samples with variable $A_{V, {\rm young}}$ is generally larger than that for the sample with low $A_{V, {\rm young}}$ of 0.5 mag.\

$\Upsilon_{\star}$($\lambda$) estimation with an optical--optical plus optical--NIR colors was first extensively studied by Z09.\ Z09 found that 
$\Upsilon_{\star}$($\lambda$) estimated with the ($g$$-$$i$)--$\log \Upsilon_{\star}$($i$) relation alone is in excellent agreement with that estimated 
with the combination of ($g$$-$$i$) and ($i-H$), except for extremely star-forming and/or dust-extincted regions of galaxies.\ In this work, we have shown 
that, not just for extremely star-forming and dust-extincted regions, combinations of ($g$$-$$i$) and an optical--NIR color are generally favoured 
over the ($g$$-$$i$)--$\log \Upsilon_{\star}$($i$) relation alone for a robust $\Upsilon_{\star}$($\lambda$) estimation.\ In addition, since dust extinction 
and reddening have a smaller effect on optical \lclrmtl($\lambda$) distributions for extinction curves shallower than our default SMC-bar extinction curves 
(Section \ref{sec: impact_dust}), the efficacy of optical--NIR colors for dusty and young populations is reduced for shallower dust extinction curves.\

Regarding the role played by an additional optical--NIR color in stellar mass estimation for extremely star-forming regions with relatively low dust 
extinction, we emphasize that the optical--NIR colors primarily help to partially remove the dependencies of optical \lclrmtl($\lambda$) distributions on [M/H], 
rather than (as often taken for granted in the literature) to recover the mass of older stellar populations that are outshined in shorter optical wavelengths 
by luminosity-dominant younger populations.\ This interpretation is also in line with the results of Section \ref{sec: sedfitting} where we will show that 
the efficacy of NIR passbands in stellar mass estimation is mainly to improve stellar mass estimation at relatively high [M/H].\

\section{Stellar Mass Estimation from the NNLS SED Fitting}\label{sec: sedfitting}

\begin{figure*}
\centering
\includegraphics[width=0.9\linewidth]{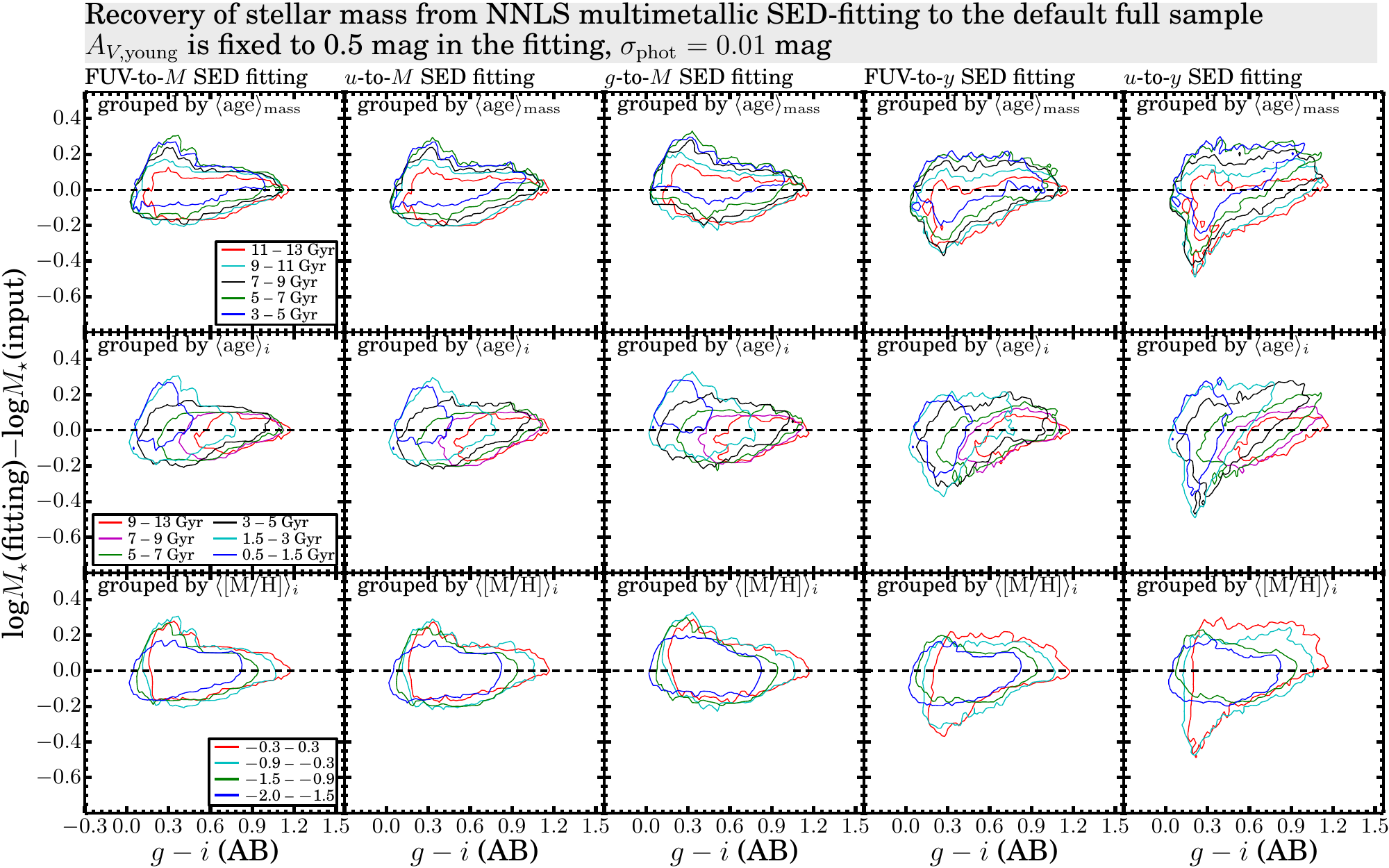}
\caption{
Recovery of the input stellar masses of our default full sample through NNLS SED fitting, where $A_{V, {\rm young}}$ is fixed to 0.5 mag.\
The logarithmic differences between the stellar mass from NNLS SED fitting and the input are plotted as a function of the $g$$-$$i$ color.\ 
From the top to bottom panels, the distributions of data points are grouped by \agem~({\it top}), $i$-band light-weighted ages ({\it middle}), 
and $i$-band light-weighted [M/H].\ From the left to right panels, the results of fitting to the full SEDs from FUV to $M$, from $u$ to $M$, from 
$g$ to $M$, from FUV to $y$, and from $u$ to $y$ are shown.\ Photometric errors of $\sigma_{\rm phot}$ = 0.01 mag are assigned 
to all passbands for SED fitting.\
\label{fig_nnls_01_gi}}
\end{figure*}

\begin{figure*}
\centering
\includegraphics[width=0.9\linewidth]{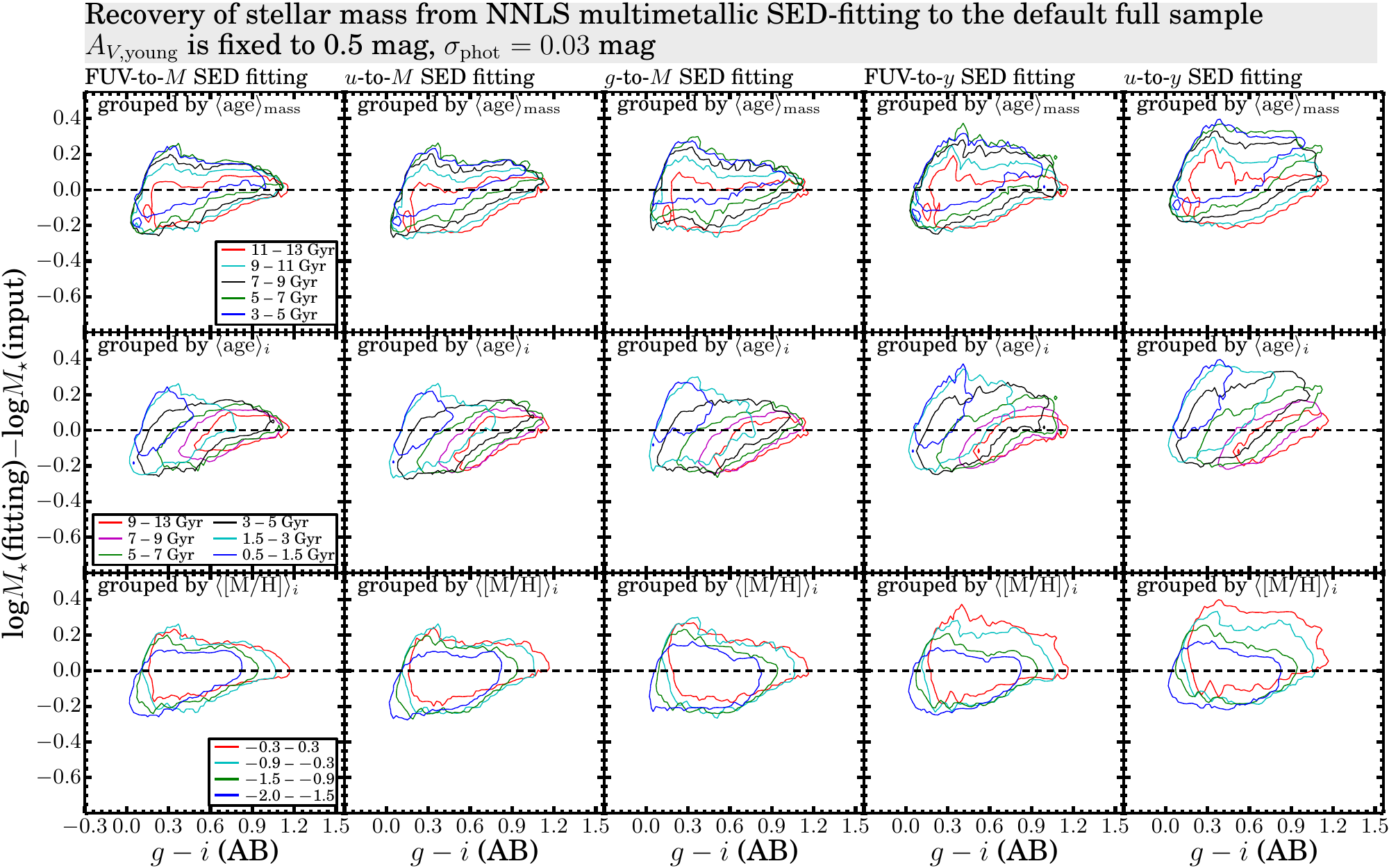}
\caption{
Same as Figure \ref{fig_nnls_01_gi}, except that the photometric errors $\sigma_{\rm phot}$ = 0.03 mag.
\label{fig_nnls_03_gi}}
\end{figure*}

\begin{figure*}
\centering
\includegraphics[width=0.9\linewidth]{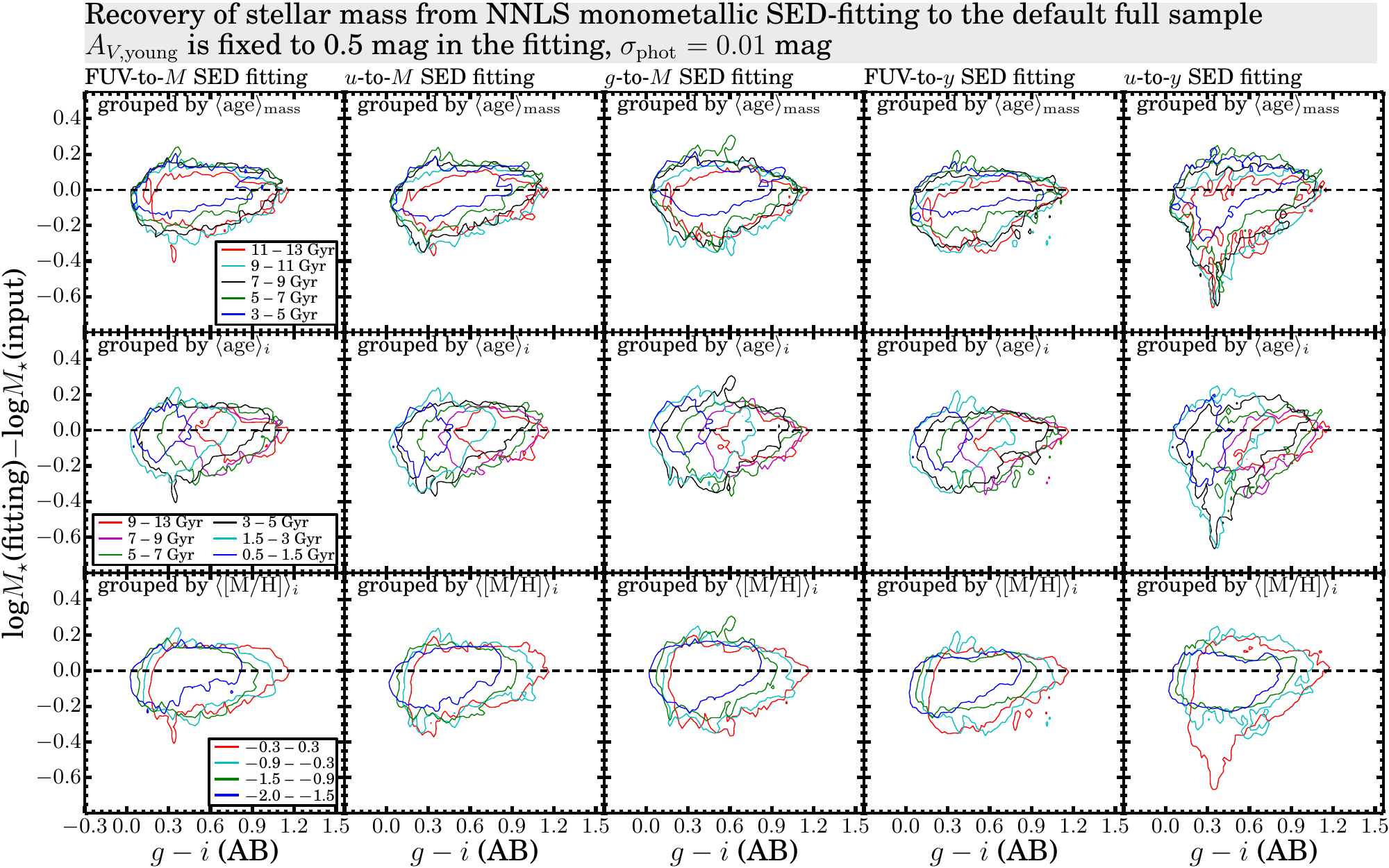}
\caption{
Same as Figure \ref{fig_nnls_01_gi}, except that the monometallic SSP templates are used for SED fitting.
\label{fig_nnls_01_gi_zfix}}
\end{figure*}

\begin{figure*}
\centering
\includegraphics[width=0.9\linewidth]{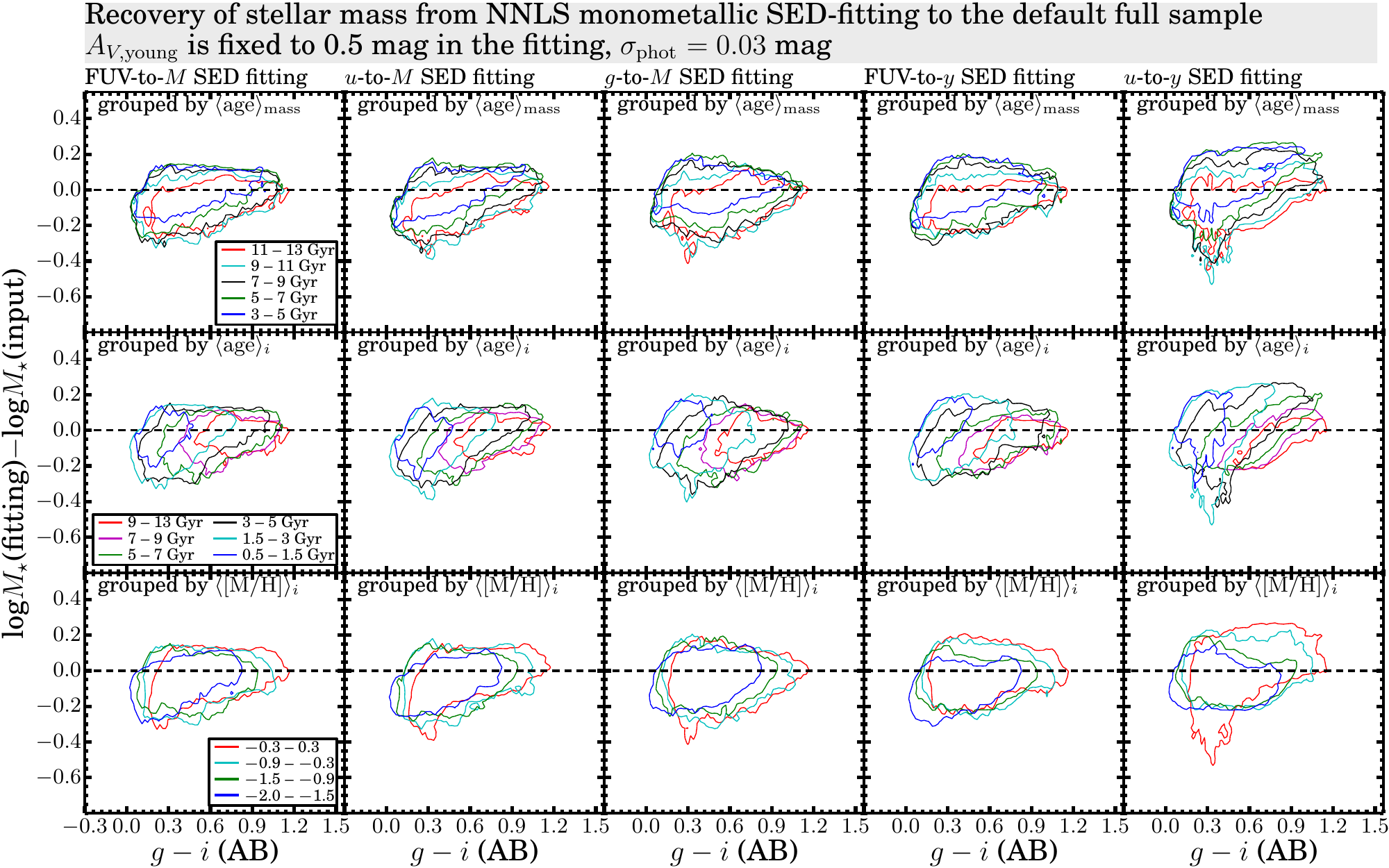}
\caption{
Same as Figure \ref{fig_nnls_03_gi}, except that the monometallic SSP templates are used for SED fitting.
\label{fig_nnls_03_gi_zfix}}
\end{figure*}

\begin{figure*}
\centering
\includegraphics[width=0.9\linewidth]{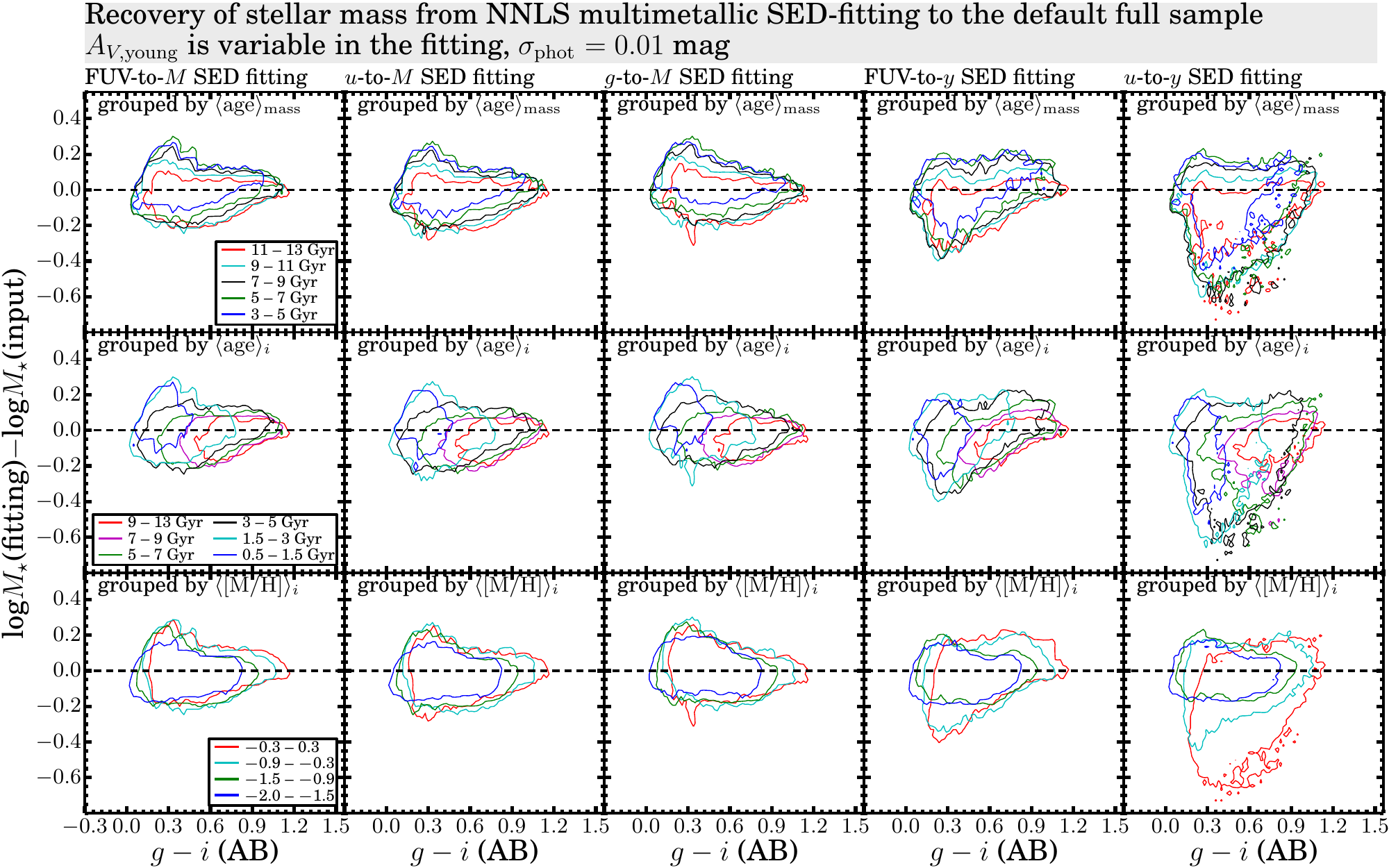}
\caption{
Same as Figure \ref{fig_nnls_01_gi}, except that the dust extinction $A_{V, {\rm young}}$ is a free parameter in the SED fitting.
\label{fig_nnls_01_gi_av}}
\end{figure*}

\begin{figure*}
\centering
\includegraphics[width=0.9\linewidth]{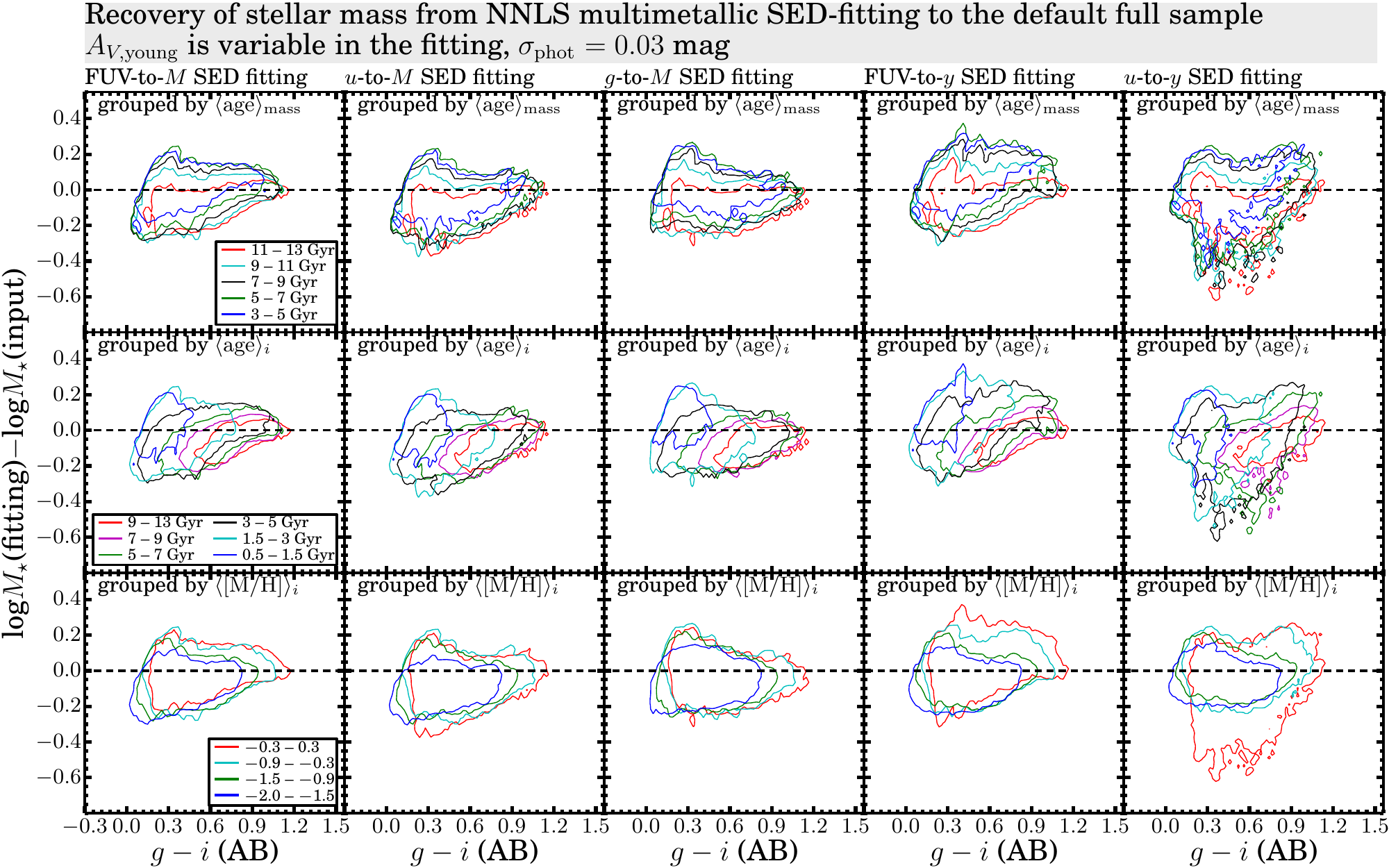}
\caption{
Same as Figure \ref{fig_nnls_03_gi}, except that the dust extinction $A_{V, {\rm young}}$ is a free parameter in the SED fitting.
\label{fig_nnls_03_gi_av}}
\end{figure*}

\begin{figure*}
\centering
\includegraphics[width=0.9\linewidth]{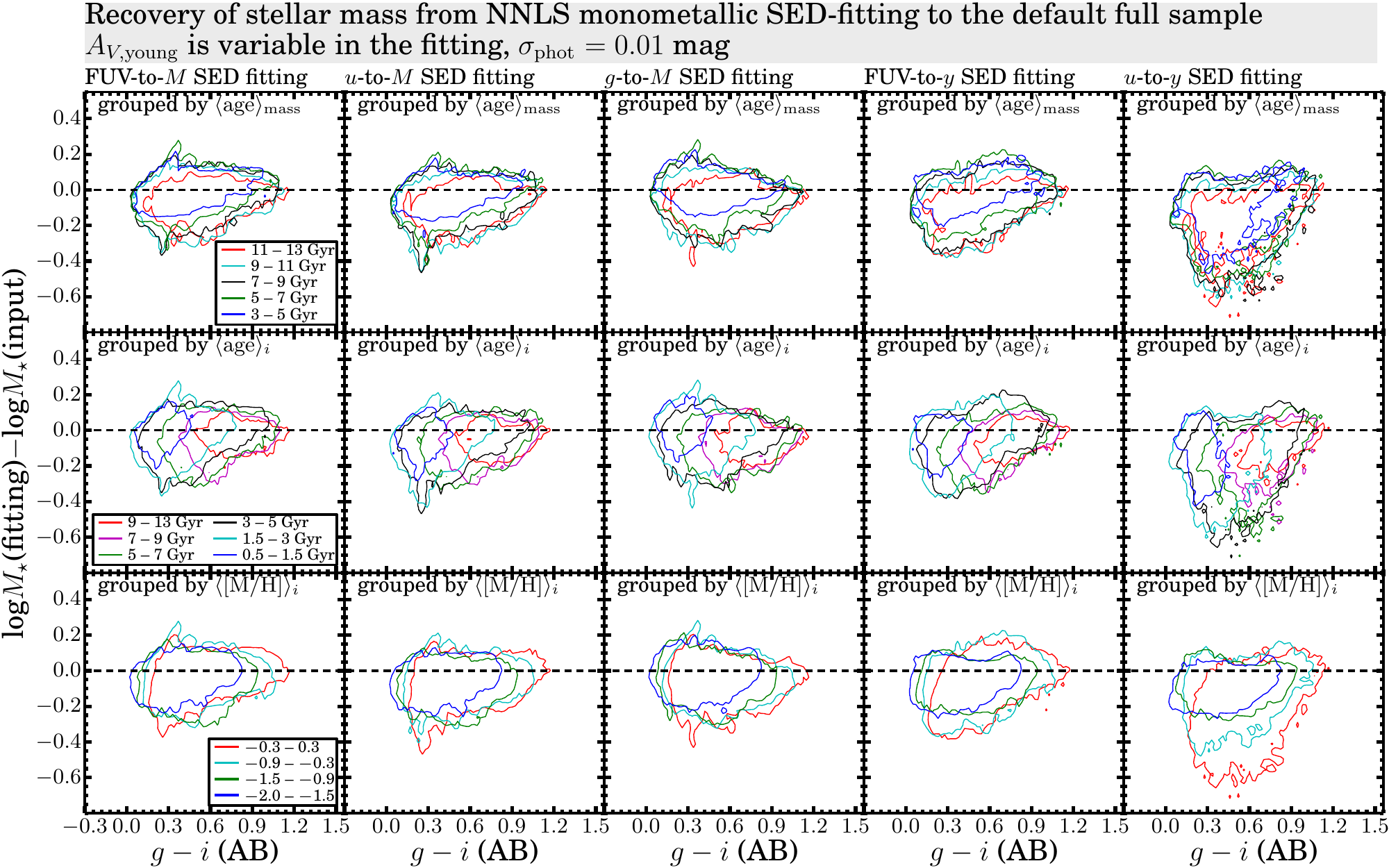}
\caption{
Same as Figure \ref{fig_nnls_01_gi_zfix}, except that the dust extinction $A_{V, {\rm young}}$ is a free parameter in the SED fitting.
\label{fig_nnls_01_gi_zfix_av}}
\end{figure*}

\begin{figure*}
\centering
\includegraphics[width=0.9\linewidth]{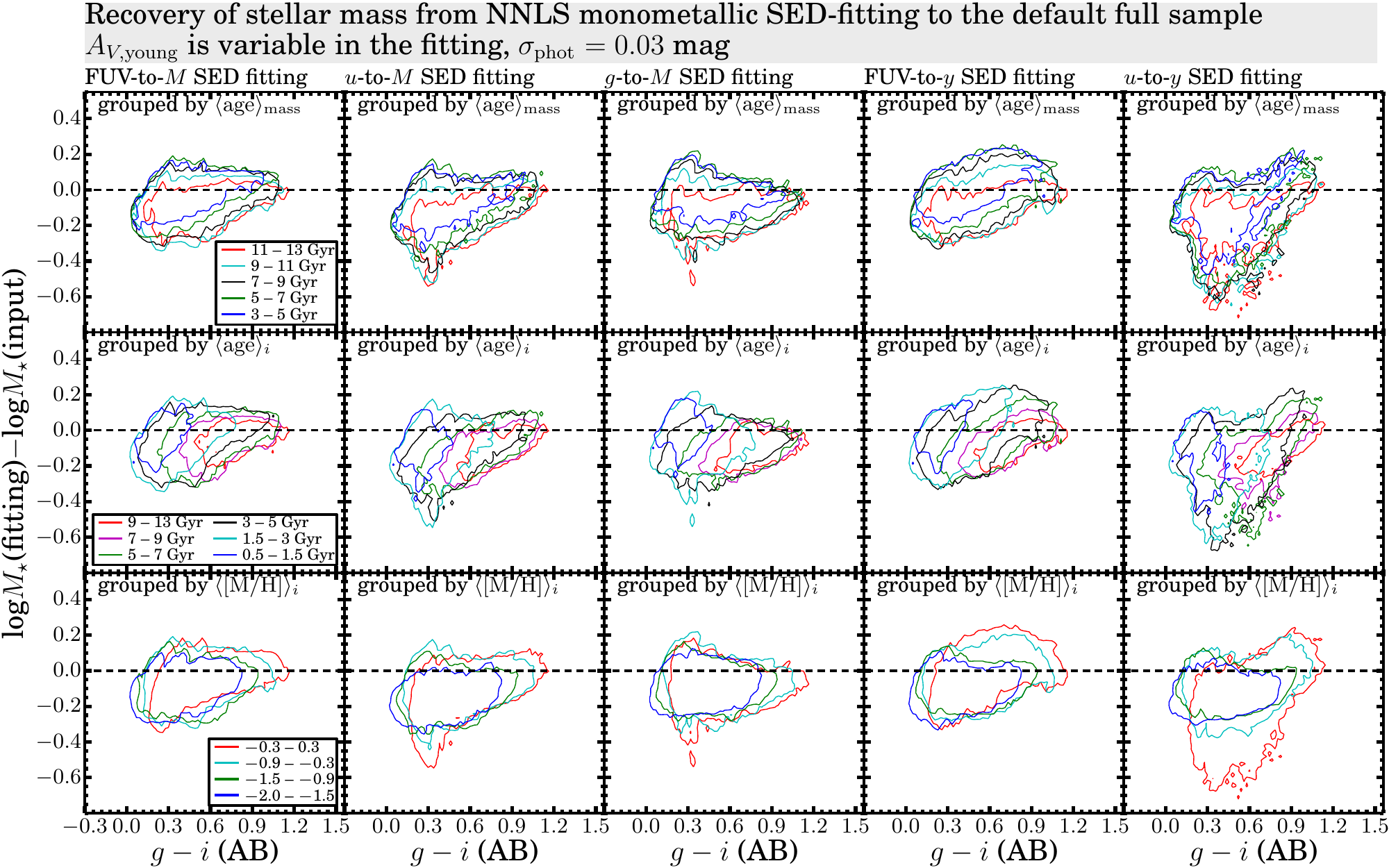}
\caption{
Same as Figure \ref{fig_nnls_03_gi_zfix}, except that the dust extinction $A_{V, {\rm young}}$ is a free parameter in the SED fitting.
\label{fig_nnls_03_gi_zfix_av}}
\end{figure*}

Following the procedure described in Section \ref{sec: nnls_method}, the broadband SEDs of our default full sample covering a variety 
of wavelength ranges from FUV to the $M$ band (i.e.\ FUV, NUV, $u$, $g$, $r$, $i$ ,$z$, $y$, $J$, $H$, $K$, $L$, and $M$) are fitted with 
the NNLS algorithm.\ Random ``observational'' photometric errors of $\sigma_{\rm phot}$ = 0.01 or 0.03 mag are assigned to all passbands involved 
in the SED fitting.\ The logarithmic differences between the stellar mass from the SED fitting and the input $\log M_{\star}$(fitting)$-$$\log M_{\star}$(input) 
as a function of the $g-i$ color are shown in Figures \ref{fig_nnls_01_gi} to \ref{fig_nnls_03_gi_zfix_av}.\ In particular, results from multimetallic SED 
fitting with $A_{V, {\rm young}}$ fixed to 0.5 mag are shown in Figures \ref{fig_nnls_01_gi} and \ref{fig_nnls_03_gi} for $\sigma_{\rm phot}$ 
= 0.01 mag and 0.03 mag, respectively, results from monometallic SED fitting with $A_{V, {\rm young}}$ fixed to 0.5 mag are shown in 
Figures \ref{fig_nnls_01_gi_zfix} and \ref{fig_nnls_03_gi_zfix} for $\sigma_{\rm phot}$ = 0.01 mag and 0.03 mag, respectively, results from 
multimetallic SED fitting with $A_{V, {\rm young}}$ being a free parameter are shown in Figures \ref{fig_nnls_01_gi_av} and \ref{fig_nnls_03_gi_av} 
for $\sigma_{\rm phot}$ = 0.01 mag and 0.03 mag, respectively, and results from monometallic SED fitting with $A_{V, {\rm young}}$ being a free parameter 
are shown in Figures \ref{fig_nnls_01_gi_zfix_av} and \ref{fig_nnls_03_gi_zfix_av} for $\sigma_{\rm phot}$ = 0.01 mag and 0.03 mag, respectively.\
To illustrate the effect of wavelength coverage (i.e.\ with/without UV and/or NIR) on the recovery accuracy of stellar mass, the results of fitting to the SEDs, 
from FUV to $M$,  from $u$ to $M$, from $g$ to $M$, from FUV to $y$, and from $u$ to $y$, are presented in different columns in the figures.\

\subsection{Stellar Mass Estimated from UV-to-NIR or Optical-to-NIR SED Fitting}
{\it \bf Multimetallic fitting}.
By fitting the full FUV-to-$M$ SEDs, stellar masses are recovered with biases (i.e.\ the vertical extent of the contours above or below 0 in the 
figures) ranging from $\sim$ $-0.2$ to 0.3 dex over the whole color ranges.\ Moreover, the uncertainties and biases are smaller at redder optical 
colors, which is in line with (1) the generally shallower and narrower \lclrmtl($\lambda$) relations and (2) a higher efficacy of NIR passbands in 
removing the [M/H] dependence of optical \lclrmtl($\lambda$) relations at redder colors.\ By fitting only the optical-to-NIR SEDs, the typical  
biases of stellar mass estimates, including their trend with colors, are very similar to those from the full SED fitting.\ The UV-to-NIR or optical-to-NIR 
SED fitting with $A_{V, {\rm young}}$ being either fixed (= 0.5 mag) or variable results in very similar quality of stellar mass estimation.\ In 
addition, increasing the photometric errors from 0.01 to 0.03 mag does not significantly affect the range of biases of stellar mass 
estimated from the UV-to-NIR or optical-to-NIR SED fitting, but does result in slightly ($\lesssim$ 0.05 dex) larger negative biases at the blue end and 
larger positive (or negative) biases at the red end when $A_{V, {\rm young}}$ is fixed (or variable) in the SED fitting.\

{\it \bf Monometallic fitting}.
Monometallic SED fitting gives stellar mass estimates with biases ranging from $\sim$ $-0.3$ to 0.15 dex.\ So monometallic fitting has a stronger 
tendency (by $\sim$ 0.1 dex on average) of underestimating stellar masses than the multimetallic fitting.\ We note that, with larger uncertainties 
of stellar mass estimates than the multimetallic fitting at the red end of colors, monometallic fitting gives more consistent uncertainties and biases 
across the whole color ranges.\ Increasing photometric errors of the monometallic fitting has an effect on stellar mass estimation that is very 
similar to the multimetallic fitting, as described above.\

\subsection{Stellar Mass Estimated from SED Fitting without NIR Passbands}

{\it \bf Multimetallic fitting}.
Without passbands longward of $y$, the SED fitting performs progressively worse in stellar mass estimation for shorter-wavelength coverages.\ 
In particular, with $A_{V, {\rm young}}$ being fixed, the optical $u$-to-$y$ SED fitting gives stellar mass estimates with biases ranging from $\sim$ 
$-0.5$ to 0.3 dex (i.e.\ being negative on average) for $\sigma_{\rm phot}$ = 0.01 mag and from $\sim$ $-0.2$ to 0.4 dex (i.e.\ being positive on 
average) for $\sigma_{\rm phot}$ = 0.03 mag.\ Moreover, with $A_{V, {\rm young}}$ being a free parameter, the optical $u$-to-$y$ SED fitting gives 
stellar mass estimates with biases ranging from $\sim$ $-0.6$ to $0.2$ dex (i.e.\ being negative on average) for either $\sigma_{\rm phot}$ = 0.01 
mag or 0.03 mag.\ Therefore, SED fitting with dust extinction as a free fitting parameter has a significantly stronger tendency of underestimating 
stellar masses if NIR passbands are not used in the fitting.\ The importance of NIR passbands in stellar mass estimation has been realized by many 
previous studies (e.g.\ Lee et al.\ 2009a; Kannappan \& Gawiser 2007; Bolzonella et al.\ 2010; Pforr et al.\ 2013; Mitchell et al.\ 2013).\

{\it \bf Monometallic fitting}.
Similar to the general trend for the full SED fitting discussed above, monometallic fitting without NIR passbands has a stronger tendency of 
underestimating stellar masses than the corresponding multimetallic fitting.\

\subsection{Dependences of Stellar Mass Estimates on Input SFHs and Metallicities}
Stellar masses tend to be underestimated at older \agem~while overestimated at younger \agem.\ The \agem-dependent biases are stronger 
for SED fitting with fewer passbands, especially when NIR passbands are not used.\ The aforementioned larger biases of stellar mass estimates 
at bluer colors are primarily driven by galaxies with relatively smaller \agelam~and higher \mhlam.\ It is noteworthy that the optical-to-NIR fitting has 
a stronger tendency of overestimating stellar masses at the youngest \agelam~than the UV-to-NIR fitting.\ Moreover, we also note that the worse 
performance of SED fitting without NIR passbands as mentioned above is primarily driven by galaxies with higher \mhlam.\

The \agem-dependent biases suggest that, as with single colors, broadband SED fitting is generally not sensitive to \agem, and the most probable 
stellar mass estimates from SED fitting are just the median over SFHs with all possible ranges of \agem~for given SEDs.\ In addition, the apparently 
larger biases of stellar mass estimates at higher \mhlam~reflect the generally broader and steeper \lclrmtl($\lambda$) relations at higher \mhlam~
(Section \ref{sec: impact_met}), rather than a poorer (or better) performance of SED fitting at higher (or lower) \mhlam.\ If the NIR passbands are 
used in SED fitting, the dependence of uncertainties and biases on \mhlam~can be partially removed.\

\subsection{Why Monometallic SED fitting Tends to Underestimate Stellar Masses?}
In a multimetallic SED fitting, SEDs with slightly bluer colors can be fitted by giving more weights to either younger SSP components of 
relatively high metallicities (leading to younger \agelam) or older SSP components of relatively low metallicities (leading to older \agelam), 
whereas in a monometallic SED fitting, SEDs with slightly bluer colors can only be fitted by giving more weights to younger SSP components 
(leading to younger \agelam).\ Since $\Upsilon_{\star}$($\lambda$) is positively correlated with \agelam, the monometallic SED fitting has a 
stronger tendency of underestimating stellar masses of real galaxies.\ 

Gallazzi \& Bell (2009) also explored the uncertainties of stellar mass estimation induced by using monometallic models to fit mock galaxies 
with multimetallic exponentially declining SFHs.\ They found that stellar masses tend to be underestimated (or overestimated) for their old (or 
young) galaxies when single $g-i$ colors are used to fit monometallic models to multimetallic input SFHs.\ The average trend of either 
overestimation or underestimation of stellar masses found by Gallazzi \& Bell (2009) is inevitably influenced by their prior assumptions on the 
SFH libraries.\ However, they found that mass estimates are barely affected when using the two age-sensitive stellar absorption features 
D4000$_{n}$ and H$\delta_{A}$, instead of $g-i$ colors, in the fitting.\ This is consistent with a tighter constraint on light-weighted ages and 
thus $\Upsilon_{\star}$ offered by the two spectral features.\

\section{Discussion}\label{sec: discuss}

\subsection{Origin of the Discrepancies between \lclrmtl($\lambda$) Relations Calibrated by Different Studies}\label{sec: clrmtl_orig}
\begin{figure*}
\centering
\includegraphics[width=0.9\linewidth]{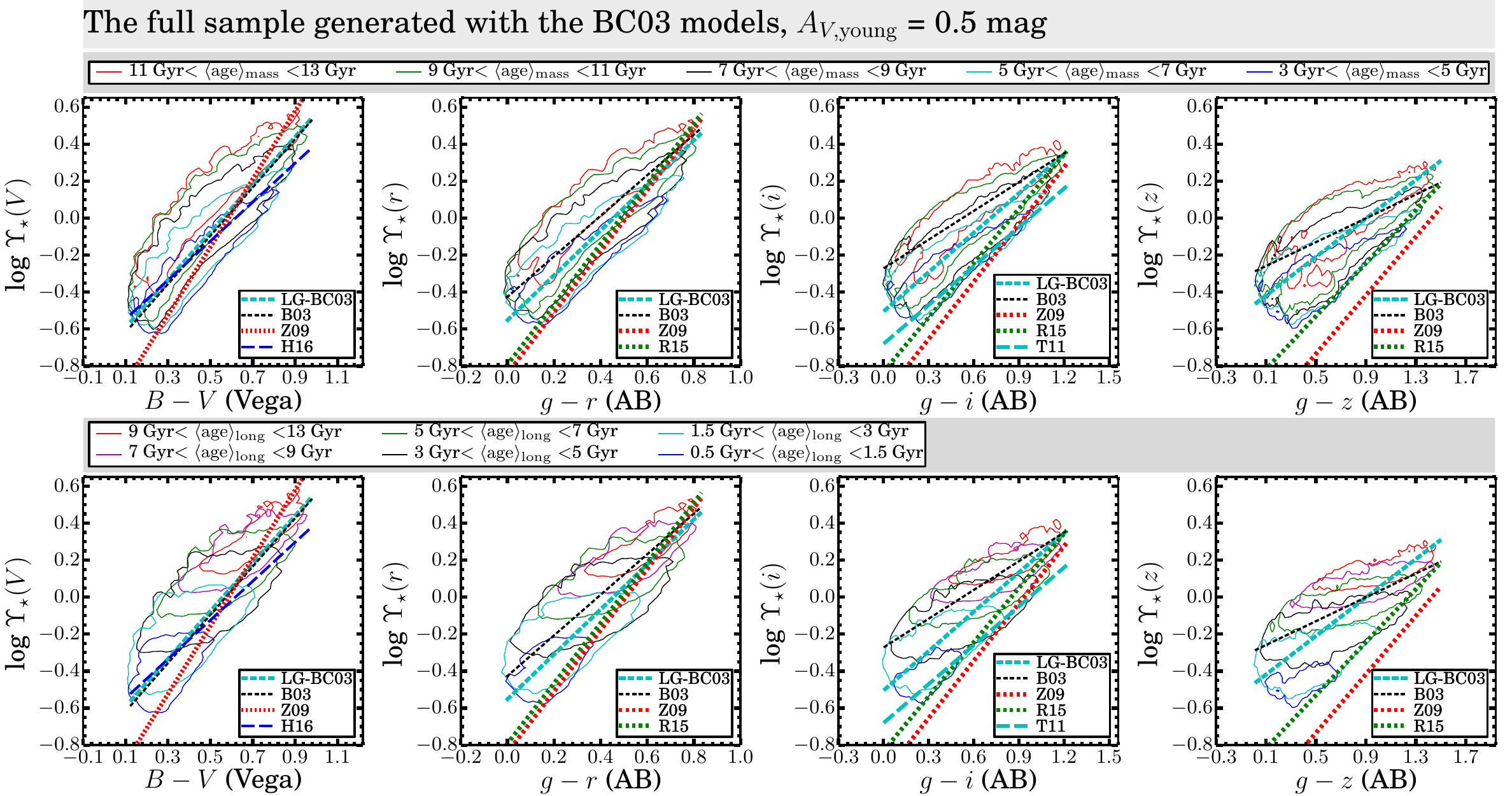}
\caption{
The four relatively tight optical \lclrmtl($\lambda$) relations of the sample with $A_{V, {\rm young}}$ = 0 mag, based on the BC03 models.\ 
The full sample is grouped by different ranges of \agem~and \agelg, respectively, in the {\it top} and {\it bottom} panels, as in 
Figures \ref{fig_m2lclr_2} and \ref{fig_m2lclr_4}.\ Several published linear \lclrmtl($\lambda$) relations (whenever available for given passbands), 
as in Figure \ref{fig_m2lclr_1}, are overplotted for comparison.
\label{fig_m2lclr_bc03}}
\end{figure*}

In light of our study of the \lclrmtl($\lambda$) distributions in this work, here we try to address the reason behind the discrepancies between 
various \lclrmtl($\lambda$) relations from previous studies.\ We focus on the four most commonly studied and tightest optical 
\lclrmtl($\lambda$) distributions, i.e.\ ($B$$-$$V$)--$\log \Upsilon_{\star}$($V$), ($g$$-$$r$)--$\log \Upsilon_{\star}$($r$), ($g$$-$$i$)--$\log \Upsilon_{\star}$($i$), 
and ($g$$-$$z$)--$\log \Upsilon_{\star}$($z$).\ Since most of the \lclrmtl($\lambda$) relations available in the literature were calibrated based on 
the BC03 models, in this subsection different literature calibrations are confronted with the distributions of our default full sample of 
$A_{V, {\rm young}}$ = 0.5 mag (Figure \ref{fig_m2lclr_bc03}), computed with the BC03 SSP models.\ We emphasize that the results based 
on the BC03 models are in general agreement with those based on the FSPS models.\

The \lclrmtl($\lambda$) distributions shown in Figure \ref{fig_m2lclr_bc03} are grouped by \agem~and \agelg~in the {\it upper} and {\it lower} 
panels respectively.\ \agem~and \agelg~are the two key parameters that differentiate the full sample into groups that follow distinctly different 
optical \lclrmtl($\lambda$) correlations (Section \ref{sec: impact_sfh}).\ The overplotted literature \lclrmtl($\lambda$) relations are as in Figure \ref{fig_m2lclr_1}.\ These relations are based on either the original or the updated but unpublished 2007 version of BC03 models (Charlot \& Bruzual 2007; hereafter 
CB07).\ The CB07 models are different from the original BC03 models primarily in a $\gtrsim$ 2$\times$ enhanced NIR luminosity contribution at 
intermediate ages from the TP-AGB evolutionary phases, following Marigo \& Girardi (2007) and Marigo et al.\ (2008).\ Among the overplotted 
\lclrmtl($\lambda$) relations, the Z09 and H16 relations are based on the CB07 models, and the T11 and R15 relations are based on the BC03 models.\ 
The most commonly used B03 relations were calibrated with the PEGASE models, which are based on the same Padova (1994) isochrones 
as the BC03 models.\ The R15 relations plotted here are the ones predicted by the BC03 version of the MAGPHYS SED fitting package 
(da Cunha et al.\ 2008), and the Z09 relations are based on SFH libraries constructed by adopting very similar prior constraints to those of  
MAGPHYS.\ Hence, it is no surprise that the R15 and Z09 relations have about the same slopes, and the discrepancies of the their zero points 
(or intercepts) are smaller for colors involving shorter-wavelength passbands that are less influenced by the different treatment of TP-AGB phases 
between the BC03 and CB07 models.\ In addition, except for the B03 relations, all of the other relations are calibrated with the Chabrier (2003) 
IMF.\ To make a fair comparison with others, we have reduced, in Figures \ref{fig_m2lclr_1} and \ref{fig_m2lclr_bc03}, the zero points of the B03 
relations by 0.12 dex to transform from a diet-Salpeter IMF, used in B03, to a Chabrier (2003) IMF.\

As shown in Figure \ref{fig_m2lclr_bc03}, the discrepancies between different relations are smaller for \lclrmtl($\lambda$) distributions covering shorter-wavelength baselines, with ($B$$-$$V$)--$\log \Upsilon_{\star}$($V$) being the one with the smallest discrepancies and ($g$$-$$z$)--$\log \Upsilon_{\star}$($z$) 
being the one with the largest discrepancies.\ This is in line with our finding that the linear slopes and intercepts of the ($B$$-$$V$)--$\log \Upsilon_{\star}$($V$) 
relations are least influenced by SFHs, metallicities, and dust extinction/reddening.\ As pointed out in Section \ref{sec: singlecolor}, \lclrmtl($\lambda$) 
relations involving $\log \Upsilon_{\star}$($V$) and colors covering wavelength baselines longer than $B$$-$$V$, such as $g$$-$$i$, are subject to larger 
systematic uncertainties induced by metallicities and dust reddening.\ This explains the finding of McGaugh \& Schombert (2014) that, for the same galaxy, 
$\Upsilon_{\star}$($\lambda$) estimated (through the $B$$-$$V$ color) with \lclrmtl($\lambda$) relations calibrated by different studies agree closely 
in the $V$ band but diverge at longer wavelengths.

\subsection{Comments on Linear \lclrmtl($\lambda$) Relations in the Literature}
Before proceeding to comment on the individual \lclrmtl($\lambda$) relations, we briefly summarize the key assumptions of SFH models 
adopted by different studies.\ The B03 relations are based on SED fitting with libraries of single exponential SFHs ($\propto$ 
$\exp(-t/\tau)$), with the $e$-folding timescale $\tau$ being a free parameter and the time $t$ since the SFHs commenced being 
fixed to 12 Gyr.\ The T11 relations are also based on single exponential SFH libraries, with both $t$ and $\tau$ being free parameters.\ 
The Z09 and R15 relations are determined by adding random bursts on top of single exponential SFHs, with $t$ and $\tau$ being free 
parameters.\ The H16 relations are determined with a library of SFHs (Zhang et al.\ 2012) that is constructed by approximating the galaxy 
lifetime (i.e.\ 14 Gyr) with 6 logarithmically divided periods of constant star formation, with the weights (or mass) of the six independent periods 
being free parameters.\ All of the above studies assumed {\it unrealistic} monometallic model SFH libraries.\

\subsubsection{Bell et al.~(2003) Relations}
B03 relations are calibrated with a large sample of SDSS galaxies in the local universe.\ In particular, B03 estimated 
$\Upsilon_{\star}$($\lambda$) of their galaxies by fitting the SDSS $ugriz$ and 2MASS $JHK$ photometry with dust-free 
single exponential model SFH libraries, with an [M/H] coverage from $-1.3$ to 0.4.\ The B03 sample covers color ranges 
(e.g.\ 0.2 $\lesssim$ $g$$-$$r$ $\lesssim$ 1.0 mag) blueward of which the optical \lclrmtl($\lambda$) distributions start strongly deviating 
(toward lower $\log \Upsilon_{\star}$($\lambda$)) from linear extrapolations at redder colors.\ Gallazzi \& Bell (2009) already 
showed that, by fixing the starting time of single exponential SFHs to 12 Gyr, the B03 model distributions are skewed toward 
older average ages and (thus) the resultant \lclrmtl($\lambda$) distributions are skewed toward larger $\Upsilon_{\star}$($\lambda$) at 
given color values, as compared to exponential SFHs with smaller starting time.\ From the comparisons of ($g$$-$$i$)--$\log \Upsilon_{\star}$($i$) 
and ($g$$-$$z$)--$\log \Upsilon_{\star}$($z$) distributions shown in Figure \ref{fig_m2lclr_bc03}, the B03 relations are largely aligned 
with the lower-$\log \Upsilon_{\star}$($\lambda$) side of the \agem~contour at 11$-$13 Gyr and at the same time the 
higher-$\log \Upsilon_{\star}$($\lambda$) side of the \agem~contour at 5$-$7 Gyr.\ Therefore, the B03 relations apply to galaxies 
with 7 Gyr $\lesssim$ \agem~$\lesssim$ 11 Gyr.\

\subsubsection{Herrmann et al.~(2016) Relations}
The H16 ($B$$-$$V$)--$\log \Upsilon_{\star}$($V$) relation is calibrated with a representative sample of nearby dwarf irregular galaxies, 
which cover a ($B$$-$$V$) color range of $\sim$ 0.1 -- 0.7 mag.\ $\log \Upsilon_{\star}$($V$) of the sample galaxies were estimated  
(Zhang et al.\ 2012) by fitting the FUV, NUV, $UBV$ and {\it Spitzer} 3.6$\mu$m multiband photometry.\ The monometallic SFH 
library used by Zhang et al.\ (2012) covers a [M/H] range from $\sim$ $-1.7$ to $-$0.4, which is appropriate for dwarf irregular 
galaxies.\ As discussed in Section~\ref{sec: impact_met}, toward the blue end of colors, galaxies with lower [M/H] follow shallower 
average \lclrmtl($\lambda$) relations than those with higher [M/H].\ Therefore, the slightly shallower H16 relation, as compared to the B03 relation, 
can be explained by the metallicity coverage of the SFH libraries.\ Indeed, H16 found that galaxies with lower measured oxygen 
abundances follow shallower \lclrmtl($\lambda$) relations.\

\subsubsection{Taylor et al.~(2011) Relations}
The T11 ($g$$-$$i$)--$\log \Upsilon_{\star}$($i$) relation is roughly aligned with the lower-$\log \Upsilon_{\star}$($i$) side of the \agem~contour 
at 7$-$9 Gyr, suggesting a typical \agem~$\lesssim$ 7 Gyr.\ The T11 relation was calibrated by fitting the SDSS $ugriz$ multiband 
photometry of a large sample of galaxies spanning a redshift range from 0.0 to 0.65 (universe age $\gtrsim$ 7.6 Gyr).\ The redshift of 
a galaxy restricts the oldest possible starting time $t$ and thus the \agem~of single exponential model SFHs, so the significantly lower 
$\log \Upsilon_{\star}$($i$) (for given color values) predicted by the T11 relation than that by B03 and the relation followed by the LG dwarf 
galaxies (LG-BC03) could be attributed to the redshift distribution of the galaxies studied by T11.\ We note that the nearly linear and particularly 
tight ($g$$-$$i$)--$\log \Upsilon_{\star}$($i$) distribution followed by the galaxies of T11 is well explained by their relatively young \agem~(Section 
\ref{sec: impact_sfh}).\

\subsubsection{Zibetti et al.~(2009) and Roediger \& Courteau (2015) Relations}
The Z09 and R15 relations are least-squared regression of the \lclrmtl($\lambda$) distributions followed by their model SFH libraries.\ The significantly 
steeper slopes and smaller intercepts of the Z09 and R15 relations can be explained by their model SFH distributions that are strongly skewed 
toward younger light-weighted ages at given color values.\ Indeed, as described in da Cunha et al.\ (2008), 50\% of their model SFHs have 
experienced a random burst in the past 2 Gyr, with the fractional mass contribution from bursts ranging from 3\% to 80\%.\ However, such 
extremely high mass fraction of recent bursts is actually a very rare phenomenon in nearby galaxies.\ For instance, Lee et al.\ (2009b) found 
that only $\sim$ 6\% of nearby dwarf galaxies are currently experiencing massive global bursts with birth-rate parameter $b\geq 2\!-\!3$, and 
only $\sim$ 23\% of the overall star formation in dwarfs occurring in the starburst mode.\ The burst frequency and strength of more massive 
galaxies are even lower (e.g.\ Kauffmann et al.\ 2003).\ 

In addition, the MAGPHYS package, as used by Z09 and R15, assumes an extinction recipe 
whereby populations younger than 10 Myr are subject to a steep extinction curve of $A_{\lambda}$ $\propto$ $\lambda^{-1.3}$ while older 
populations are subject to an extinction curve of $A_{\lambda}$ $\propto$ $\lambda^{-0.7}$.\ Although da Cunha et al.\ (2008) claim that adopting 
a steeper extinction curve solely for the youngest populations has a negligible influence on the emergent integrated SEDs, the steep extinction 
curve for the youngest populations that may account for an excessively large mass fraction results in a larger spread of $\log \Upsilon_{\star}$($\lambda$) 
toward the lower right part of the \lclrmtl($\lambda$) planes (Section \ref{sec: impact_dust}), and thus steepens the average \lclrmtl($\lambda$) relations.\

\subsection{Remarks on the Generality of Our Results}
Throughout this work, we have quantified and discussed all of the relevant parameters impacting the determination of stellar 
masses by using the half max--min ranges, instead of the usually used standard deviations.\ By doing so, our general conclusions 
are not limited by the exact shape of the underlying distributions of various physical parameters characterizing the SFHs and metallicity 
evolution of our samples.\

Our samples span a relatively uniform range of \agelam~from $\sim$ 0.5 to 13 Gyr and of \agem~from $\sim$ 3 to 13 Gyr (corresponding to an 
average mass-assembly redshift range from $\sim$ 0.25 to 7).\ These age ranges cover the average ages measured for local universe galaxies with 
stellar masses up to 10$^{12}$ M$_{\odot}$ (Gallazzi et al.\ 2005).\ Our finding that the ``concentration'' of SFHs in time does not significantly affect 
the \lclrmtl($\lambda$) relations gives us further confidence that the SFHs of our full sample are adequate for exploring the issue of stellar mass 
estimation for local universe galaxies in general.\ In addition, observations suggest that there exists a nearly universal relation between metallicities, 
SFRs, and stellar masses for galaxies with different stellar masses and redshift (e.g.\ Hunt et al.\ 2016).\ This indicates that the ``shape'' 
of metallicity evolution histories is, on average, closely linked to that of the SFHs, while the absolute values of SFRs and stellar masses set 
the absolute metallicity scales.\

Given the above considerations, our full samples are expected to adequately cover the physically plausible range of SFHs and 
metallicity evolution histories for ordinary galaxies with different masses in the local universe.\

\section{Summary}\label{sec: summary}
Taking advantage of the relatively accurate SFHs, the associated metallicity evolution, and age-dependent dust extinction recipe self-consistently 
determined through resolved CMD modeling of 40 LG dwarf galaxies (Weisz et al.\ 2014), we derived their expected broadband SEDs and studied their 
\lclrmtl($\lambda$) distributions in various broad passbands from the UV to NIR.\ By expanding the parameter coverages of SFHs in the recent 
1 Gyr, average metallicities ([M/H]), and dust extinction of the LG dwarf galaxies, we systematically studied the dependence of \lclrmtl($\lambda$) 
distributions on SFHs, mass-weighted ages \agem, monochromatic light-weighted ages \agelam~and metallicities \mhlam, metallicity evolution, 
and dust extinction.\ In addition, we explored the efficacy of nonparametric broadband SED fitting in recovering the input stellar masses of our 
expanded sample.\ The primary results from this work are summarized below.\

\begin{enumerate}

\item
The 40 LG dwarf galaxies follow \lclrmtl($\lambda$) relations that fall in between the ones calibrated by previous studies.\ 
Moreover, the discrepancies between the \lclrmtl($\lambda$) relations of the LG dwarf galaxies and those from previous studies 
are generally smaller for \lclrmtl($\lambda$) relations covering shorter-wavelength baselines, with the ($B$$-$$V$)--$\log \Upsilon_{\star}$($V$) and, 
to a lesser extent, the ($g$$-$$r$)--$\log \Upsilon_{\star}$($r$) relations being the ones that agree most closely.\ In addition, we found that 
the classic BC03 stellar population models give \lclrmtl$\lambda$) distributions with larger scatter than the FSPS models, but the 
two models give average \lclrmtl($\lambda$) relations with consistent slopes and intercepts.

\item 
The shape of SFHs, as quantified by their concentration in time (i.e.\ $\log$(\tma/\tmb)), does not significantly affect the slopes and 
intercepts of the \lclrmtl($\lambda$) relations, except for the most concentrated SFHs (i.e.\ $\log$(\tma/\tmb) $\gtrsim$ 0) which have 
systematically higher $\Upsilon_{\star}$($\lambda$) at given colors.\ \agelam~is the physical parameter most closely correlated 
with $\Upsilon_{\star}$($\lambda$), with its correlation strength peaking at the $J$ and $H$ bands, in analogy to similar correlations 
between ages and $\Upsilon_{\star}$($\lambda$) of SSPs.\ Instead of being correlated with $\Upsilon_{\star}$($\lambda$), \agem~alone 
sets the upper and, to a lesser extent, lower limits of $\Upsilon_{\star}$($\lambda$) at given color values.\ For the SFHs with associated 
metallicity evolution, \agelam~and \mhlam~together constrain $\log \Upsilon_{\star}$($\lambda$) with uncertainties, as quantified by the 
half max--min ranges, ranging from $\lesssim$ 0.1 dex for the $J$ and longer-wavelength passbands and up to $\lesssim$ 0.2 dex for the 
$r$ and shorter-wavelength passbands.\

\item
SFHs without associated metallicity evolution result in \lclrmtl($\lambda$) relations that are slightly shallower than but still well consistent with that 
of SFHs with associated metallicity evolution.\ Moreover, on the \lclrmtl($\lambda$) planes, monometallic SFHs with [M/H] fixed to \mhlam~match 
their multimetallic version of SFHs better than monometallic SFHs with [M/H] fixed to \mhm.\ The half max--min differences between 
$\log \Upsilon_{\star}$($\lambda$) resulting from given SFHs with and without metallicity evolution are $\lesssim$ 0.1 dex.\ Metallicity 
evolution has a stronger effect on optical--NIR colors ($\lesssim$ 0.3 mag) than optical--optical colors ($\lesssim$ 0.1 mag).\ By using monometallic 
SFH models to infer $\log \Upsilon_{\star}$($\lambda$) of real galaxies, one tends to underestimate the uncertainties by $\lesssim$ 0.1 dex.

\item
The influence of dust extinction and reddening on \lclrmtl($\lambda$) distributions depends on the stellar population mixes and the 
adopted dust extinction curves.\ In the context of a physically plausible age-dependent dust extinction recipe, dust extinction and 
reddening barely affect the \lclrmtl($\lambda$) distributions if the relatively shallow Charlot \& Fall (2000) extinction curve ($A_{\lambda}$ $\propto$ $\lambda^{-0.7}$) is adopted, while they can flatten the slopes (by up to 30\%) and increase the scatter (by up to 50\%) 
of the resultant optical \lclrmtl($\lambda$) distributions if the steep average SMC-bar extinction curve is adopted.\ 

\item
Regarding the practice of using single NIR passbands as proxies for stellar masses, the $J$ and $H$ bands have smaller max--min 
ranges of $\log \Upsilon_{\star}$($\lambda$) ($\sim$ 0.7 dex) than either the shorter- or longer-wavelength passbands.\ However, 
because $\Upsilon_{\star}$($\lambda$) is strongly correlated with \agelam, especially in the $J$ and $H$ passbands, using luminosities 
of the $JH$ or other passbands as indicators of stellar masses of galaxies of different types is subject to systematic uncertainties as 
large as the max--min ranges of $\Upsilon_{\star}$($\lambda$).\

\item
We searched for the optimal \lclrmtl($\lambda$) relation for $\Upsilon_{\star}$($\lambda$) estimation in general and find that the $V$ band 
is the optimal luminance passband for estimating $\Upsilon_{\star}$($\lambda$) based on single colors.\ In particular, the combinations of 
$\Upsilon_{\star}$($V$) and optical colors such as $B$$-$$V$, $g$$-$$r$ and $B$$-$$R$ exhibit weaker {\it systematic} dependences 
on SFHs, [M/H], and dust extinction than the combinations involving longer-wavelength luminance passbands.\ In addition, by partially 
removing the dependences of optical \lclrmtl($\lambda$) distributions on [M/H] and dust extinction, optical--NIR colors help to improve the 
constraints on \agelam~and thus $\Upsilon_{\star}$($\lambda$) in general.\

\item
With respect to the limitations of recovering stellar masses $\mathcal{M_{\star}}$ of galaxies with relatively low dust extinction ($A_{V}$ $\lesssim$ 0.5 mag) 
from fitting the integrated broadband SEDs, without any prior assumption on SFHs and metallicity evolution, $\log M_{\star}$ can be constrained with 
maximum biases of $\lesssim$ 0.3 dex when UV-to-NIR or optical-to-NIR passbands are used in SED fitting, regardless of whether dust extinction is fixed or 
variable in SED fitting.\ Optical passbands alone can constrain $\log M_{\star}$ with maximum biases $\lesssim$ 0.4 dex when dust extinction is fixed 
or $\lesssim$ 0.6 dex when dust extinction is variable in SED fitting.\ The uncertainties and biases of $\log M_{\star}$ estimation are generally smaller 
at redder optical colors.\ Passbands blueward of $g$ are helpful for $\log M_{\star}$ estimation only when NIR passbands are not used in the SED fitting.\ 
NIR passbands are more important for improving the $\log M_{\star}$ estimation at higher [M/H], probably due to the generally broader and steeper 
optical \lclrmtl($\lambda$) relations.\ Even with the full FUV-to-NIR SED fitting, $\log M_{\star}$ of galaxies with the oldest (or youngest) \agem~tend to 
be underestimated (or overestimated).\ Such \agem-dependent biases are significantly stronger when NIR passbands are not used for SED fitting.\ 
By using purely monometallic SFH models to fit the broadband SEDs of real galaxies, $\log M_{\star}$ has a strong tendency to be underestimated.

\end{enumerate}

\begin{acknowledgements}
\noindent 
We thank the anonymous referee for a prompt and helpful report.\
This project is supported by FONDECYT Postdoctoral Grant (No.~3160538), FONDECYT Regular Project Grant (No.~1161817), and the BASAL 
Center for Astrophysics and Associated Technologies (PFB-06).\ D.R.W. is supported by a fellowship from the Alfred P. Sloan Foundation.\ H.-X.Z. acknowledges an earlier support from the Chinese Academy of Sciences through a CAS-CONICYT Postdoctoral Fellowship administered by the CAS South America Centre for Astronomy (CASSACA) in Santiago, Chile.\ H.-X.Z. also acknowledges a support from the National Natural Science Foundation of China 
(NSFC, Grant No. 11390373).\ We thank Deidre Hunter for giving comments on an earlier draft version of this paper.\ We thank Andrew Cole for sharing his CMD-based SFHs of DDO 210 and  Leo A.\ We also thank Sebastian Hidalgo for sharing his CMD-based SFH of LGS 3.\ 

\end{acknowledgements}

%{\it Facilities: }\facility{CFHT/MegaCam} %, \facility{AAT/AAOmega}, \facility{MMT/Hectospec}
 
\appendix 
\section{Linear Least-squares Fitting of the \lclrmtl($\lambda$) Relations}\label{sec: appen}
Following the common practice, we determine the linear least-squares fitting parameters for the optical \lclrmtl($\lambda$) 
relations of the 40 LG dwarf galaxies, the default full sample, and the full sample expanded with different $A_{V, {\rm young}}$.\ 
For the 40 LG dwarf galaxies, the individual data points are directly used in the linear least-squares fitting, while for the default 
full sample and the expanded sample, the data points are first binned in color intervals of 0.05 mag, and then the mean of the maximum   
and minimum $\log \Upsilon_{\star}$($\lambda$) in each color bin is fitted as a function of colors.\ Note that, for the expanded sample 
with variable $A_{V, {\rm young}}$, here we only consider ``galaxies'' with $A_{V, {\rm young}}$ $>$ 0 mag, because no star formation  
in the present day takes place in a dust-free environment.\ The fitted slopes $b_{\lambda}$ and intercepts $a_{\lambda}$ are given in 
Tables \ref{tab_clrmtl_fsps} and \ref{tab_clrmtl_bc03} for the samples generated based on the FSPS and BC03 models, respectively.\

%Treu, T. 2010, ARA\&A, 48, 87

\clearpage
\begin{turnpage}
%\Distribution of the best-fit surface brightness profile models.
\begin{deluxetable*}{lrrrrrccccccccccr}
\tabletypesize{\footnotesize}
\tablecolumns{17}
\setlength{\columnsep}{0.01pt}
\tablewidth{0pt}
\tablecaption{Linear least-squares fitting to $\log$$\Upsilon_{\star}$($\lambda$) = $a_{\lambda}$ + ($b_{\lambda}$ $\times$ color) based on FSPS models}
\tablehead{
\colhead{Color}
&\colhead{$a_{B}$}
&\colhead{$b_{B}$}
&\colhead{$a_{g}$}
&\colhead{$b_{g}$}
&\colhead{$a_{V}$}
&\colhead{$b_{V}$}
&\colhead{$a_{r}$}
&\colhead{$b_{r}$}
&\colhead{$a_{R}$}
&\colhead{$b_{R}$}
&\colhead{$a_{i}$}
&\colhead{$b_{i}$}
&\colhead{$a_{I}$}
&\colhead{$b_{I}$}
&\colhead{$a_{z}$}
&\colhead{$b_{z}$}
}
\startdata

\multicolumn{17}{c}{The 40 LG Dwarf Galaxies}\\
\noalign{\vskip 0.5mm}
\hline
\noalign{\vskip 0.5mm}
$g$$-$$r$  &  $-$0.778  &  1.747  &  $-$0.684  &  1.575  &  $-$0.555  &  1.322  &  $-$0.491  &  1.175  &  $-$0.481  &  1.127  &  $-$0.457  &  1.012  &  $-$0.448  &  0.963  &  $-$0.447  &  0.911\\
\noalign{\vskip 0.4mm}
$g$$-$$i$  &  $-$0.802  &  1.242  &  $-$0.706  &  1.120  &  $-$0.573  &  0.941  &  $-$0.508  &  0.836  &  $-$0.497  &  0.802  &  $-$0.471  &  0.720  &  $-$0.461  &  0.685  &  $-$0.459  &  0.647\\
\noalign{\vskip 0.4mm}
$g$$-$$z$  &  $-$0.774  &  1.030  &  $-$0.680  &  0.928  &  $-$0.550  &  0.778  &  $-$0.486  &  0.690  &  $-$0.476  &  0.661  &  $-$0.451  &  0.592  &  $-$0.440  &  0.561  &  $-$0.438  &  0.528\\
\noalign{\vskip 0.4mm}
$r$$-$$i$  &  $-$0.849  &  4.249  &  $-$0.749  &  3.835  &  $-$0.610  &  3.224  &  $-$0.541  &  2.866  &  $-$0.529  &  2.748  &  $-$0.499  &  2.466  &  $-$0.487  &  2.341  &  $-$0.483  &  2.207\\
\noalign{\vskip 0.4mm}
$r$$-$$z$  &  $-$0.729  &  2.397  &  $-$0.639  &  2.157  &  $-$0.514  &  1.804  &  $-$0.453  &  1.596  &  $-$0.443  &  1.527  &  $-$0.419  &  1.361  &  $-$0.408  &  1.282  &  $-$0.404  &  1.196\\
\noalign{\vskip 0.4mm}
$B$$-$$V$  &  $-$0.921  &  1.632  &  $-$0.811  &  1.470  &  $-$0.661  &  1.232  &  $-$0.585  &  1.094  &  $-$0.571  &  1.049  &  $-$0.537  &  0.942  &  $-$0.524  &  0.896  &  $-$0.519  &  0.847\\
\noalign{\vskip 0.4mm}
$B$$-$$R$  &  $-$1.110  &  1.123  &  $-$0.982  &  1.012  &  $-$0.805  &  0.849  &  $-$0.713  &  0.754  &  $-$0.693  &  0.723  &  $-$0.648  &  0.650  &  $-$0.629  &  0.618  &  $-$0.618  &  0.584\\
\noalign{\vskip 0.4mm}
$B$$-$$I$  &  $-$1.254  &  0.885  &  $-$1.112  &  0.797  &  $-$0.913  &  0.669  &  $-$0.809  &  0.594  &  $-$0.786  &  0.569  &  $-$0.729  &  0.511  &  $-$0.706  &  0.485  &  $-$0.690  &  0.457\\
\noalign{\vskip 0.4mm}
$V$$-$$R$  &  $-$1.518  &  3.581  &  $-$1.352  &  3.231  &  $-$1.117  &  2.715  &  $-$0.991  &  2.414  &  $-$0.960  &  2.315  &  $-$0.888  &  2.080  &  $-$0.857  &  1.978  &  $-$0.834  &  1.869\\
\noalign{\vskip 0.4mm}
$V$$-$$I$  &  $-$1.626  &  1.906  &  $-$1.448  &  1.719  &  $-$1.196  &  1.442  &  $-$1.059  &  1.280  &  $-$1.025  &  1.227  &  $-$0.943  &  1.100  &  $-$0.907  &  1.042  &  $-$0.877  &  0.981\\
\noalign{\vskip 0.4mm}
$R$$-$$I$  &  $-$1.710  &  3.992  &  $-$1.523  &  3.596  &  $-$1.256  &  3.012  &  $-$1.111  &  2.669  &  $-$1.074  &  2.556  &  $-$0.983  &  2.283  &  $-$0.941  &  2.156  &  $-$0.905  &  2.017\\
\noalign{\vskip 0.4mm}
\hline
\noalign{\vskip 0.5mm}
\multicolumn{17}{c}{The Default Full Sample with $A_{V, {\rm young}} = 0.5$ mag and an SMC-bar Extinction Curve}\\
\noalign{\vskip 0.5mm}
\hline
\noalign{\vskip 0.5mm}
$g$$-$$r$  &  $-$0.629  &  1.520  &  $-$0.535  &  1.355  &  $-$0.407  &  1.109  &  $-$0.345  &  0.962  &  $-$0.336  &  0.913  &  $-$0.317  &  0.795  &  $-$0.313  &  0.728  &  $-$0.318  &  0.647\\
\noalign{\vskip 0.4mm}
$g$$-$$i$  &  $-$0.680  &  1.085  &  $-$0.582  &  0.972  &  $-$0.448  &  0.801  &  $-$0.383  &  0.698  &  $-$0.373  &  0.663  &  $-$0.351  &  0.578  &  $-$0.343  &  0.529  &  $-$0.345  &  0.471\\
\noalign{\vskip 0.4mm}
$g$$-$$z$  &  $-$0.683  &  0.864  &  $-$0.584  &  0.770  &  $-$0.452  &  0.632  &  $-$0.388  &  0.551  &  $-$0.378  &  0.524  &  $-$0.355  &  0.457  &  $-$0.346  &  0.418  &  $-$0.347  &  0.375\\
\noalign{\vskip 0.4mm}
$r$$-$$i$  &  $-$0.714  &  3.437  &  $-$0.614  &  3.072  &  $-$0.479  &  2.537  &  $-$0.414  &  2.218  &  $-$0.403  &  2.110  &  $-$0.378  &  1.845  &  $-$0.367  &  1.690  &  $-$0.365  &  1.511\\
\noalign{\vskip 0.4mm}
$r$$-$$z$  &  $-$0.680  &  1.841  &  $-$0.587  &  1.651  &  $-$0.458  &  1.363  &  $-$0.396  &  1.189  &  $-$0.386  &  1.131  &  $-$0.364  &  0.987  &  $-$0.355  &  0.904  &  $-$0.354  &  0.807\\
\noalign{\vskip 0.4mm}
$B$$-$$V$  &  $-$0.747  &  1.440  &  $-$0.638  &  1.281  &  $-$0.490  &  1.046  &  $-$0.417  &  0.906  &  $-$0.404  &  0.860  &  $-$0.375  &  0.747  &  $-$0.364  &  0.682  &  $-$0.360  &  0.603\\
\noalign{\vskip 0.4mm}
$B$$-$$R$  &  $-$0.937  &  0.997  &  $-$0.811  &  0.890  &  $-$0.634  &  0.730  &  $-$0.544  &  0.634  &  $-$0.525  &  0.602  &  $-$0.481  &  0.524  &  $-$0.461  &  0.479  &  $-$0.449  &  0.426\\
\noalign{\vskip 0.4mm}
$B$$-$$I$  &  $-$1.072  &  0.765  &  $-$0.931  &  0.684  &  $-$0.735  &  0.563  &  $-$0.632  &  0.489  &  $-$0.609  &  0.464  &  $-$0.554  &  0.403  &  $-$0.527  &  0.368  &  $-$0.506  &  0.327\\
\noalign{\vskip 0.4mm}
$V$$-$$R$  &  $-$1.267  &  3.021  &  $-$1.108  &  2.702  &  $-$0.885  &  2.232  &  $-$0.766  &  1.948  &  $-$0.737  &  1.853  &  $-$0.670  &  1.620  &  $-$0.636  &  1.486  &  $-$0.606  &  1.328\\
\noalign{\vskip 0.4mm}
$V$$-$$I$  &  $-$1.393  &  1.587  &  $-$1.219  &  1.416  &  $-$0.974  &  1.164  &  $-$0.844  &  1.015  &  $-$0.813  &  0.966  &  $-$0.736  &  0.845  &  $-$0.698  &  0.777  &  $-$0.664  &  0.698\\
\noalign{\vskip 0.4mm}
$R$$-$$I$  &  $-$1.359  &  2.955  &  $-$1.191  &  2.641  &  $-$0.950  &  2.163  &  $-$0.819  &  1.873  &  $-$0.787  &  1.775  &  $-$0.708  &  1.537  &  $-$0.667  &  1.399  &  $-$0.630  &  1.239\\
\noalign{\vskip 0.4mm}
\hline
\noalign{\vskip 0.5mm}
\multicolumn{17}{c}{The Full Sample with Variable $A_{V, {\rm young}}$ and an SMC-bar Extinction Curve}\\
\noalign{\vskip 0.5mm}
\hline
\noalign{\vskip 0.5mm}
$g$$-$$r$  &  $-$0.635  &  1.469  &  $-$0.539  &  1.311  &  $-$0.408  &  1.064  &  $-$0.346  &  0.912  &  $-$0.337  &  0.860  &  $-$0.315  &  0.727  &  $-$0.308  &  0.647  &  $-$0.309  &  0.542\\
\noalign{\vskip 0.4mm}
$g$$-$$i$  &  $-$0.661  &  0.995  &  $-$0.564  &  0.891  &  $-$0.431  &  0.726  &  $-$0.366  &  0.624  &  $-$0.356  &  0.588  &  $-$0.330  &  0.494  &  $-$0.322  &  0.438  &  $-$0.320  &  0.366\\
\noalign{\vskip 0.4mm}
$g$$-$$z$  &  $-$0.621  &  0.709  &  $-$0.529  &  0.631  &  $-$0.402  &  0.509  &  $-$0.340  &  0.433  &  $-$0.330  &  0.406  &  $-$0.305  &  0.335  &  $-$0.295  &  0.292  &  $-$0.291  &  0.236\\
\noalign{\vskip 0.4mm}
$r$$-$$i$  &  $-$0.631  &  2.637  &  $-$0.542  &  2.363  &  $-$0.416  &  1.918  &  $-$0.354  &  1.636  &  $-$0.344  &  1.533  &  $-$0.318  &  1.263  &  $-$0.307  &  1.098  &  $-$0.299  &  0.868\\
\noalign{\vskip 0.4mm}
$r$$-$$z$  &  $-$0.554  &  1.254  &  $-$0.474  &  1.126  &  $-$0.360  &  0.912  &  $-$0.306  &  0.776  &  $-$0.298  &  0.726  &  $-$0.277  &  0.593  &  $-$0.268  &  0.509  &  $-$0.262  &  0.389\\
\noalign{\vskip 0.4mm}
$B$$-$$V$  &  $-$0.736  &  1.407  &  $-$0.627  &  1.250  &  $-$0.479  &  1.013  &  $-$0.407  &  0.871  &  $-$0.394  &  0.823  &  $-$0.366  &  0.702  &  $-$0.355  &  0.630  &  $-$0.350  &  0.535\\
\noalign{\vskip 0.4mm}
$B$$-$$R$  &  $-$0.926  &  0.971  &  $-$0.800  &  0.866  &  $-$0.621  &  0.705  &  $-$0.530  &  0.607  &  $-$0.511  &  0.574  &  $-$0.465  &  0.489  &  $-$0.444  &  0.439  &  $-$0.427  &  0.374\\
\noalign{\vskip 0.4mm}
$B$$-$$I$  &  $-$1.036  &  0.717  &  $-$0.900  &  0.643  &  $-$0.706  &  0.525  &  $-$0.604  &  0.452  &  $-$0.580  &  0.426  &  $-$0.520  &  0.360  &  $-$0.491  &  0.320  &  $-$0.462  &  0.269\\
\noalign{\vskip 0.4mm}
$V$$-$$R$  &  $-$1.202  &  2.755  &  $-$1.049  &  2.463  &  $-$0.825  &  2.002  &  $-$0.704  &  1.716  &  $-$0.673  &  1.614  &  $-$0.596  &  1.354  &  $-$0.558  &  1.198  &  $-$0.513  &  0.991\\
\noalign{\vskip 0.4mm}
$V$$-$$I$  &  $-$1.210  &  1.302  &  $-$1.053  &  1.158  &  $-$0.825  &  0.932  &  $-$0.701  &  0.794  &  $-$0.669  &  0.744  &  $-$0.587  &  0.616  &  $-$0.543  &  0.540  &  $-$0.495  &  0.441\\
\noalign{\vskip 0.4mm}
$R$$-$$I$  &  $-$1.130  &  2.288  &  $-$0.989  &  2.047  &  $-$0.777  &  1.653  &  $-$0.661  &  1.408  &  $-$0.631  &  1.319  &  $-$0.555  &  1.089  &  $-$0.513  &  0.949  &  $-$0.465  &  0.760\\
\noalign{\vskip 0.4mm}

\noalign{\vskip -3mm} 
  
\enddata
\tablecomments{
Linear least-squares fitting to optical \lclrmtl($\lambda$) distributions of our samples generated based on the FSPS models and the 
Chabrier (2003) IMF.\ The monochromatic mass-to-light ratios $\Upsilon_{\star}$($\lambda$) are in solar units.\ The SDSS $griz$ filters 
are in the AB magnitude system.\ The Johnson--Cousins $BVRI$ filters are in the Vega magnitude system.
}
\label{tab_clrmtl_fsps}
\end{deluxetable*}
\clearpage
%\Distribution of the best-fit surface brightness profile models.
\begin{deluxetable*}{lrrrrrccccccccccr}
\tabletypesize{\footnotesize}
\tablecolumns{17}
\setlength{\columnsep}{0.05pt}
\tablewidth{0pt}
\tablecaption{Linear least-squares fitting to $\log$$\Upsilon_{\star}$($\lambda$) = $a_{\lambda}$ + ($b_{\lambda}$ $\times$ color) based on BC03 models}
\tablehead{
\colhead{Color}
&\colhead{$a_{B}$}
&\colhead{$b_{B}$}
&\colhead{$a_{g}$}
&\colhead{$b_{g}$}
&\colhead{$a_{V}$}
&\colhead{$b_{V}$}
&\colhead{$a_{r}$}
&\colhead{$b_{r}$}
&\colhead{$a_{R}$}
&\colhead{$b_{R}$}
&\colhead{$a_{i}$}
&\colhead{$b_{i}$}
&\colhead{$a_{I}$}
&\colhead{$b_{I}$}
&\colhead{$a_{z}$}
&\colhead{$b_{z}$}
}
\startdata

\multicolumn{17}{c}{The 40 LG Dwarf Galaxies}\\
\noalign{\vskip 0.5mm}
\hline
\noalign{\vskip 0.5mm}
$g$$-$$r$  &  $-$0.842  &  1.786  &  $-$0.745  &  1.616  &  $-$0.617  &  1.367  &  $-$0.552  &  1.216  &  $-$0.540  &  1.164  &  $-$0.511  &  1.038  &  $-$0.504  &  0.988  &  $-$0.501  &  0.937\\
\noalign{\vskip 0.4mm}
$g$$-$$i$  &  $-$0.837  &  1.227  &  $-$0.739  &  1.109  &  $-$0.612  &  0.937  &  $-$0.547  &  0.833  &  $-$0.534  &  0.797  &  $-$0.505  &  0.709  &  $-$0.497  &  0.674  &  $-$0.494  &  0.638\\
\noalign{\vskip 0.4mm}
$g$$-$$z$  &  $-$0.808  &  1.020  &  $-$0.712  &  0.920  &  $-$0.587  &  0.776  &  $-$0.523  &  0.687  &  $-$0.511  &  0.657  &  $-$0.482  &  0.582  &  $-$0.474  &  0.552  &  $-$0.470  &  0.520\\
\noalign{\vskip 0.4mm}
$r$$-$$i$  &  $-$0.808  &  3.844  &  $-$0.711  &  3.467  &  $-$0.586  &  2.924  &  $-$0.523  &  2.592  &  $-$0.511  &  2.476  &  $-$0.481  &  2.192  &  $-$0.474  &  2.076  &  $-$0.469  &  1.956\\
\noalign{\vskip 0.4mm}
$r$$-$$z$  &  $-$0.725  &  2.270  &  $-$0.634  &  2.039  &  $-$0.517  &  1.709  &  $-$0.459  &  1.507  &  $-$0.448  &  1.435  &  $-$0.422  &  1.260  &  $-$0.415  &  1.186  &  $-$0.410  &  1.107\\
\noalign{\vskip 0.4mm}
$B$$-$$V$  &  $-$0.981  &  1.689  &  $-$0.869  &  1.526  &  $-$0.721  &  1.289  &  $-$0.643  &  1.145  &  $-$0.627  &  1.095  &  $-$0.587  &  0.975  &  $-$0.576  &  0.927  &  $-$0.568  &  0.878\\
\noalign{\vskip 0.4mm}
$B$$-$$R$  &  $-$1.160  &  1.140  &  $-$1.031  &  1.030  &  $-$0.858  &  0.871  &  $-$0.766  &  0.773  &  $-$0.743  &  0.740  &  $-$0.691  &  0.659  &  $-$0.675  &  0.626  &  $-$0.662  &  0.593\\
\noalign{\vskip 0.4mm}
$B$$-$$I$  &  $-$1.288  &  0.879  &  $-$1.146  &  0.794  &  $-$0.954  &  0.670  &  $-$0.850  &  0.595  &  $-$0.823  &  0.569  &  $-$0.761  &  0.505  &  $-$0.740  &  0.479  &  $-$0.722  &  0.453\\
\noalign{\vskip 0.4mm}
$V$$-$$R$  &  $-$1.528  &  3.502  &  $-$1.365  &  3.166  &  $-$1.141  &  2.678  &  $-$1.018  &  2.381  &  $-$0.985  &  2.278  &  $-$0.906  &  2.029  &  $-$0.880  &  1.929  &  $-$0.856  &  1.828\\
\noalign{\vskip 0.4mm}
$V$$-$$I$  &  $-$1.610  &  1.822  &  $-$1.435  &  1.643  &  $-$1.197  &  1.386  &  $-$1.065  &  1.229  &  $-$1.028  &  1.174  &  $-$0.940  &  1.040  &  $-$0.909  &  0.986  &  $-$0.880  &  0.929\\
\noalign{\vskip 0.4mm}
$R$$-$$I$  &  $-$1.662  &  3.716  &  $-$1.478  &  3.343  &  $-$1.228  &  2.809  &  $-$1.088  &  2.483  &  $-$1.048  &  2.368  &  $-$0.952  &  2.085  &  $-$0.916  &  1.968  &  $-$0.882  &  1.845\\
\noalign{\vskip 0.4mm}
\hline
\noalign{\vskip 0.5mm}
\multicolumn{17}{c}{The Default Full Sample with $A_{V, {\rm young}} = 0.5$ mag and an SMC-bar Extinction Curve}\\
\noalign{\vskip 0.5mm}
\hline
\noalign{\vskip 0.5mm}
$g$$-$$r$  &  $-$0.690  &  1.566  &  $-$0.601  &  1.410  &  $-$0.475  &  1.169  &  $-$0.411  &  1.020  &  $-$0.402  &  0.969  &  $-$0.382  &  0.848  &  $-$0.380  &  0.793  &  $-$0.381  &  0.721\\
\noalign{\vskip 0.4mm}
$g$$-$$i$  &  $-$0.738  &  1.128  &  $-$0.646  &  1.021  &  $-$0.517  &  0.855  &  $-$0.451  &  0.751  &  $-$0.440  &  0.716  &  $-$0.416  &  0.630  &  $-$0.414  &  0.590  &  $-$0.413  &  0.538\\
\noalign{\vskip 0.4mm}
$g$$-$$z$  &  $-$0.742  &  0.914  &  $-$0.651  &  0.829  &  $-$0.523  &  0.696  &  $-$0.455  &  0.611  &  $-$0.444  &  0.582  &  $-$0.420  &  0.511  &  $-$0.416  &  0.478  &  $-$0.414  &  0.435\\
\noalign{\vskip 0.4mm}
$r$$-$$i$  &  $-$0.770  &  3.381  &  $-$0.674  &  3.071  &  $-$0.546  &  2.607  &  $-$0.480  &  2.314  &  $-$0.469  &  2.206  &  $-$0.446  &  1.946  &  $-$0.442  &  1.819  &  $-$0.439  &  1.650\\
\noalign{\vskip 0.4mm}
$r$$-$$z$  &  $-$0.707  &  1.722  &  $-$0.621  &  1.570  &  $-$0.503  &  1.341  &  $-$0.442  &  1.192  &  $-$0.432  &  1.133  &  $-$0.409  &  0.987  &  $-$0.405  &  0.913  &  $-$0.401  &  0.811\\
\noalign{\vskip 0.4mm}
$B$$-$$V$  &  $-$0.842  &  1.527  &  $-$0.735  &  1.371  &  $-$0.585  &  1.134  &  $-$0.510  &  0.992  &  $-$0.498  &  0.946  &  $-$0.468  &  0.831  &  $-$0.462  &  0.779  &  $-$0.457  &  0.709\\
\noalign{\vskip 0.4mm}
$B$$-$$R$  &  $-$1.030  &  1.057  &  $-$0.906  &  0.951  &  $-$0.731  &  0.793  &  $-$0.638  &  0.696  &  $-$0.619  &  0.663  &  $-$0.576  &  0.584  &  $-$0.564  &  0.548  &  $-$0.550  &  0.500\\
\noalign{\vskip 0.4mm}
$B$$-$$I$  &  $-$1.185  &  0.821  &  $-$1.050  &  0.742  &  $-$0.854  &  0.621  &  $-$0.747  &  0.545  &  $-$0.721  &  0.519  &  $-$0.663  &  0.455  &  $-$0.643  &  0.426  &  $-$0.621  &  0.388\\
\noalign{\vskip 0.4mm}
$V$$-$$R$  &  $-$1.340  &  3.131  &  $-$1.185  &  2.822  &  $-$0.965  &  2.357  &  $-$0.847  &  2.075  &  $-$0.819  &  1.980  &  $-$0.756  &  1.752  &  $-$0.736  &  1.648  &  $-$0.711  &  1.509\\
\noalign{\vskip 0.4mm}
$V$$-$$I$  &  $-$1.463  &  1.621  &  $-$1.302  &  1.467  &  $-$1.068  &  1.231  &  $-$0.936  &  1.082  &  $-$0.901  &  1.029  &  $-$0.821  &  0.902  &  $-$0.790  &  0.842  &  $-$0.753  &  0.763\\
\noalign{\vskip 0.4mm}
$R$$-$$I$  &  $-$1.415  &  2.888  &  $-$1.266  &  2.629  &  $-$1.052  &  2.240  &  $-$0.929  &  1.988  &  $-$0.894  &  1.888  &  $-$0.810  &  1.639  &  $-$0.774  &  1.513  &  $-$0.728  &  1.339\\
\noalign{\vskip 0.4mm}
\hline
\noalign{\vskip 0.5mm}
\multicolumn{17}{c}{The Full Sample with Variable $A_{V, {\rm young}}$ and an SMC-bar Extinction Curve}\\
\noalign{\vskip 0.5mm}
\hline
\noalign{\vskip 0.5mm}
$g$$-$$r$  &  $-$0.701  &  1.550  &  $-$0.608  &  1.394  &  $-$0.479  &  1.148  &  $-$0.417  &  0.998  &  $-$0.410  &  0.948  &  $-$0.390  &  0.818  &  $-$0.389  &  0.756  &  $-$0.389  &  0.667\\
\noalign{\vskip 0.4mm}
$g$$-$$i$  &  $-$0.732  &  1.049  &  $-$0.639  &  0.948  &  $-$0.509  &  0.788  &  $-$0.444  &  0.685  &  $-$0.432  &  0.647  &  $-$0.406  &  0.552  &  $-$0.401  &  0.506  &  $-$0.396  &  0.439\\
\noalign{\vskip 0.4mm}
$g$$-$$z$  &  $-$0.695  &  0.764  &  $-$0.606  &  0.691  &  $-$0.484  &  0.573  &  $-$0.421  &  0.498  &  $-$0.410  &  0.468  &  $-$0.383  &  0.393  &  $-$0.376  &  0.354  &  $-$0.368  &  0.296\\
\noalign{\vskip 0.4mm}
$r$$-$$i$  &  $-$0.664  &  2.540  &  $-$0.575  &  2.298  &  $-$0.456  &  1.910  &  $-$0.397  &  1.661  &  $-$0.386  &  1.555  &  $-$0.360  &  1.285  &  $-$0.353  &  1.141  &  $-$0.343  &  0.924\\
\noalign{\vskip 0.4mm}
$r$$-$$z$  &  $-$0.597  &  1.187  &  $-$0.518  &  1.084  &  $-$0.409  &  0.908  &  $-$0.357  &  0.794  &  $-$0.347  &  0.741  &  $-$0.326  &  0.602  &  $-$0.319  &  0.524  &  $-$0.309  &  0.403\\
\noalign{\vskip 0.4mm}
$B$$-$$V$  &  $-$0.811  &  1.473  &  $-$0.702  &  1.317  &  $-$0.552  &  1.078  &  $-$0.478  &  0.935  &  $-$0.468  &  0.889  &  $-$0.440  &  0.770  &  $-$0.435  &  0.713  &  $-$0.432  &  0.634\\
\noalign{\vskip 0.4mm}
$B$$-$$R$  &  $-$1.016  &  1.030  &  $-$0.890  &  0.926  &  $-$0.712  &  0.764  &  $-$0.621  &  0.667  &  $-$0.604  &  0.635  &  $-$0.559  &  0.551  &  $-$0.546  &  0.510  &  $-$0.529  &  0.453\\
\noalign{\vskip 0.4mm}
$B$$-$$I$  &  $-$1.147  &  0.773  &  $-$1.015  &  0.699  &  $-$0.821  &  0.581  &  $-$0.715  &  0.506  &  $-$0.688  &  0.479  &  $-$0.625  &  0.409  &  $-$0.601  &  0.375  &  $-$0.570  &  0.326\\
\noalign{\vskip 0.4mm}
$V$$-$$R$  &  $-$1.259  &  2.820  &  $-$1.118  &  2.554  &  $-$0.906  &  2.115  &  $-$0.787  &  1.834  &  $-$0.756  &  1.729  &  $-$0.679  &  1.467  &  $-$0.650  &  1.338  &  $-$0.610  &  1.153\\
\noalign{\vskip 0.4mm}
$V$$-$$I$  &  $-$1.285  &  1.340  &  $-$1.138  &  1.209  &  $-$0.926  &  1.003  &  $-$0.805  &  0.871  &  $-$0.771  &  0.818  &  $-$0.685  &  0.684  &  $-$0.646  &  0.614  &  $-$0.591  &  0.511\\
\noalign{\vskip 0.4mm}
$R$$-$$I$  &  $-$1.147  &  2.166  &  $-$1.019  &  1.974  &  $-$0.828  &  1.659  &  $-$0.722  &  1.445  &  $-$0.690  &  1.352  &  $-$0.609  &  1.112  &  $-$0.572  &  0.983  &  $-$0.516  &  0.789\\
\noalign{\vskip 0.4mm}

\noalign{\vskip -3mm} 
  
\enddata
\tablecomments{
Same as Table \ref{tab_clrmtl_fsps}, with the exception that here the optical \lclrmtl($\lambda$) distributions of our samples are generated 
based on the BC03 models.
}
\label{tab_clrmtl_bc03}
\end{deluxetable*}
\end{turnpage}


\vspace{1cm}
\begin{thebibliography}{}

\bibitem[Alongi et al.(1993)]{alongi93} Alongi, M., Bertelli, G., Bressan, A., et al., 1993, A\&AS, 97, 851
\bibitem[Acquaviva et al.(2011)]{acquaviva11} Acquaviva, V., Gawiser, E., Guaita, L., 2011, \apj, 737, 47
\bibitem[Arnouts \& IIbert(2011)]{arnouts11} Arnouts, S., Ilbert, O., 2011, LePHARE: Photometric Analysis for Redshift Estimate, astrophysics Source Code Library, ascl: 1108.009
\bibitem[Baldwin et al.(2017)]{baldwin17} Baldwin, C.~M., McDermid, R.~M., Kuntschner, H., Maraston, C., Conroy, C., arXiv:1709.09300
\bibitem[Bressan et al.(1993)]{bressan93} Bressan, A., Fagotto, F., Bertelli, G., Chiosi, C., 1993, A\&AS, 100, 647
\bibitem[Brinchmann \& Ellis(2000)]{brinchmann00} Brinchmann, J., Ellis, R. S., 2000, ApJL, 536, 77
\bibitem[Bershady et al.(2000)]{bershady00} Bershady, M.A., Jangren, J.A., Conselice, C.J. 2000, \aj, 119, 2645
\bibitem[Bolzonella et al.(2000)]{bolzonella00} Bolzonella, M., Miralles, J.M., Pell\'o, R., 2000, A\&A, 363, 476
\bibitem[Bolzonella et al.(2010)]{bolzonella10} Bolzonella, M., Kovac, K., Pozzetti, L., Zucca, E., Cucciati, O., et al., 2010, A\&A, 524, 76 
\bibitem[Bell \& de Jong(2001)]{bell01} Bell, E. F. \& de Jong, R. S. 2001, \apj, 550, 212
\bibitem[Bell et al.(2003)]{bell03} Bell, E. F., McIntosh, D. H., Katz, N., Weinberg, M. D., 2003, \apjs, 149, 289
\bibitem[Bruzual \& Charlot(2003)]{bruzual03} Bruzual, G., Charlot, S., 2003, \mnras, 344, 1000 (BC03)
\bibitem[Borch et al.(2006)]{borch06} Borch, A., Meisenheimer, K., Bell, E.F. et al., 2006, A\&A, 453, 869
\bibitem[Blanton \& Roweis(2007)]{blanton07} Blanton, M. R., Roweis, S. 2007, \aj, 133, 734
\bibitem[Brammer et al.(2008)]{brammer08} Brammer, G.B., van Dokkum, P.G., Coppi, P., 2008, \apj, 686, 1503
\bibitem[Bendo et al.(2015)]{bendo15} Bendo, G.J., Baes, M., Bianchi, S., et al. 2015, \mnras, 448, 135
\bibitem[Battisti et al.(2016)]{battisti16} Battisti, A.J., Calzetti, D., Chary, R.-R., 2016, \apj, 818, 13
\bibitem[Charlot \& Fall(2000)]{charlot00} Charlot, S., Fall, S.M., 2000, \apj, 539, 718
\bibitem[Chabrier(2003)]{chabrier03} Chabrier, G. 2003, \pasp, 115, 763
%\bibitem[Cariulo et al.(2004)]{cariulo04} Cariulo, P., Degl'Innocenti, S., Castellani, V. 2004, A\&A, 421, 1121
\bibitem[Cid Fernandes et al.(2005)]{cid05} Cid Fernandes, R., Mateus, A., Sodre, L., et al. 2005, \mnras, 358, 363
\bibitem[Conroy et al.(2009)]{conroy09} Conroy, C., Gunn, J. E., White, M., 2009, \apj, 699, 486
\bibitem[Conroy \& Gunn(2010)]{conroy10} Conroy, C., Gunn, J. E. 2010, \apj, 712, 833
\bibitem[Conroy(2013)]{conroy13} Conroy, C., 2013, ARA\&A, 51, 393
\bibitem[Cole et al.(2007)]{cole07} Cole, A. A., Skillman, E. D., Tolstoy, E., et al. 2007, \apjl, 659, 17
\bibitem[Cole et al.(2014)]{cole14} Cole, A. A., Weisz, D. R., Dolphin, A. E., et al. 2014, \apj, 795, 54
\bibitem[Chen et al.(2012)]{chen12} Chen, Y. M., Kauffmann, G., Tremonti, C., et al., 2012, \mnras, 421, 314
\bibitem[Chevallard et al.(2013)]{chevallard13} Chevallard, J., Charlot, S., Wandelt, B., Wild, V., 2013, \mnras, 432, 2061
\bibitem[Calzetti et al.(1994)]{calzetti94} Calzetti, D., Kinney, A.L., Storchi-Bergmann, T., 1994, \apj, 429, 582
%\bibitem[Calzetti et al.(2000)]{calzetti00} Calzetti, D., Armus, L., Bohlin, R.C., et al. 2000, \apj, 533, 682
\bibitem[Chevallard \& Charlot(2016)]{chevallard16} Chevallard, J., Charlot, S., 2016, \mnras, 462, 1415
\bibitem[Cappellari(2017)]{cappellari17} Cappellari, M. 2017, \mnras, 466, 798
\bibitem[Dolphin et al.(2003)]{dolphin03} Dolphin, A. E., Saha, A., Skillman, E. D., Dohm-Palmer, R. C. 2003, \aj, 2003, 126, 187
\bibitem[Dabringhausen et al.(2008)]{dabringhausen08} Dabringhausen, J., Hilker, M., \& Kroupa, P. 2008, \mnras, 386, 864
\bibitem[da Cunha et al.(2008)]{dacunha08} da Cunha, E., Charlot, S., \& Elbaz, D. 2008, \mnras, 388, 1595
\bibitem[Eldridge \& Stanway(2009)]{eldridge09} Eldridge, J. J., Stanway, E. R., 2009, \mnras, 400, 1019
\bibitem[El-Badry et al.(2017)]{ElBadry17} El-Badry, K., Weisz, D. R., Quataert, E., 2017, \mnras, 468, 319
\bibitem[Fagotto et al.(1994a)]{fagotto94a} Fagotto, F., Bressan, A., Bertelli, G., Chiosi, C., 1994a, A\&AS, 104, 365
\bibitem[Fagotto et al.(1994b)]{fagotto94b} Fagotto, F., Bressan, A., Bertelli, G., Chiosi, C., 1994b, A\&AS, 105, 29
\bibitem[Fioc \& Rocca-Volmerange(1997)]{fioc97} Fioc, M., Rocca-Volmerange, B., 1997, A\&A, 326, 950
\bibitem[Ford et al.(1998)]{ford98} Ford, H. C., Bartko, F., Bely, P. Y., et al. 1998, Proc. SPIE, 3356, 234
\bibitem[Flaugher(2005)]{flaugher05} Flaugher, B., 2005, International Journal of Modern Physics A. 20, 3121
\bibitem[Geha et al.(2013)]{Geha13} Geha, M., Brown, T. M., Tumlinson, J., et al. 2013, \apj, 771, 29
\bibitem[Girardi et al.(1996)]{girardi96} Girardi, L., Bressan, A., Chiosi, C., et al., 1996, A\&AS, 117, 113
\bibitem[Girardi et al.(2000)]{girardi00} Girardi, L., Bressan, A., Bertelli, G., et al., 2000, A\&AS, 141, 371
\bibitem[Gordon et al.(2003)]{gordon03} Gordon, K.D., Clayton, G.C., Misselt, K.A., et al.\ 2003, \apj, 594, 279
\bibitem[Gallazzi et al.(2005)]{Gallazzi05} Gallazzi, A., Charlot, S., Brinchmann, J., et al., 2005, \mnras, 362, 41
\bibitem[Gallazzi \& Bell(2009)]{gallazzi09} Gallazzi, A., Bell, E.F., 2009, ApJS, 185, 253
\bibitem[Holtzman et al.(1995)]{holtzman95} Holtzman, J. A., Burrows, C. J., Casertano, S., et al. 1995, \pasp, 107, 1065
\bibitem[Hunter \& Hoffman(1999)]{hunter99} Hunter, D. A., Hoffman, L., 1999, \aj, 117, 2789
\bibitem[Hidalgo et al.(2011)]{hidalgo11} Hidalgo, S. L., Aparicio, A., Skillman, E., et al. 2011, \apj, 730, 14
\bibitem[Han \& Han(2014)]{han14} Han, Y., Han, Z., 2014, ApJS, 215, 2
\bibitem[Hayward \& Smith(2015)]{hayward15} Hayward, C.C., Smith, D.J.B. 2015, \mnras, 446, 1512
\bibitem[Heavens et al.(2000)]{heavens00} Heavens, A. F., Jimenez, R., Lahav, O., 2000, \mnras, 317, 965
%\bibitem[Hewett et al.(2006)]{hewett06} Hewett, P. C., Warren, S. J., Leggett, S. K., Hodgkin, S. T. 2006, \mnras, 367, 454
\bibitem[Herrmann et al.(2016)]{herrmann16} Herrmann, K. A., Hunter, D. A., Zhang, H.-X., Elmegreen, B. G. 2016, \aj, 152, 177
\bibitem[Hunt et al.(2016)]{Hunt16} Hunt, L., Dayal, P., Magrini, L., Ferrara, A., et al. 2016, \mnras, 463, 2002
\bibitem[Ivezic et al.(2008)]{ivezic08} Ivezic, Z., Tyson, J. A., Abel, B., et al. 2008, arXiv:0805.2366
\bibitem[Into \& Portinari(2013)]{into13} Into, T., Portinari, L. 2013, \mnras, 430, 2715
\bibitem[Iyer \& Gawiser(2017)]{iyer17} Iyer, K., Gawiser, E., 2017, \apj, 838, 127
\bibitem[Jarrett et al.(2000)]{jarrett00} Jarrett, T.H., Chester, T., Cutri, R. et al., 2000, \aj, 119, 2498
\bibitem[Johnson et al.(2013a)]{johnson13a} Johnson, B. D., Weisz, D. R., Dalcanton, J. J. et al. 2013a, ApJ, 772, 8
\bibitem[Johnson et al.(2013b)]{johnson13b} Johnson, S.P., Wilson, G.W., Tang, Y., Scott, K.S., 2013b, \mnras, 436, 2535
\bibitem[Kroupa(2001)]{kroupa01} Kroupa, P. 2001, \mnras, 322, 231
\bibitem[Kauffmann et al.(2003)]{kauffmann03} Kauffmann, G., Heckman, T.M., White, S.D.M. et al. 2003, \mnras, 341, 33
\bibitem[Kannappan \& Gawiser(2007)]{kannappan07} Kannappan, S. J., Gawiser, E., 2007, \apj, 657, 5
\bibitem[Kotulla et al.(2009)]{kotulla09} Kotulla, R., Fritze, U., Weilbacher, P., Anders, P., 2009, \mnras, 396, 462
\bibitem[Karachentsev et al.(2013)]{karachentsev13} Karachentsev, I. D., Makarov, D. I., Kaisina,, E. I., 2013, \aj, 145, 101
\bibitem[Lawson \& Hanson(1974)]{lawson74} Lawson, C. L., \& Hanson, R. J., {\it Solving Least Squares Problems}, First Edition, Prentice-Hall, Englewood Cliffs, N. J., 1974
\bibitem[Leitherer et al.(1999)]{leitherer99} Leitherer, C., Schaerer, D., Goldader, J. D., et al. 1999, \apjs, 123, 3
%\bibitem[Lawrence et al.(2007)]{lawrence07} Lawrence, A. et al., 2007, \mnras, 379, 1599
\bibitem[Lee et al.(2009a)]{lee09a} Lee, S.-K., Idzi, R., Ferguson, H.C. et al., 2009a, ApJS, 184, 100
\bibitem[Lee et al.(2009b)]{lee09b} Lee, J.C., Kennicutt, R.C. Jr., Funes, J.G. et al. 2009b, \apj, 692, 1305
\bibitem[Leja et al.(2017)]{leja17} Leja, J., Johnson, B.D., Conroy, C. et al., 2017, \apj, 837, 170
\bibitem[Lo Faro et al.(2017)]{lo17}Lo Faro, B., Buat, V., Roehlly, Y. et al., 2017, \mnras, 472, 1472
%\bibitem[Martin et al.(2005)]{martin05} Martin, D.C., Seibert, M., Buat, V., et al. 2005, ApJL, 619, 59
\bibitem[Martin et al.(2005)]{martin05} Martin, D. C., Fanson, J., Schiminovich, D. et al., 2005, \apjl, 619, 1 
\bibitem[Marigo \& Girardi(2007)]{marigo07} Marigo, P., Girardi, L., 2007, A\&A, 469, 239
\bibitem[Marigo et al.(2008)]{marigo08} Marigo, P., Girardi, L., Bressan, A., et al. 2008, A\&A, 482, 883
%\bibitem[Markwardt(2009)]{markwardt09} Markwardt, C. B.\ 2009, Astronomical Data Analysis Software and Systems XVIII, 411, 251
\bibitem[Madau \& Dickinson(2014)]{madau14} Madau, P., Dickinson, M., 2014, \araa, 52, 415
\bibitem[Mobasher et al.(2015)]{mobasher15} Mobasher, B., Dahlen, T., Ferguson, H.C. et al., 2015, \apj, 808, 101
\bibitem[Maraston(2005)]{maraston05} Maraston, C., 2005, \mnras, 362, 799
\bibitem[Maraston et al.(2006)]{maraston06} Maraston, C., Daddi, E., Renzini, A., et al.\ 2006, \apj, 652, 85
\bibitem[Maraston et al.(2013)]{maraston13} Maraston, C., Pforr, J., Henriques, B.M. et al., 2013, \mnras, 435, 2764
\bibitem[McQuinn et al.(2010)]{mcquinn10} McQuinn, K. B. W., Skillman, E. D., Cannon, J. M., et al. 2010, \apj, 724, 49
\bibitem[Moultaka \& Pelat(2000)]{moultaka00} Moultaka, J., Pelat, D. 2000, \mnras, 314, 409
\bibitem[Moustakas et al.(2013)]{moustakas13} Moustakas, J., Coil, A. L., Aird, J.,  et al., 2013, \apj, 767, 50
\bibitem[Mitchell et al.(2013)]{mitchell13} Mitchell, P.D., Lacey, C.G., Baugh, C.M., Cole, S. 2013, \mnras, 435, 87
\bibitem[Meinshausen(2013)]{meinshausen13} Meinshausen, N. 2013, {\it Electronic Journal of Statistics}, 7, 1607
\bibitem[McGaugh \& Schombert(2014)]{mcgaugh14} McGaugh, S. S., Schombert, J. M. 2014, \aj, 148, 77
\bibitem[Micha{\l}owski et al.(2014)]{michalowski14} Micha{\l}owski, M., Hayward, C.C., Dunlop, J.S. et al., 2014, A\&A, 571, 75
\bibitem[McConnell et al.(2016)]{McConnell16} McConnell, N. J., Lu, J. R., Mann, A. W., 2016, \apj, 821, 39
\bibitem[Naab \& Ostriker(2017)]{naab17} Naab, T., Ostriker, J.P. 2017, \araa, 55, 59
\bibitem[Noll et al.(2009)]{noll09} Noll, S., Burgarella, D., Giovannoli, E., et al. 2009, A\&A, 507, 1793
\bibitem[Ocvirk et al.(2006a)]{ocvirk06a} Ocvirk, P., Pichon, C., Lancon, A., Thiebaut, E. 2006a, \mnras, 365, 74
\bibitem[Ocvirk et al.(2006b)]{ocvirk06b} Ocvirk, P., Pichon, C., Lancon, A., et al. 2006b, \mnras, 365, 46
\bibitem[Offner(2016)]{Offner16} Offner, S. S. R., 2016, IAUS, 315, 73
\bibitem[Press et al.(1992)]{press92} Press, W. H., Teukolsky, S. A., Vetterling, W. T., Flannery, B. P., 1992, Numerical Recipes in FORTRAN. The Art of Scientific Computing. Cambridge Univ. Press, Cambridge
\bibitem[Papovich et al.(2001)]{papovich01} Papovich, C., Dickinson, M., Ferguson, H.C., 2001, \apj, 559, 620
\bibitem[Panter et al.(2007)]{panter07} Panter, B., Jimenez, R., Heavens, A. F., Charlot, S., 2007, \mnras, 378, 1550
\bibitem[Portinari et al.(2004)]{portinari04} Portinari, L., Sommer-Larsen, J., Tantalo, R. 2004, \mnras, 347, 691
\bibitem[Pietrinferni et al.(2004)]{pietrinferni04} Pietrinferni, A., Cassisi, S., Salaris, M., Castelli, F., 2004, \apj, 612, 168
\bibitem[Pozzetti et al.(2007)]{pozzetti07} Pozzetti, L., Bolzonella, M., Lamareille, F. et al., 2007, A\&A, 474, 443
\bibitem[Popescu et al.(2011)]{popescu11} Popescu, C.C., Tuffs, R.J., Dopita, M.A., et al. 2011, A\&A, 527, 109
\bibitem[Powalka et al.(2017)]{powalka17} Powalka, M., Lan{\c c}on, A., Puzia, T.~H., et al.\ 2017, \apj, 844, 104
\bibitem[Pacifici et al.(2012)]{pacifici12} Pacifici, C., Charlot, S., Blaizot, J., Brinchmann, J. 2012, \mnras, 421, 2002
\bibitem[Pforr et al.(2012)]{pforr12} Pforr, J., Maraston, C., Tonini, C., 2012, \mnras, 422, 3285
\bibitem[Pforr et al.(2013)]{pforr13} Pforr, J., Maraston, C., Tonini, C., 2013, \mnras, 435, 1389
\bibitem[Peacock et al.(2014)]{Peacock14} Peacock, M. B., Zepf, S. E., Maccarone, T. J., et al. 2014, \apj, 784, 162
\bibitem[Rich et al.(1997)]{rich97} Rich, R. M., Sosin, C., Djorgovski, S., et al. 1997, \apj, 484, L25
\bibitem[Reddy et al.(2015)]{reddy15} Reddy, N.A., Kriek, M., Shapley, A. et al., 2015, \apj, 806, 259
\bibitem[Roediger \& Courteau(2015)]{roediger15} Roediger, J. C., Courteau, S., 2015, \mnras, 452, 3209
\bibitem[Scalo(1986)]{scalo86} Scalo, J. M. 1986, Fundam. Cosm. Phys. 11, 1
\bibitem[Sawicki \& Yee(1998)]{sawicki98} Sawicki, M., Yee, H.K.C., 1998, \aj, 115, 1329
\bibitem[Silva et al.(1998)]{silva98} Silva, L, Granato, G. L., Bressan, A., Danese, L. 1998, \apj, 509, 103 
\bibitem[Schaye et al.(2015)]{schaye15} Schaye, J., Crain, R. A., Bower, R. G., et al., 2015, \mnras, 446, 521
\bibitem[Smith et al.(2015)]{Smith15} Smith, R. J., Lucey, J. R., Conroy, C. 2015, \mnras, 449, 3441
\bibitem[Salim et al.(2016)]{salim16} Salim, S., Lee, J. C., Janowiecki, S., et al. 2016, \apjs, 227, 2
\bibitem[Tojeiro et al.(2007)]{tojeiro07} Tojeiro, R., Heavens, A. F., Jimenez, R., et al. 2007, \mnras, 381, 1252
\bibitem[Tolstoy et al.(2009)]{tolstoy09}Tolstoy, E., Hill, V., Tosi, M. et al., 2009, ARAA, 47, 371
\bibitem[Taylor et al.(2011)]{taylor11} Taylor, E. N., Hopkins, A. M., Baldry, I. K., et al. 2011, \mnras, 418, 1587
\bibitem[Vogelsberger et al.(2014)]{vogelsberger14} Vogelsberger, M., Genel, S.; Springel, V., et al., 2014, Nature, 509, 177
\bibitem[van den Bergh(1999)]{van99} van den Bergh, S., 1999, A\&ARv, 9, 273
\bibitem[van Dokkum \& Conroy(2012)]{van12} van Dokkum, P. G., Conroy, C., 2012, \apj, 760, 70 
\bibitem[Worthey(1994)]{worthey94} Worthey, G., 1994, ApJS, 95, 107
\bibitem[Worthey et al.(1994)]{worthey94} Worthey, G., Faber, S.M., Gonzalez, J.J., Burstein, D. 1994, \apjs, 94, 687
%%Wuyts, S. et al. 2007, ApJ, 655, 51
\bibitem[Wyse et al.(2002)]{Wyse02} Wyse, R. F. G., Gilmore, G., Houdashelt, M. L., et al. 2002, NewA, 7, 395
\bibitem[Wuyts et al.(2009)]{wuyts09} Wuyts, S., Franx, M., Cox, T.J. et al., 2009, \apj, 696, 348
\bibitem[Wuyts et al.(2011)]{wuyts11} Wuyts, S., Schreiber, N. M. F., Lutz, D., et al. 2011, \apj, 738, 106
\bibitem[Wright et al.(2010)]{wright10} Wright, E.L., Eisenhardt, P. R. M., Mainzer, A. K., et al., 2010, \aj, 140, 1868
\bibitem[Weisz et al.(2011)]{weisz11} Weisz, D. R., Dalcanton, J. J., Williams, B. F. et al., 2011, \apj, 739, 5
\bibitem[Weisz et al.(2014)]{weisz14} Weisz, D. R., Dolphin, A. E., Skillman, E. D. et al., 2014, \apj, 789, 148 (W14)
\bibitem[Wild et al.(2011)]{wild11} Wild, V., Charlot, S., Brinchmann, J. et al., 2011, \mnras, 417, 1760
\bibitem[Walcher et al.(2011)]{walcher11} Walcher, J., Groves, B., Budav\'ari, T., et al. 2011, Ap\&SS, 331, 1
\bibitem[Xilouris et al.(1999)]{xilouris99} Xilouris, E. M., Byun, Y.I., Kylafis, N. D., et al. 1999, A\&A, 344, 868
\bibitem[York et al.(2000)]{york00}York, D. G., Adelman, J., Anderson, J. E. et al., 2000, \aj, 120, 1579
\bibitem[Yoachim \& Dalcanton(2006)]{yoachim06} Yoachim, P., Dalcanton, J. J. 2006, \aj, 131, 226
\bibitem[Zibetti et al.(2009)]{zibetti09}Zibetti, S., Charlot, S., Rix, H.-W., 2009, \mnras, 400, 1181
\bibitem[Zhao et al.(2011)]{zhao11} Zhao, Y., Gu, Q., Gao, Y., 2011, \aj, 141, 68
\bibitem[Zhang et al.(2012)]{zhang12} Zhang, H.-X., Hunter, D. A., Elmegreen, B. G., Gao, Y., Schruba, A., 2012, \aj, 143, 47
\end{thebibliography}
\end{document}